\documentclass[journal,final,twoside]{IEEEtran}
\usepackage[font=small]{caption}
\usepackage{subcaption}

\usepackage{array,amsfonts,amssymb,bm,paralist}
\usepackage[cmex10]{amsmath}
\usepackage{graphicx}
\graphicspath{{./Figures/}{./Figures/CRB/}}

\usepackage{cite}
\usepackage[normalem]{ulem} 
\usepackage{textcomp}
\usepackage[colorlinks=true, allcolors=blue]{hyperref}
\usepackage[inline]{enumitem}
\usepackage{cleveref}
\usepackage{mathtools}
\usepackage[usenames,dvipsnames]{xcolor}

\Crefname{myThm}{Theorem}{Theorems}

\Crefname{myLem}{Lemma}{Lemmas}

\Crefname{myExample}{Example}{Examples}

\Crefname{myProp}{Proposition}{Propositions}
\newtheorem{myRem}{Remark}

\Crefname{figure}{Fig.}{Figs.}
\Crefname{tabular}{Tab.}{Tabs.}

\def\area{\mathop{\mathrm{area}}}
\def\argmin{\mathop{\mathrm{arg\,min}}}

\def\diag{\mathop{\mathrm{diag}}}
\newcommand{\norm}[1]{\left\lVert #1\right\rVert}

\def\CR{Cram{\'e}r--Rao}
\def\CRB{Cram{\'e}r--Rao bound}

\mathchardef\mhyphen="2D
\def\nocc{\rm no\mhyphen EO}
\def\CRBcc{\mathop{\mathrm{CRB}_{\rm EO}}}
\def\CRBnocc{\mathop{\mathrm{CRB}_{\nocc}}}

\newcommand{\cs}{c_{\rm s}}
\newcommand{\Li}{L_{\rm i}}
\newcommand{\Lo}{L_{\rm o}}
\newcommand{\Nt}{N_{\rm t}}
\newcommand{\phis}{\phi_{\rm s}}
\newcommand{\rhos}{\rho_{\rm s}}
\newcommand{\rhoh}{\rho_{\rm h}}
\newcommand{\rhovec}{\bm{\rho}}
\newcommand{\rhovech}{\bm{\rho}_{\rm h}}

\newcommand{\Sh}{S_{\rm h}}
\newcommand{\Shbar}{\bar{S}_{\rm h}}
\newcommand{\sh}{{s}_{\rm h}}

\newcommand{\avec}{\mathbf{a}}
\newcommand{\cvec}{\mathbf{c}}

\newcommand{\fvec}{\mathbf{f}}

\newcommand{\svec}{\mathbf{s}}
\newcommand{\svech}{\svec_{\rm h}}

\newcommand{\yvec}{\mathbf{y}}

\newcommand{\Amat}{\mathbf{A}}
\newcommand{\Bmat}{\mathbf{B}}

\newcommand{\Dmat}{\mathbf{D}}

\newcommand{\Fmat}{\mathbf{F}}

\newcommand{\Imat}{\mathbf{I}}

\newcommand{\Vmat}{\mathbf{V}}
\newcommand{\Wmat}{\mathbf{W}}
\newcommand{\zvec}{\mathbf{z}}

\newcommand{\overbar}[1]{\mkern 1.5mu\overline{\mkern-1.5mu#1\mkern-2mu}\mkern 2mu}
\newcommand{\overbarbm}[1]{\mkern 2.75mu\overline{\mkern-2.75mu#1\mkern 0.5mu}\mkern-0.5mu}

\newcommand{\Dmatbar}{\overbar{\mathbf{D}}}
\newcommand{\rhobarbm}{\overbarbm{\bm{\rho}}}

\newcommand{\svecbar}{\bar{\svec}}
\newcommand{\svecbarh}{\bar{\svec}_{\rm h}}

\def\transpose{^{\!\mathsf{T}}}

\newcommand{\defeq}{\stackrel{\text{\tiny{def}}}{=}}

\newcommand{\Frac}[2]{{#1}/{#2}}

\begin{document}
\title{Two-Dimensional Non-Line-of-Sight Scene Estimation from a Single Edge Occluder}
\author{Sheila~W.~Seidel,
        John Murray-Bruce,
        Yanting Ma,
        Christopher Yu,
        William T. Freeman,
        and~Vivek K Goyal

\thanks{This work was supported in part by
a Draper Fellowship,
by the Defense Advanced Research Projects Agency REVEAL Program under contract number HR0011-16-C-0030,
and by the US National Science Foundation under Grant 1815896.}
\thanks{S. W. Seidel and V. K. Goyal  are with the Department of Electrical and Computer Engineering, Boston University, Boston, MA 02215 USA (sseidel@bu.edu; goyal@bu.edu).}
\thanks{J. Murray-Bruce is with the Department of Computer Science and Engineering, University of South Florida, Tampa, FL 33620 USA (murraybruce@usf.edu).}
\thanks{Y. Ma is with Mitsubishi Electric Research Laboratories, Cambridge, MA 02139 USA (yma@merl.com).}
\thanks{C. Yu is with Charles Stark Draper Laboratory, Cambridge, MA 02139 USA (cyu@draper.com).}
\thanks{W. T. Freeman is with the Department of Electrical Engineering and Computer Science, Massachusetts Institute of Technology, Cambridge, MA 02139 USA (billf@mit.edu).}
}

\markright{Seidel \MakeLowercase{\textit{et al.}}: NLOS Scene Estimation}

\maketitle

\begin{abstract}
Passive non-line-of-sight imaging methods are often faster and stealthier than their active counterparts, requiring less complex and costly equipment.
However, many of these methods exploit motion of an occluder or the hidden scene, or require knowledge or calibration of complicated occluders.
The edge of a wall is a known and ubiquitous occluding structure that may be used as an aperture to image the region hidden behind it.
Light from around the corner is cast onto the floor forming a fan-like
penumbra rather than a sharp
shadow.
Subtle variations in the penumbra contain a remarkable amount of information about the hidden scene.
Previous work has leveraged the vertical nature of the edge to demonstrate 1D (in angle measured around the corner) reconstructions of moving and
stationary
hidden scenery from as little as a single photograph of the penumbra.
In this work, we introduce a second reconstruction dimension: range measured from the edge.
We derive a new forward model, accounting for radial falloff, and propose two inversion algorithms to form 2D reconstructions from a single photograph of the penumbra.
Performances of both algorithms are demonstrated on experimental data corresponding to several different hidden scene configurations.
A \CR~bound analysis further demonstrates the feasibility (and utility) of the 2D corner camera.
\end{abstract}

\begin{IEEEkeywords}
corner camera, non-line-of-sight imaging, computational photography, remote sensing, computer vision.
\end{IEEEkeywords}

\IEEEpeerreviewmaketitle

\section{Introduction}
\label{sec:Intro}
\IEEEPARstart{T}{he}
ability to form non-line-of-sight (NLOS) images would be useful in a variety of situations.
Current NLOS imaging methods may be \textit{active}, based predominantly on the transient imaging framework first proposed in~\cite{Kirmani2012,Velten2012}
and requiring control of hidden scene illumination, or \textit{passive}, where only light sources already present are used.
The earliest active NLOS imaging systems combined a femtosecond laser with a 2 picosecond resolution streak camera~\cite{Velten2012,Gupta2012}; newer systems using single-photon avalanche diode (SPAD) detectors and time-correlated single photon counting (TCSPC) modules provide a less expensive alternative. These systems have been used extensively for both line of sight imaging~\cite{Kirmani2014,Shin2015,Rapp2017} and NLOS applications~\cite{Xu2013,Laurenzis2015,Laurenzis2015a,Buttafava2015,Gariepy2016,Klein2016a,Chan2017,Tsai2017a,HeideOZLDW:19,OToole2018,Liu2019_pField,Lindell2019_Wavebased,Musarra_2019}. Recently, SPAD-based NLOS imaging systems have demonstrated faster processing using confocal scanning~\cite{OToole2018}, reconstruction algorithms based on wave properties~\cite{Liu2019_pField,Lindell2019_Wavebased}, and color reconstructions using multiple wavelengths of illumination~\cite{Musarra_2019}.  

\begin{figure}
\centering
\includegraphics[width=.7\linewidth]{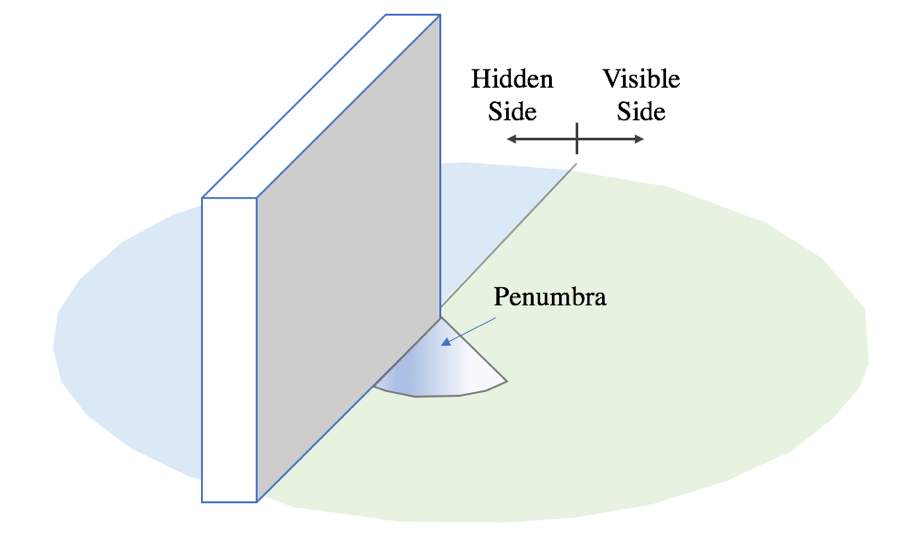}
\caption{Light from the hidden side is cast onto the floor on the visible side of the occluding edge.}
\label{fig:penumbra}
\end{figure}

Compared to active methods, passive NLOS imaging techniques may be less expensive and stealthier, with lower power requirements and faster data acquisition times. These passive methods leverage occluding structures and light sources already present in the environment~\cite{Thrampoulidis2018_exploiting}.
Useful structures may be the aperture formed by a partially open window or door, or the `accidental pinhole' formed when a once present object is moved~\cite{Cohen1982,Torralba2014}.
Using an ordinary digital camera, Saunders \textit{et al.} formed NLOS color reconstructions when the form of the occluder was known~\cite{Saunders2019computational}.
Other methods use the motion of the hidden scene to discern the shape of an unknown occluder~\cite{Yedidia2019_using}, or deep matrix factorization to simultaneously reconstruct an unknown hidden scene and occluder~\cite{Aittala2019ComputationalMB}.
Unlike other occluders used in NLOS imaging systems~\cite{Saunders2019computational,Murray-Bruce2019_Occlusion,Yedidia2019_using, Aittala2019ComputationalMB}, a wall edge has a known shape and is ubiquitous.
In this case, light is cast onto the visible floor around the occluding edge forming a \emph{penumbra}, as shown in \Cref{fig:penumbra}.
Photographs of the penumbra may be used to produce angularly resolved reconstructions of the hidden scene.
This was first shown in~\cite{Bouman2017}, where smoothed differences between  consecutive video frames were used to form one-dimensional reconstructions of hidden objects in \emph{motion}.
Our previous work demonstrated 1D reconstruction of both moving and stationary hidden scene components from a single photograph, while simultaneously estimating unknown nonuniform floor albedo~\cite{Seidel2019_corner}.

In this paper, we explore the addition of a second dimension: \emph{range}.
Although the corner induces high angular resolution, it indiscriminately passes light from all different ranges.
Instead, our coarser range resolution arises from $\Frac{1}{r^2}$ intensity falloff observed across the measured photograph.
The challenge of reconstructing scene range becomes more tractable when the scene is composed of a few objects, each with a single unknown range.
In this work, we exploit the high angular resolution provided by the corner to form an initial estimate of the scene as a function of angle, which allows us to count the few objects contained in the hidden space.
Our inversion algorithm alternates between estimating a range for each hidden object and updating the angular estimate of the hidden scene to ultimately form a 2D plan-view reconstruction of the hidden scene.

The edge occluder may be better understood by first considering some well-known occluders.
For example, a pinhole opposite a vertical flat plane maps light from each incident direction to a unique position on the observation plane.
In this case, direction (i.e., azimuth and elevation angles) of incident light is completely recoverable, but range of origin is not.
When the occluder is a vertical slit opposite a flat plane, a slice of the 3D world is mapped to a line on the observation plane.
Here, the azimuthal angle of incident light is well-conditioned for recovery.
Although very challenging, recovery of higher-dimensional information is not impossible due to path length differences between different points on the line in the observation plane.

The edge occluder may be thought of as `half' of a slit occluder, with an observation plane (i.e., the floor) that is perpendicular, rather than parallel, to the the occluding edge.
With the observation plane oriented in this way, path length differences for targets near the ground plane at different ranges become more pronounced for hidden objects resting on the floor.
Unlike a vertical slit, a vertical edge integrates incident light on the observation plane from all unoccluded directions, meaning a single hidden point source may affect a multitude of pixels in the observation plane.
The radial falloff pattern across these affected pixels emanates out from the hidden source rather than from the edge,
a difference that becomes more pronounced for targets in the near-field.

In this work, we leverage these small variations in the measurement to add a second reconstruction dimension.
Our key contributions include:
\begin{itemize}
    \item A new forward model that describes a single photograph as a combination of light originating from the hidden scene and unknown scene depth (Section~\ref{sec:ForwardModel}).
    \item \CRB~(CRB) analysis (Section~\ref{sec:CRB}) to demonstrate the limits of exploiting measurement of visible penumbrae for 2D hidden scene reconstruction.
    Our analysis shows that while range estimation is possible, it is inherently difficult relative to angle estimation.
    \item Two different inversion algorithms, proposed in Section~\ref{sec:Inverse}). 
    \item Experimental demonstration of our 2D reconstruction algorithm on a variety of colored hidden scenes (Section~\ref{sec:experimental_results}).
\end{itemize}

\section{Forward Model}
\label{sec:ForwardModel}

\subsection{Light Transport}
\begin{figure}
\centering
\includegraphics[width=1.0\linewidth]{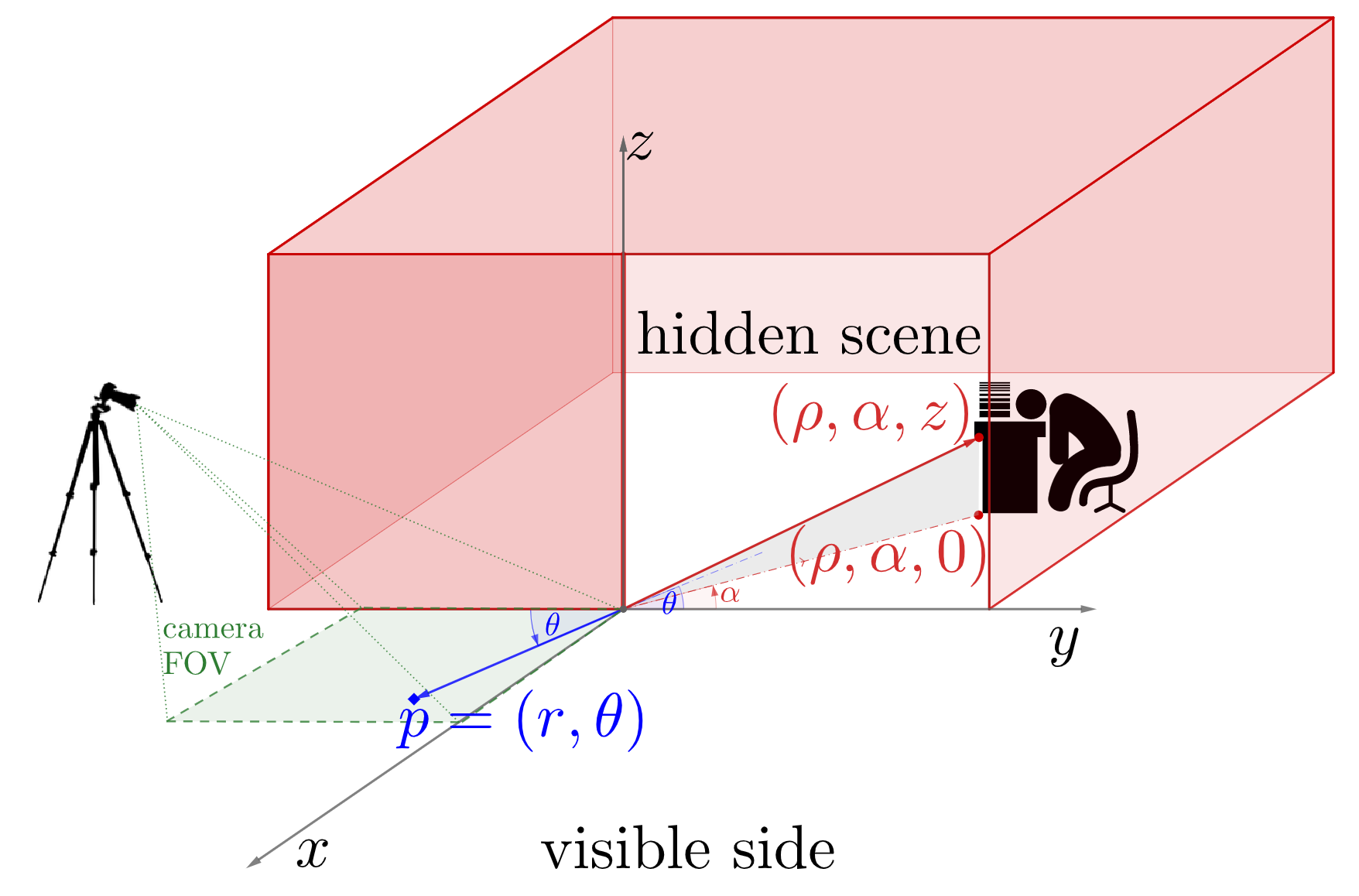}
\caption{Acquisition setup and depiction of problem geometry. A point $p$ in the camera's field of view is represented in polar coordinates, while the 3D hidden scene is represented in cylindrical coordinates.}
\label{fig:Geo}
\end{figure}

Consider the NLOS imaging scenario in \Cref{fig:Geo}, where a distressed researcher works in the hidden scene.
We parameterize the hidden scene in cylindrical coordinates with range $\rho$, angle $\alpha$, and height $z$.
A point $\mathbf{p} = (r,\theta)$ on the floor in the camera field of view is described by its range $r$ and angle measured from the wall $\theta$.
Assuming the camera looks straight down at a Lambertian floor, and the effects of all forshortening terms are negligible,  the radiosity $\Lo(\mathbf{p})$ of point $ \mathbf{p}$ on the floor is the albedo at point $\mathbf{p}$, $f(\mathbf{p})$, multiplied by the sum of all incident light:
\begin{equation}
    \Lo(\mathbf{p}) = f(\mathbf{p})\left( L_{\rm v} (\mathbf{p}) + L_{\rm h}(\mathbf{p}) \right),
    \label{eq:Lo_polar}
\end{equation}
where $L_{\rm v}(\mathbf{p})$  is the incident light originating from the visible side, and $L_{\rm h}(\mathbf{p})$ is the incident light originating from the hidden side.

The measured photograph is an array of size $M_x \times M_y$, with $M = M_x M_y$ total pixels.
The measurement $y_m$ of camera pixel $m$ is proportional to the total radiosity of floor patch $\mathcal{P}_m$, which consists of all points $\mathbf{p}$ on the measurement plane that are focused on camera pixel $m$.
Thus,
\begin{align*}
    y_m &\propto  \kappa \int_{\mathbf{p} \in \mathcal{P}_m} \Lo(\mathbf{p}) \, \mathrm{d}\mathbf{p},
\end{align*}
where $\kappa$ is the constant of proportionality
associated with various camera scale factors---e.g., shutter speed and gain control---that lead to (dimensionless) pixel values.
Because all camera pixels have equal projected area $\kappa_{\rm cam} \defeq {\area \left({\mathcal{P}_m}\right)}$ on the measurement plane, using \eqref{eq:Lo_polar}, we can write
\begin{align}
    y_m &\approx \kappa_{\rm cam} \kappa  \Lo(r_m, \theta_m) \nonumber \\
        &= \kappa_{\rm cam} \kappa f(r_m,\theta_m)\left( L_{\rm v} (r_m,\theta_m) + L_{\rm h}(r_m,\theta_m) \right),
    \label{eq:Meas_scaled}
\end{align}
where $(r_m,\theta_m)$ is the center of floor patch $\mathcal{P}_m$.

By adopting a cylindrical coordinate parameterization of the hidden scene, the hidden scene contribution $L_{\rm h}(r,\theta)$ becomes
\begin{align}
    L_{\rm h}(r,\theta) &= \int_{0}^{\theta} \!\!\! \int_{0}^{\infty} \!\!\! \int_{0}^{\infty} \!
                            \Li(\rho, \alpha, z) \rho  \, \mathrm{d}z \, \mathrm{d}\rho \, \mathrm{d}\alpha \nonumber \\
                        &= \int_{0}^{\theta} \!\!\! \int_{0}^{\infty} \!\!\! \int_{0}^{\infty} \!
                            \frac{\Sh(\rho, \alpha, z)}{d^2 + z^2} \rho  \, \mathrm{d}z \, \mathrm{d}\rho \, \mathrm{d}\alpha,
    \label{eq:Lh_cylindrical}
\end{align}
where $\Sh(\rho, \alpha, z)$ is the radiosity of a hidden scene location $(\rho, \alpha, z)$, and
\begin{equation}
\label{eq:distance-mapping}
  d^2(r,\theta,\rho,\alpha) = r^2 + \rho^2 - 2 r\rho\cos\!\left(\pi - \theta + \alpha\right)
\end{equation}
is the distance between point $p$ on the visible floor and a hidden scene (floor) point $(\rho, \alpha, 0)$.
While a fully 3D recovery of the hidden scene from a single 2D digital photograph of a visible floor surface is hopelessly ill-conditioned, the presence of the vertical edge occluding our view of the hidden scene can be exploited to faithfully recover a 2D (plan view) representation of the hidden scene. To this end, we rewrite \eqref{eq:Lh_cylindrical} as
\begin{align}
    L_{\rm h}(r,\theta) &{=} \int_{0}^{\theta} \!\!\! \int_{0}^{\infty}   \frac{\rho}{d^2}  \left( \int_{0}^{\infty} \!
                            \frac{\Sh(\rho, \alpha, z)}{1 + \left(\Frac{z}{d}\right)^2} \, \mathrm{d}z \right) \mathrm{d}\rho \, \mathrm{d}\alpha \nonumber\\
                        &{=} \int_{0}^{\theta} \!\!\! \int_{0}^{\infty}   \frac{\rho}{d^2} \, \Shbar(\rho,\alpha)  \mathrm{d}\rho \, \mathrm{d}\alpha,
    \label{eq:Lh_cyl1}
\end{align}
where
\begin{equation}
  \Shbar(\rho,\alpha) \defeq \int_{0}^{\infty} \frac{\Sh(\rho, \alpha, z)}{1 + \left( \Frac{z}{d} \right)^2} \mathrm{d}z
\end{equation}
is the unknown \textit{height-adjusted 2D radiosity} of the hidden scene. This is the plan view that we ultimately seek to reconstruct in Section~\ref{sec:Inverse}.

To further model occlusion in the hidden scene, we assume that all the contributions to our measurement from a given angle $\alpha$ come from a single range $\rho$.
This roughly corresponds to a hidden scene composed of opaque vertical objects resting on the ground.
Under this assumption, we write
$\Shbar(\rho,\alpha) = \delta(\rho - \rhoh(\alpha))\sh(\alpha)$
as a separable function of range
$\rhoh(\alpha) \geq 0$
and angle $\alpha \in (0, \pi]$,
where $\sh(\alpha)$ denotes the dependence of scene radiosity on $\alpha$,
and $\delta(\cdot)$ is the Dirac delta function.
Then~\eqref{eq:Lh_cyl1} becomes
\begin{align}
    L_{\rm h}(r,\theta)
        &= \!\int_{0}^{\theta} \!\!\! \int_{0}^{\infty}   \frac{\rho}{d^2} \delta(\rho - \rhoh(\alpha))\sh(\alpha)  \, \mathrm{d}\rho \, \mathrm{d}\alpha \nonumber \\
        &= \!\int_{0}^{\theta} \!\!\! \frac{\rhoh(\alpha) \sh(\alpha) }{d^2} \, \mathrm{d}\alpha. \nonumber
\end{align}
Thus, substituting
\begin{align}
    L_{\rm h}(r_m,\theta_m) &= \int_{0}^{\theta_m} \!\!\! \frac{\rhoh(\alpha) }{d^2(r_m,\theta_m,\rhoh(\alpha),\alpha)} \sh(\alpha)  \, \mathrm{d}\alpha 
    \label{eq:Lh_pm}
\end{align}
into~\eqref{eq:Meas_scaled} and assuming $\kappa_{\rm cam}\kappa = 1$,\footnote{This is without loss of generality because we are not attempting to estimating a physically meaningful overall scaling factor for the hidden scene radiosity.} we obtain the model 
\begin{align}
    y_m \approx f(&r_m,\theta_m) \Bigg( L_{\rm v}(r_m,\theta_m) \nonumber \\
    &+ \int_{0}^{\theta_m} \!\!\! \frac{\rhoh(\alpha) }{d^2(r_m,\theta_m,\rhoh(\alpha),\alpha)} \sh(\alpha)\, \mathrm{d}\alpha \Bigg)
    \label{eq:Meas}
\end{align}
for the hidden scene and visible scene contributions to camera measurement $m$.

\subsection{Discrete Forward Model}
\label{sec:discrete-model}
We discretize the hidden region into $N$ equiangular wedges identified by the angles
$\{\alpha_n\}_{n=1}^N \subset  (0, \Frac{\pi}{2})$,
and associate a single unknown range value $\rhoh(\alpha_n)$ with each wedge.
Then the pair $(\rhoh(\alpha_n), \alpha_n)$ defines a (unique) position in the hidden space for each $n=1,\ldots,N$. Now gathering these variables into the hidden-scene radiosity vector
$\svech = [\sh(\alpha_1), \sh(\alpha_2), \ldots, \sh(\alpha_N) ]\transpose$
and range vector $\rhovech = [\rhoh(\alpha_1), \rhoh(\alpha_2), \ldots, \rhoh(\alpha_N) ]\transpose$
gives the discrete, nonlinear forward model
\begin{equation}
    \yvec = \avec \odot \fvec + \fvec \odot (\Vmat \odot \Dmat(\rhovech))\svech + \bm{\epsilon},
    \label{eq:discreteforwardmodel_nonlinear}
\end{equation}
where
$\yvec = [y_1, y_2, \ldots, y_m] \in \mathbb{R}^M$ denotes the vectorized camera photograph,
$\avec \in \mathbb{R}^M$ is the discretization of ambient light contribution $L_{\rm v}$,
$\fvec \in \mathbb{R}^M$ is the floor albedo,
$\Vmat \in \mathbb{R}^{M \times N}$ is a binary-valued visibility matrix (with the entry $[\Vmat]_{m,n}$ equalling 0 if the path joining $p_m$ and $(\rho(\alpha_n), \alpha_n)$ is occluded by the wall, otherwise it is equal to 1).
The matrix $\Dmat(\rhovech) \in \mathbb{R}^{M \times N}$ has elements $$[\Dmat(\rhovech)]_{m,n} = \frac{\rhoh(\alpha_n)}{d^2(r_m,\theta_m,\rhoh(\alpha_n),\alpha_n)},$$
and $\bm{\epsilon}$ models the effect of noise and other possible model mismatch.

\textit{Inverse Problem:} Our goal is to recover a 2D (plan view) reconstruction $(\svech,\rhovech)$ of a hidden scene $\Sh$ from a single photograph $\yvec$ of the penumbra created on a visible floor surface using \eqref{eq:discreteforwardmodel_nonlinear}.

Before presenting our approaches for solving \eqref{eq:discreteforwardmodel_nonlinear},
we study the feasibility (and certain limits) of realizing the 2D corner camera.
Specifically, by evaluating the CRBs for hypothetical cases where the hidden scene comprises only a few hidden point targets, we demonstrate the merits of the occluding wall (or corner occluder) for hidden scene recovery.

\section{\CR~Bound for Hidden Target Estimation}
\label{sec:CRB}
In the subsections that follow, we present CRB analysis to demonstrate the merit and challenge of an edge occluder for 2D plan-view reconstruction of a hidden scene. In order to truly understand the effect of the edge, we perform our analysis both for the edge occluder scenario, and the scenario where no edge is in place. We start with the former. 

In \eqref{eq:Lh_pm}, measurement $y_m$, with the edge in place, is approximated by the intensity  at the center of the pixel. Now, we leave the more precise integral across floor patch $\mathcal{P}_m$ in place and assume no ambient light contributions, i.e. $L_{\rm v} = 0$. Under an additive white Gaussian noise (AWGN) model, the noisy camera measurement is given by

\begin{equation}
    \begin{split}
        y_m 
            =& \int_{\mathcal{P}_m} \int_{0}^{\theta_m} \!\!\! \frac{\rhoh(\alpha) }{d^2(r,\theta,\rhoh(\alpha),\alpha)} \sh(\alpha)\, \mathrm{d}\alpha \, \mathrm{d}p
            + \epsilon,
    \end{split}
    \label{eq:NoisyMeas1}
\end{equation}
where $\epsilon \sim \mathcal{N}(0, \sigma^2)$.

\subsection{Single Hidden Target}
\label{ssec:CRB_pointtarget}
Assume the hidden target is a hypothetical point emitter, located at the point $(\rho_{\rm s}, \phi_{\rm s},0)$ on the ground, i.e.
$\Sh(\rho,\alpha,z) = c_{\rm s} \delta(\rho - \rho_{\rm s} )\delta(\alpha - \phis)\delta(z)$, where $\phis \in (0,\pi]$. Evaluating \eqref{eq:Lh_cylindrical}, the
outgoing radiosity from a point $p = (r,\theta)$ is
\begin{equation}
    \Lo(p) = f(p) \frac{c_s H(\theta - \phis)}{r^2 + \rhos^2 - 2r\rho_s\cos(\phis + \pi - \theta)},
\label{eq:Lop_PointTarget}
\end{equation}
where $H(x)$ is the Heaviside step function.

Assuming a uniform albedo $f(p) =1$, and $\kappa = 1$ without loss of generality, the measurement at pixel $m$ is $y_m = i_m + \kappa_{\rm cam}\epsilon$, with
\begin{equation}
    i_m = \int_{p \in \mathcal{P}_m} \frac{c_s H( \theta - \phis)}{r^2 + \rhos^2 - 2r\rho_s\cos(\phis + \pi - \theta)} \, \mathrm{d}p.
    \label{eq:MeasurementPtSource}
\end{equation}
Then
\begin{align}
\label{eq:di_dc}
    \frac{\partial i_m}{\partial \cs} &= \int_{p \in \mathcal{P}_m} \frac{H(\theta - \phis)}{r^2 + \rhos^2 - 2r\rho_s\cos(\phis + \pi - \theta)} \, \mathrm{d}p, \\
\label{eq:di_drho}
    \frac{\partial i_m}{\partial \rhos} &= -2\cs \int_{p \in \mathcal{P}_m} \frac{(\rhos - r\cos(\pi-\theta +\phis)) H(\theta - \phis)}{r^2 + \rhos^2 - 2r\rho_s\cos(\phis + \pi - \theta)} \, \mathrm{d}p,
\end{align}
and
\begin{equation}
\begin{split}
    \frac{\partial i_m}{\partial \phis} = -\cs\int_{p \in \mathcal{P}_m} & \frac{ \delta(\theta - \phis)}{r^2 + \rhos^2 - 2r\rho_s\cos(\phis + \pi - \theta)} \\
    +& \frac{2 r \rhos \sin(\pi-\theta +\phis) H( \theta - \phis)}{(r^2 + \rhos^2 - 2r\rho_s\cos(\phis + \pi - \theta))^2} \mathrm{d}p.
\end{split}
\label{eq:di_dphi}
\end{equation}
Interchanging the integral and derivative is justified since the definite integral $i_m$ is finite. 
We define the following matrix:  
\begin{equation}
    \nabla \Imat = 
    \left[ \begin{array}{ccc}
        \frac{\partial i_1}{\partial \cs} & \frac{\partial i_1}{\partial \rhos} & \frac{\partial i_1}{\partial \phis} \\
        \frac{\partial i_2}{\partial \cs} & \frac{\partial i_2}{\partial \rhos} & \frac{\partial i_2}{\partial \phis} \\ 
        \vdots                            &    \vdots      &      \vdots    \\
        \frac{\partial i_M}{\partial \cs} & \frac{\partial i_2}{\partial \rhos} & \frac{\partial i_M}{\partial \phis}
    \end{array} \right]
    \label{eq:derivs_meas}
\end{equation}
and note that under our Gaussian model,  the Fisher information matrix for estimating $(\cs, \rhos, \phis)$ from the noisy measurements $\{y_m\}_{m=1}^M$ is given by
\begin{align}
    \Fmat &= \frac{1}{\sigma^2}\left(\nabla\Imat{\transpose} \nabla\Imat\right), \nonumber \\
          &= \frac{1}{\sigma^2} \left[ \begin{array}{ccc}
        \sum_{m} \left(\frac{\partial i_m}{\partial \cs}\right)^2 & \sum_m\frac{\partial i_m}{\partial \cs}\frac{\partial i_m}{\partial \rhos} & \sum_{m}\frac{\partial i_m}{\partial \cs}\frac{\partial i_m}{\partial \phis} \\
        \sum_m \frac{\partial i_m}{\partial \rhos}\frac{\partial i_m}{\partial \cs} & \sum_m \left(\frac{\partial i_m}{\partial \rhos}\right)^2 & \sum_{m}\frac{\partial i_m}{\partial \rhos}\frac{\partial i_m}{\partial \phis} \\
        \sum_m \frac{\partial i_m}{\partial \phis}\frac{\partial i_m}{\partial \cs} & \sum_m \frac{\partial i_m}{\partial \phis}\frac{\partial i_m}{\partial \rhos} & \sum_{m} \left(\frac{\partial i_m}{\partial \phis}\right)^2 
    \end{array} \right].
    \label{eq:FIM}
\end{align}
Therefore CRBs of the unknown parameters $\cs,\rhos,$ and $\phis$, respectively, follow from \eqref{eq:FIM}:
\begin{subequations}
\label{eq:CRBcc}
\begin{align}
 \CRBcc(\cs)   &= \sigma^2 [\Fmat^{-1}]_{1,1},\\
 \CRBcc(\rhos) &= \sigma^2 [\Fmat^{-1}]_{2,2},\\
 \CRBcc(\phis) &= \sigma^2 [\Fmat^{-1}]_{3,3},
\end{align}
\end{subequations}
where the EO subscript indicates that these CRB results are for the edge occluder scenario.

Without the occluding edge, corresponding CRBs
($\CRBnocc(\cs)$, $\CRBnocc(\rhos)$, and $\CRBnocc(\phis)$)
for estimating the same ``out-of-view'' target parameters follow similarly.
Without the occlusion described in \eqref{eq:NoisyMeas1}, the measurement by the $m$th camera pixel is
\begin{align*}
   y^{\nocc}_m &\defeq \int_{p \in \mathcal{P}_m} \frac{c_s}{r^2 + \rhos^2 - 2r\rho_s\cos(\phis + \theta)} \, \mathrm{d}p + \epsilon \\
               & = i_m^{\nocc} +  \epsilon.
\end{align*}
Using  the derivatives of $i_m^{\nocc}$ with respect to the hidden target's parameters: 
\begin{align*}
    \frac{\partial i^{\nocc}_m}{\partial \cs} &= \int_{p \in \mathcal{P}_m} \frac{1}{r^2 + \rhos^2 - 2r\rho_s\cos(\phis + \theta)} \, \mathrm{d}p, \\
    \frac{\partial i^{\nocc}_m}{\partial \rhos} &= -2\cs \int_{p \in \mathcal{P}_m} \frac{(\rhos - r\cos(\theta + \phis))}{r^2 + \rhos^2 - 2r\rho_s\cos(\phis + \theta)} \, \mathrm{d}p,
\end{align*}
and
\begin{equation*}
    \frac{\partial i^{\nocc}_m}{\partial \phis} = -\cs\int_{p \in \mathcal{P}_m} \frac{2 r \rhos \sin(\theta + \phis) }{r^2 + \rhos^2 - 2r\rho_s\cos(\phis + \theta)} \mathrm{d}p,
\label{eq:di_dphi_nocc}
\end{equation*}
the Fisher information matrix $\Fmat_{\nocc}$, along with CRBs ($\CRBnocc(\cs)$, $\CRBnocc(\rhos)$, and $\CRBnocc(\phis)$) may be computed using the approach outlined in (\ref{eq:derivs_meas}, \ref{eq:FIM}, \ref{eq:CRBcc}).

\begin{figure}
    \centering
    \begin{subfigure}{0.45\linewidth}
      \includegraphics[width=\linewidth]{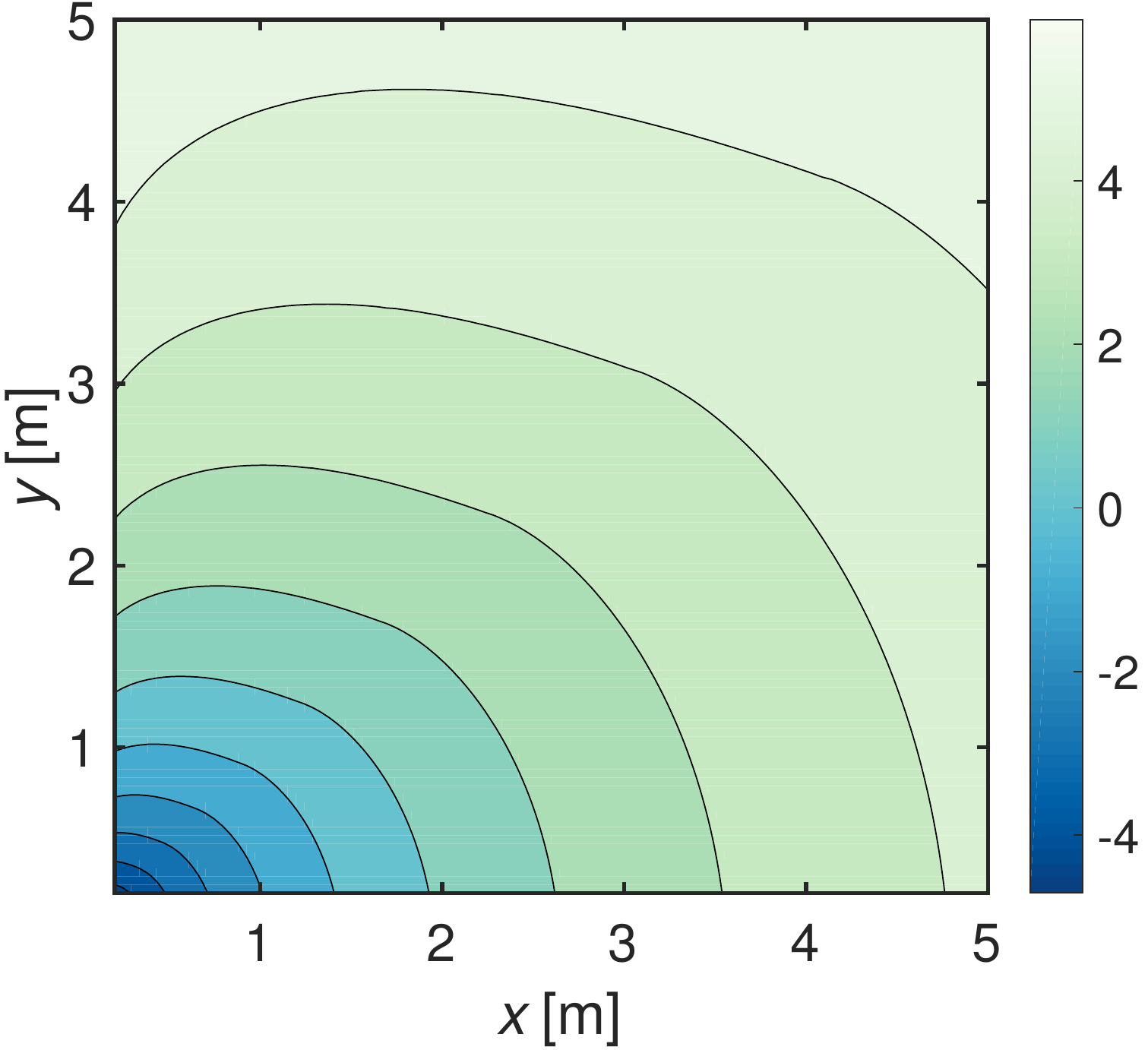}
      \captionsetup{justification=centering}
      \caption{\footnotesize $\log_{10}\mathrm{CRB_{EO}}(\rhos)$}
      \label{fig:CRBrho_CC_surf}
    \end{subfigure}
    \begin{subfigure}{0.45\linewidth}
      \includegraphics[width=\linewidth]{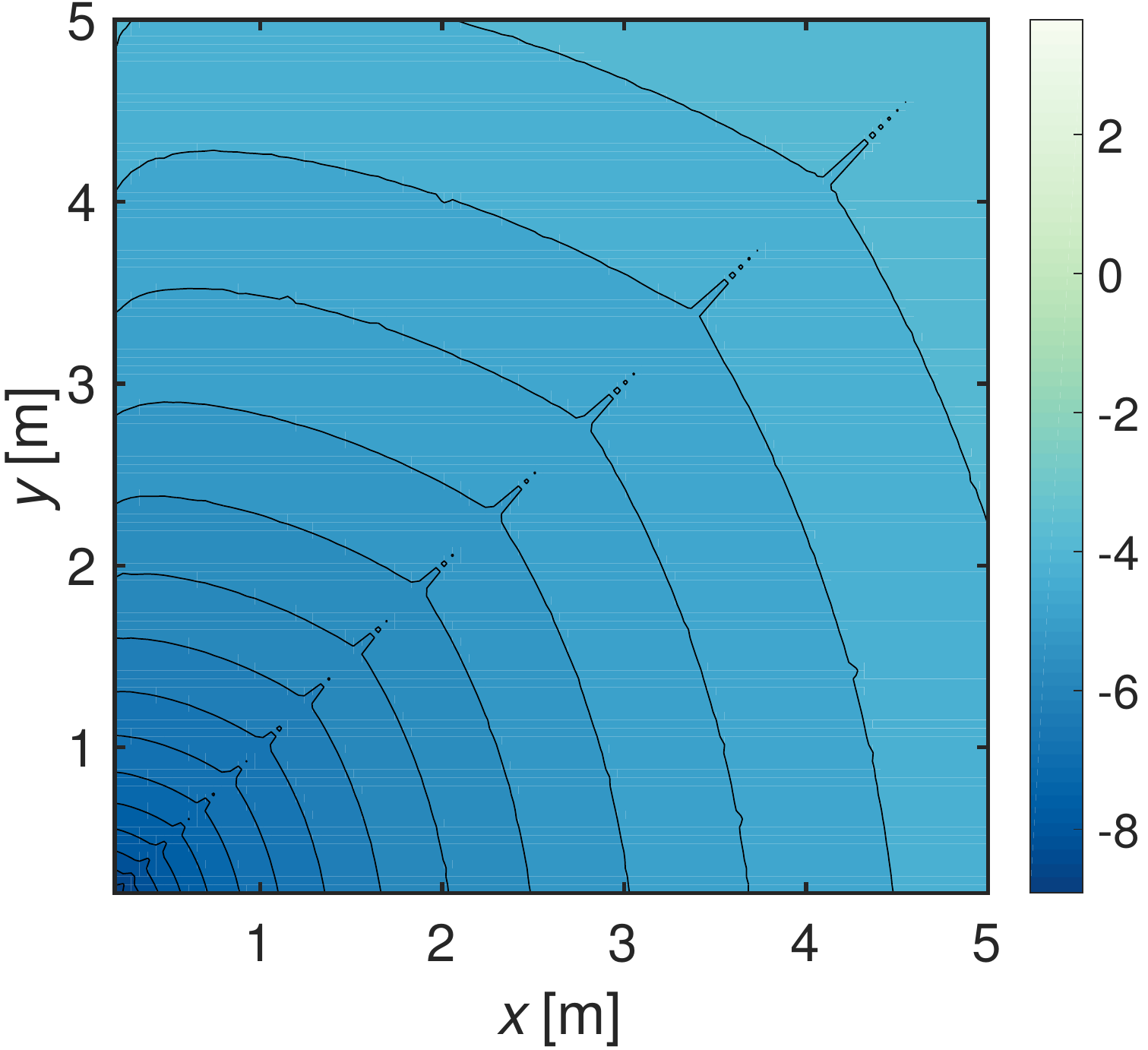}
      \captionsetup{justification=centering}
      \caption{\footnotesize $\log_{10}\mathrm{CRB_{EO}}(\phis)$}
      \label{fig:CRBphi_CC_surf}
    \end{subfigure} \\[3mm]
    \begin{subfigure}{0.45\linewidth}
      \includegraphics[width=\linewidth]{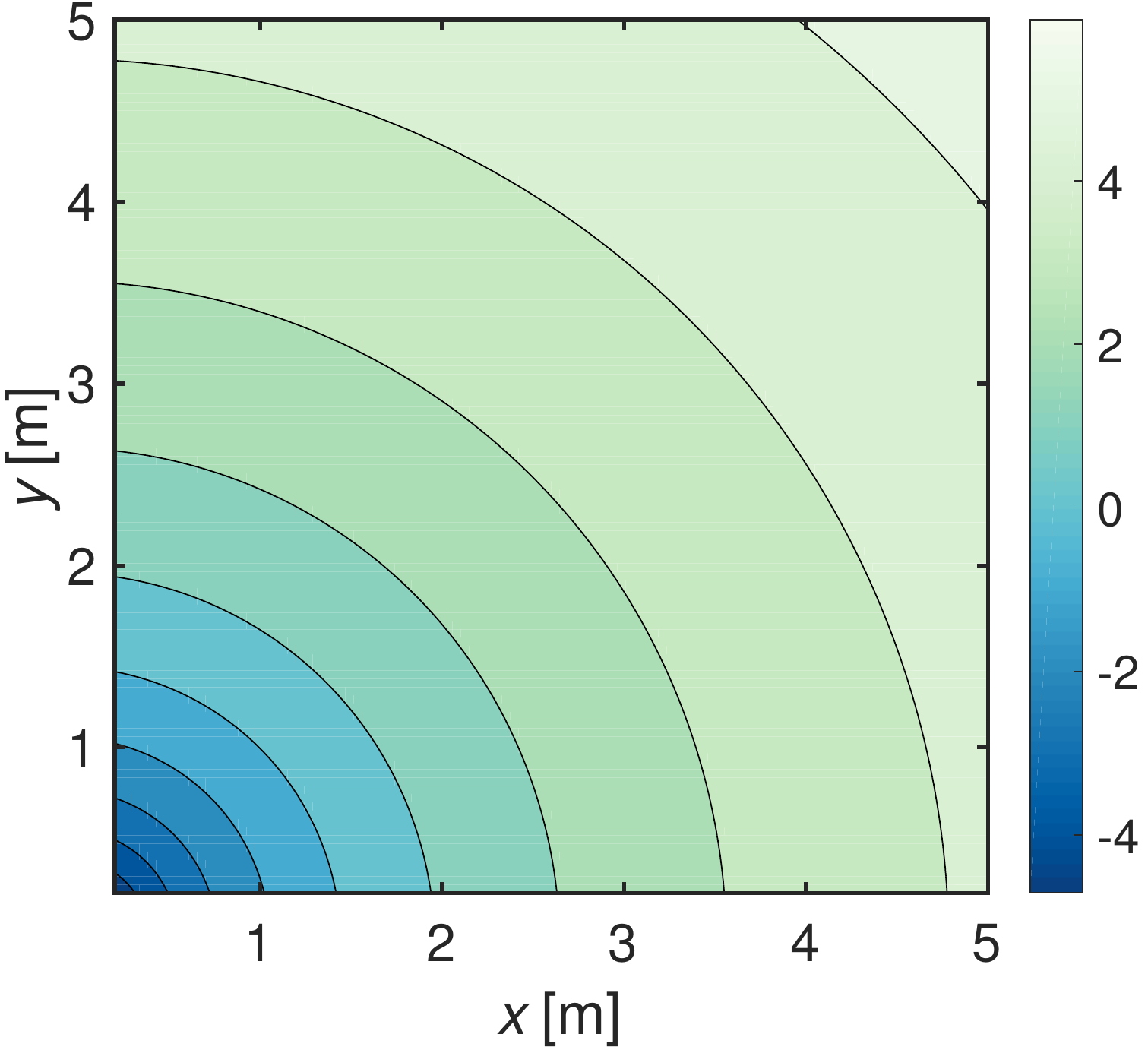}
      \captionsetup{justification=centering}
      \caption{\footnotesize $\log_{10}\CRBnocc(\rhos)$}
      \label{fig:CRBrho_noCC_surf}
    \end{subfigure}
    \begin{subfigure}{0.45\linewidth}
      \includegraphics[width=\linewidth]{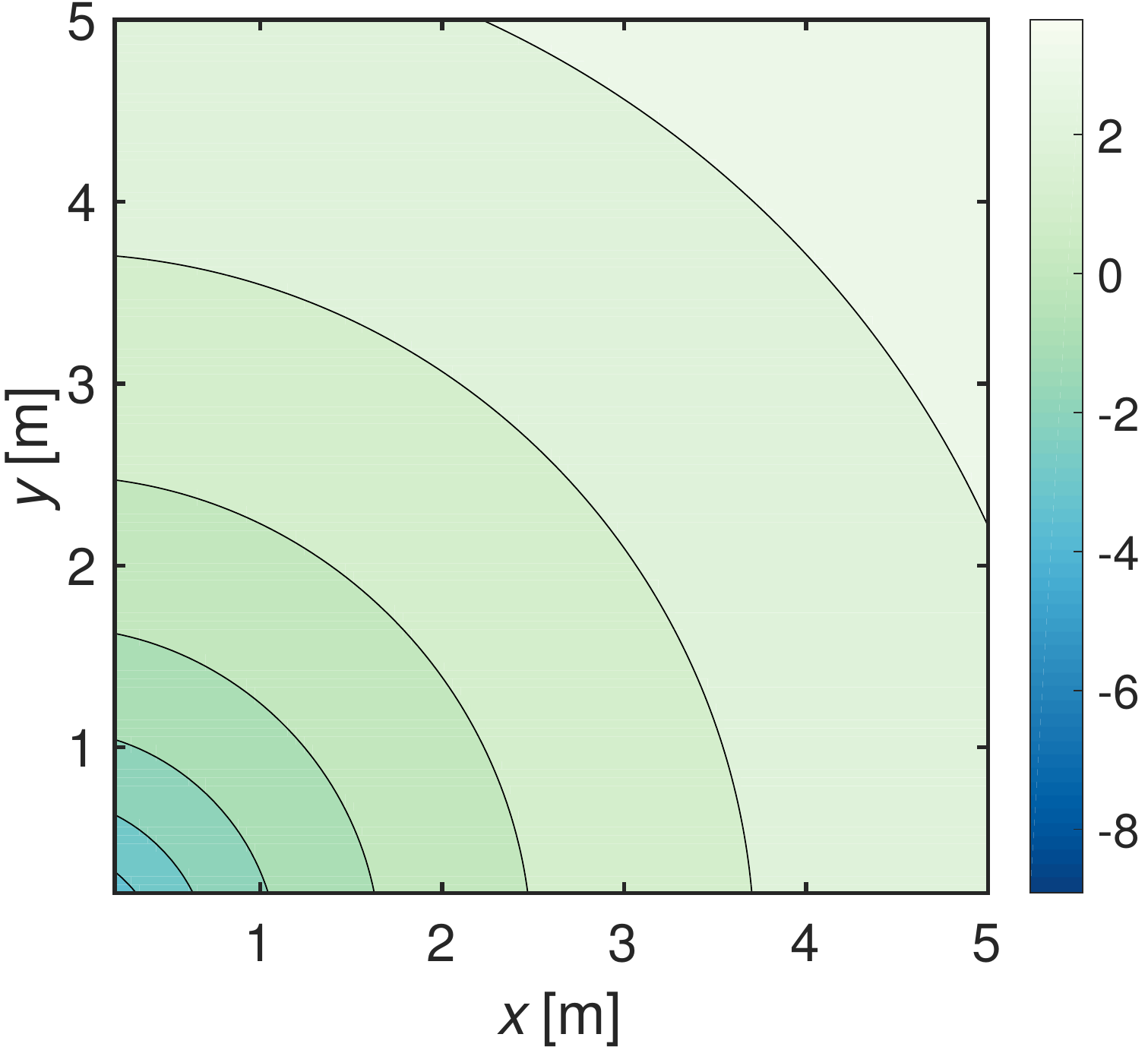}
      \captionsetup{justification=centering}
      \caption{\footnotesize $\log_{10}\CRBnocc(\phis)$}
      \label{fig:CRBphi_noCC_surf}
    \end{subfigure}
    \caption{Variation of the CRBs for estimating a single hidden target for different target locations.
    The number of measurement pixels is $M=155^2$ and the measurement FOV is $0.2\,{\rm m} \times 0.2\,{\rm m}$, with fixed noise variance $\sigma^2 = 10$.}
    \label{fig:CRB_surface}
\end{figure}

\begin{figure}
    \centering
    \begin{subfigure}{0.49\linewidth}
      \includegraphics[width=\linewidth]{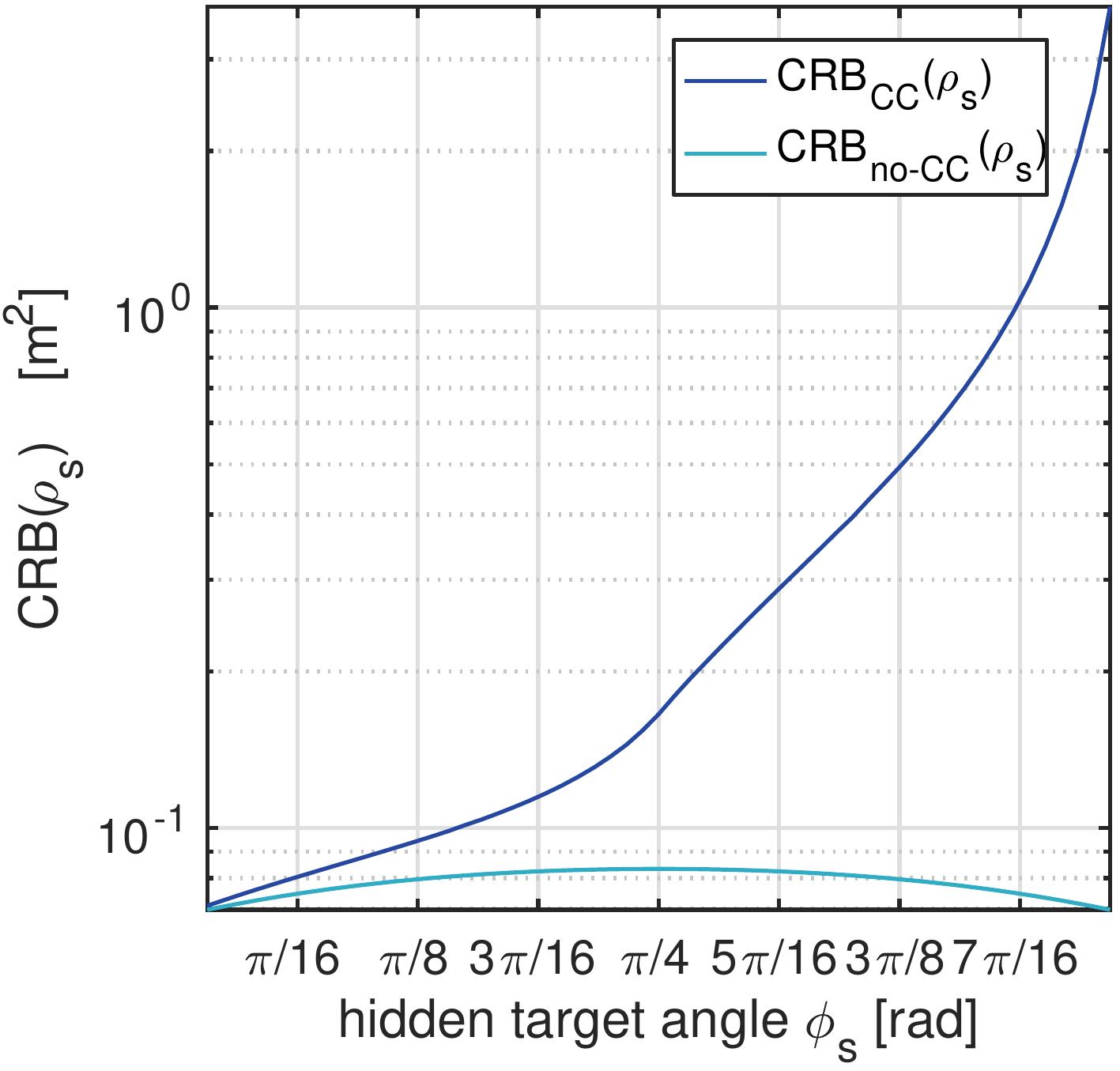}
      \captionsetup{justification=centering}
      \caption{\footnotesize ${\rm CRB}(\rhos)$}
      \label{fig:CRBrhovsTargetAngle_single}
    \end{subfigure}
    \begin{subfigure}{0.49\linewidth}
      \includegraphics[width=\linewidth]{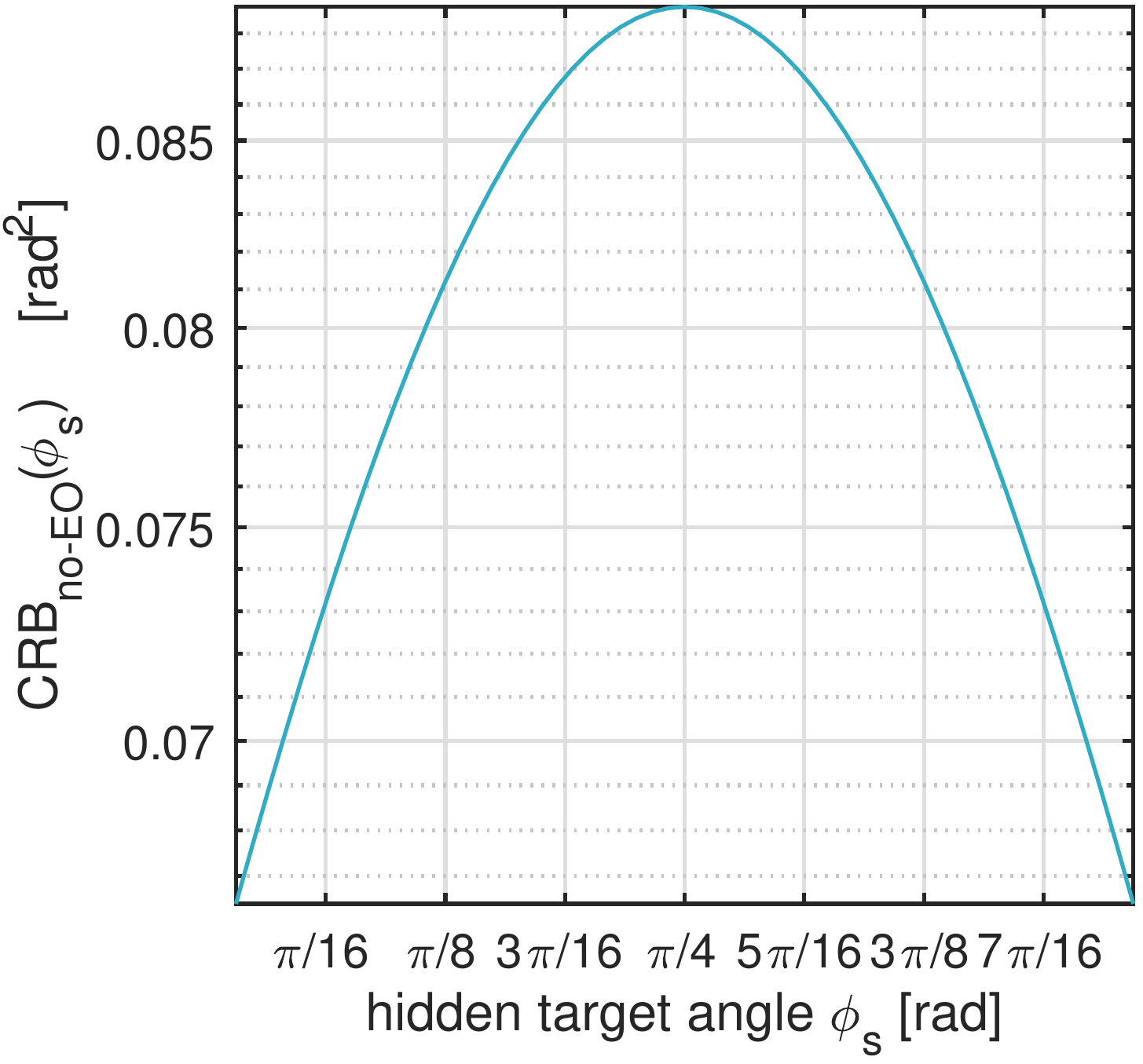}
      \captionsetup{justification=centering}
      \caption{\footnotesize $\CRBnocc(\phis)$}
      \label{fig:CRBphinocc_vsTargetAngle_single}
    \end{subfigure}
    \begin{subfigure}{0.6\linewidth}
      \includegraphics[width=\linewidth]{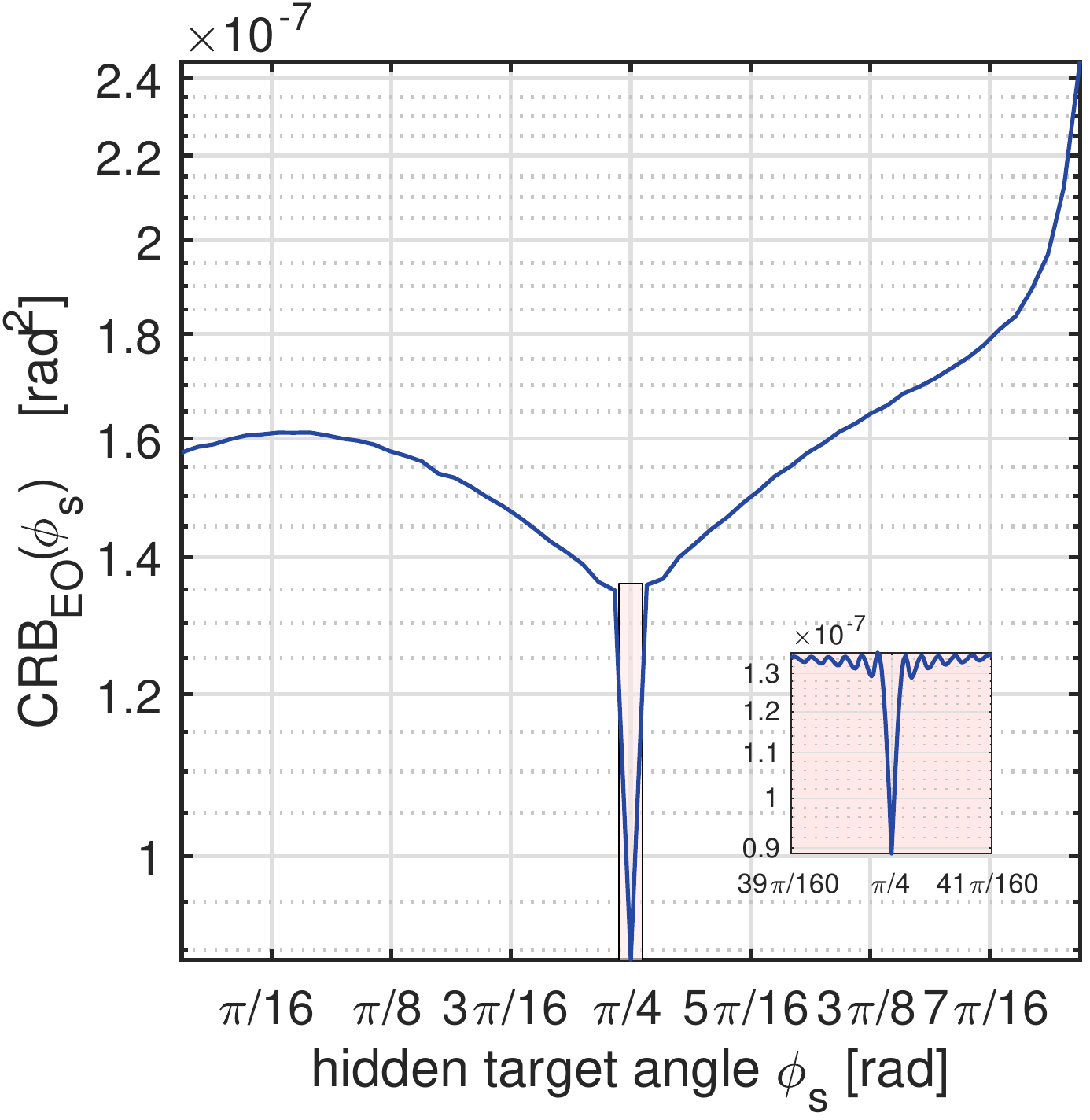}
      \captionsetup{justification=centering}
      \caption{\footnotesize $\CRBcc(\phis)$}
      \label{fig:CRBphicc_vsTargetAngle_single}
    \end{subfigure}
    \caption{Variation of CRB of the hidden target estimate in response to varying the target's angular position.
    (a) CRBs for range estimates with and without a corner.
    (b) CRB for angle estimates without a corner.
    (c) CRB for angle estimates with a corner camera.
    Camera FOV = $0.2\,{\rm m} \times 0.2\,{\rm m}$,
    $\sigma^2 = 10$, and $\rhos = 1\,{\rm m}$.}
    \label{fig:CRBvsTargetAngle_single}
\end{figure}

Contour plots of computed CRBs for various ground truth target positions with respect to the origin (corner) are shown in \Cref{fig:CRB_surface}, for the corner ($\CRBcc$) and no corner cases ($\CRBnocc$). Comparing \Cref{fig:CRBrho_CC_surf,fig:CRBrho_noCC_surf}, achievable target range estimates MSE has marginal dependence on the presence of a corner, when estimating a single point target. On the other hand, \Cref{fig:CRBphi_CC_surf,fig:CRBphi_noCC_surf}
suggest that CRBs for angle estimates with the corner are around five to seven orders of magnitude smaller when compared to the no-corner case.

Fixing the target's range at $\rhos = 1~{\rm m}$, \Cref{fig:CRBvsTargetAngle_single} summarizes the dependence of the computed CRBs on $\phis \in [\Frac{\pi}{64},\Frac{63\pi}{64}]~{\rm rads}$.
First, \Cref{fig:CRBrhovsTargetAngle_single} shows that $\CRBcc(\rhos)$ and $\CRBnocc(\rhos)$ are nearly equal at very shallow target angles, because the shadowed region in the occluded case is very small (the measurements for the corner and no corner cases are almost the same).
However, with measurement noise variance fixed and $\phis$ increasing, $\CRBcc(\rhos)$ diverges because the in-shadow region---which cannot possibly be informative about the occluded target's distance---grows, while $\CRBnocc(\rhos)$ changes only marginally (reaching a maximum at $\Frac{\pi}{4}~{\rm rads}$ before decreasing again).
At the deepest angle, $\CRBcc(\rhos)$ is roughly 28 times $\CRBnocc(\rhos)$.
\Cref{fig:CRBphinocc_vsTargetAngle_single,fig:CRBphicc_vsTargetAngle_single} indicate that $\CRBnocc(\phis)$ has relatively mild dependence on the true target angle $\phis$, with symmetry around $\Frac{\pi}{4}$.
The observed partial symmetry, in \Cref{fig:CRBphicc_vsTargetAngle_single}, about $\Frac{\pi}{4}$, with $\phis \in [\Frac{\pi}{8},\Frac{3\pi}{8}]$ is because, in contrast to range estimation, the in-shadow region is also informative (subject to prevalent noise levels) about the target's angular position.
The asymmetry (for $\phis \notin [\Frac{\pi}{8},\Frac{3\pi}{8}]$) is explained by a fixed noise variance (i.e., measurement SNR reduces with increasing target angle). Overall, the variation in $\CRBcc(\phis)$ is small relative to the roughly five orders of magnitude improvement due to the occluding wall.

\begin{figure}
    \centering
    \begin{subfigure}{0.49\linewidth}
      \includegraphics[width=\linewidth]{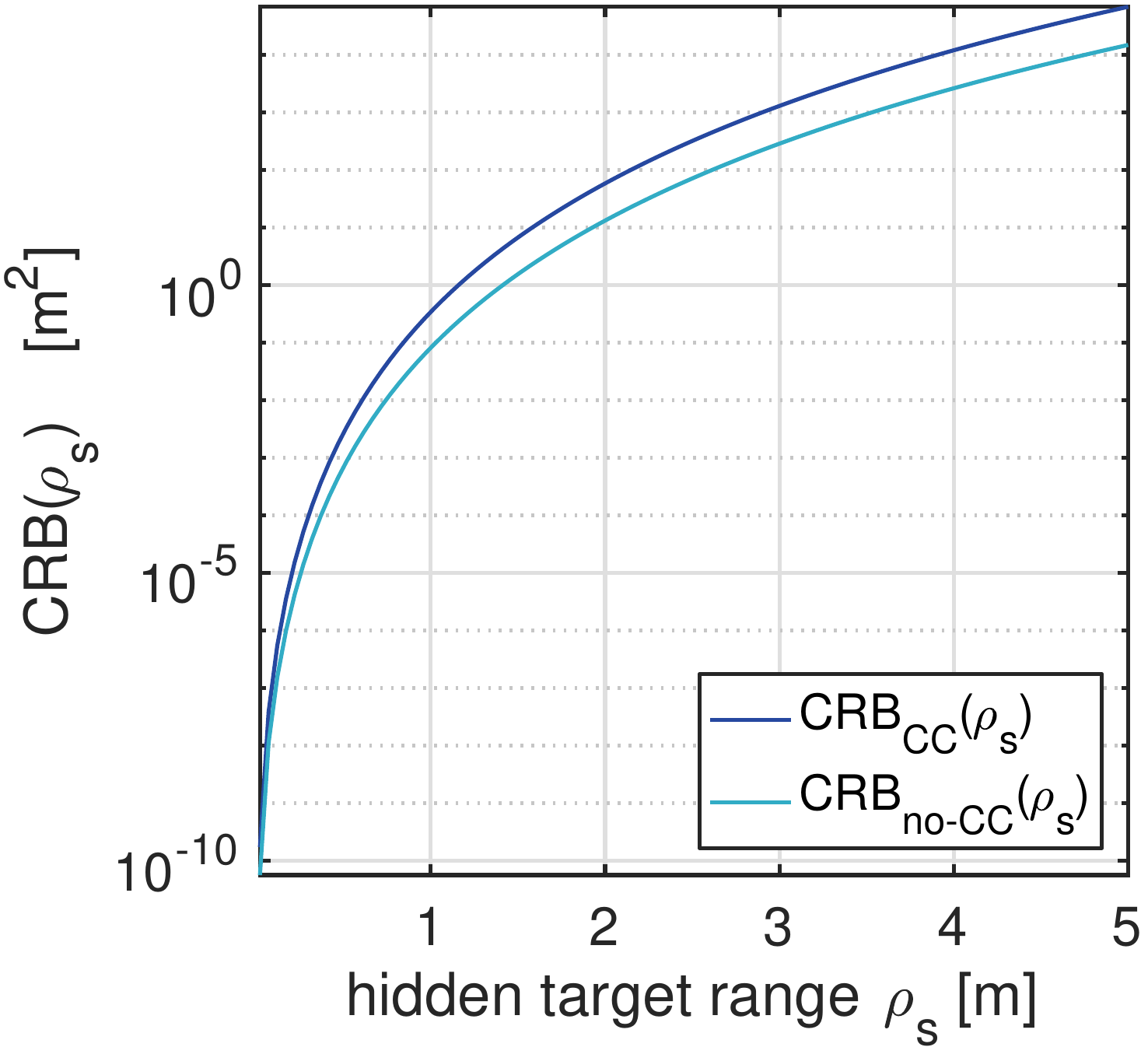}
      \captionsetup{justification=centering}
      \caption{\footnotesize ${\rm CRB}(\rhos)$}
      \label{fig:CRBrhovsTargetRange_single}
    \end{subfigure}
    \begin{subfigure}{0.49\linewidth}
      \includegraphics[width=\linewidth]{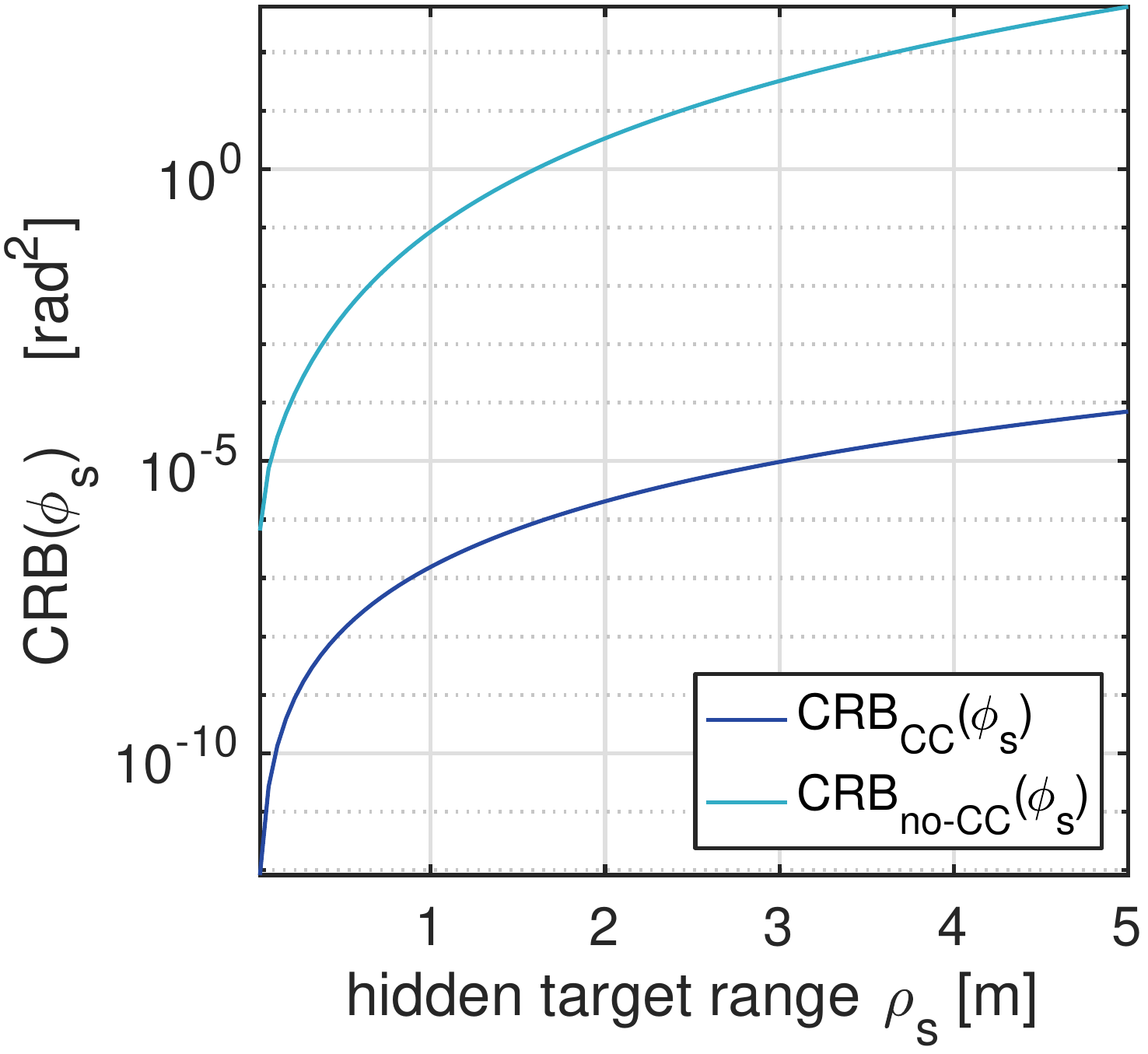}
      \captionsetup{justification=centering}
      \caption{\footnotesize ${\rm CRB}(\phis)$}
      \label{fig:CRBphivsTargetRange_single}
    \end{subfigure}
    \caption{Variation of CRB of the hidden target estimate in response to varying the target's distance from the corner $\rhos$. FOV $= [0.2~{\rm m} \times 0.2~{\rm m}]$, $\sigma^2 = 10$, $\phis = \pi/3~{\rm rads}$.}
    \label{fig:CRBvsTargetRange_single}
\end{figure}

Second, with the target's angle is held constant ($\phis = \Frac{\pi}{3}~{\rm rads}$) while its distance from the corner increases from zero, \Cref{fig:CRBrhovsTargetRange_single} shows that $\CRBcc(\rhos)$ and $\CRBnocc(\rhos)$ are both small for a close target, but increase dramatically with target's distance.
The uninformativeness of the in-shadow measurements for range estimation causes $\CRBcc(\rhos)$ to be higher than $\CRBnocc(\rhos)$, whereas the presence of the corner makes $\CRBcc(\phis)$ at least five orders of magnitude lower than $\CRBnocc(\phis)$ (see \Cref{fig:CRBphivsTargetRange_single}).
Under our measurement scenario, a target
$3\,{\rm m}$
from the corner (with $\phis = \Frac{\pi}{3}$) for instance has
$\sqrt{\CRBcc(\phis)} \approx 10^{-\Frac{5}{2}} = 0.003\,{\rm rads}$,
while
$\sqrt{\CRBnocc(\phis)} \approx 10^{-\Frac{3}{4}} = 0.178\,{\rm rads}$.

Our study for a single point target demonstrates overwhelming improvement in the estimation of $\phis$ due to the occluding wall, with marginal negative impact on the expected estimation quality of $\rhos$.
This is because the occluding wall effectively separates light paths arising from different angles in the hidden region.
Phrased differently, the exact proportion of shadowed-to-nonshadowed regions within the camera's FOV is informative about the angular location of the hidden target.

\begin{figure}
    \centering
    \begin{subfigure}{\linewidth}
    \begin{minipage}{.32\linewidth}
        \centering
        \includegraphics[width=\linewidth]{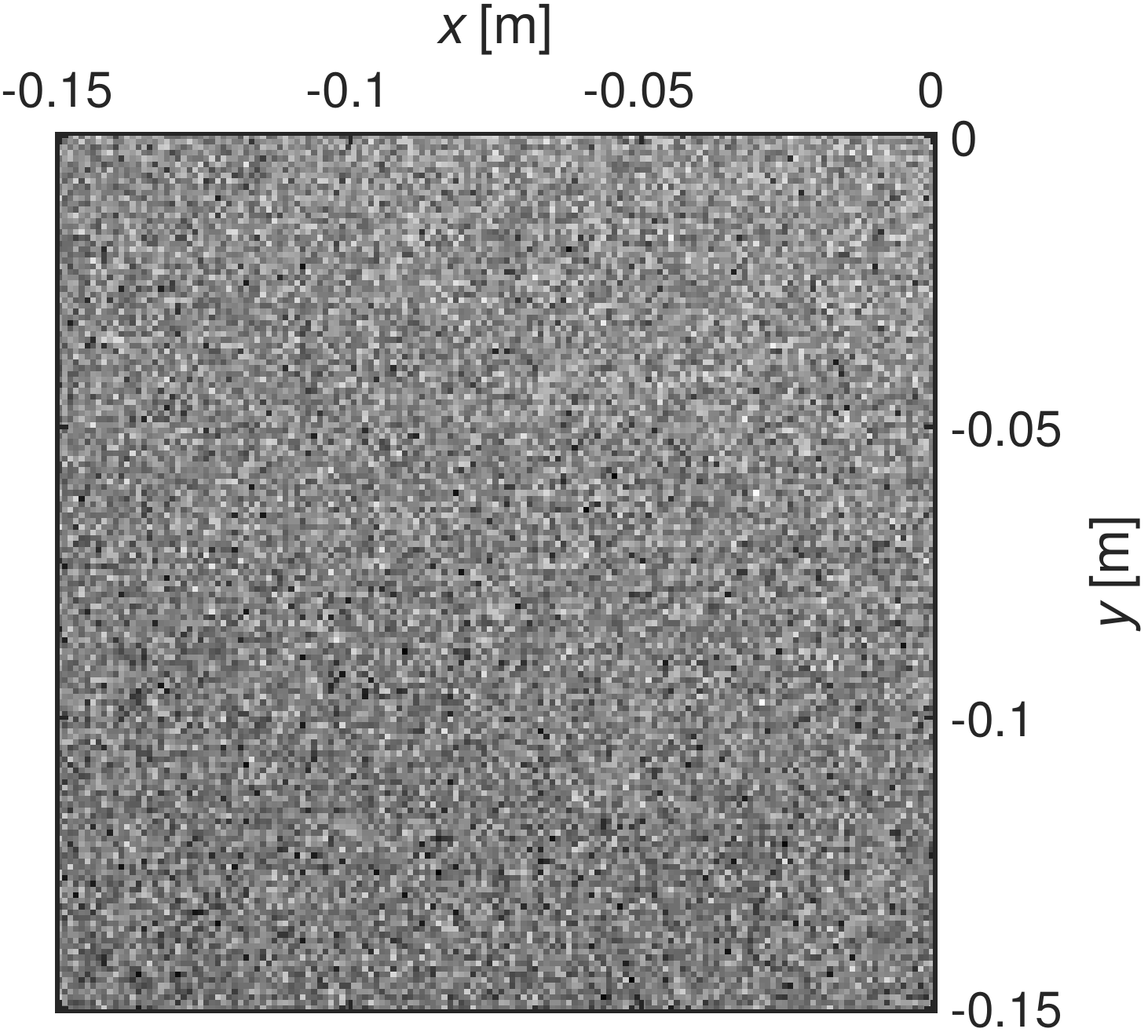}\\[0.5em]
        \includegraphics[width=\linewidth]{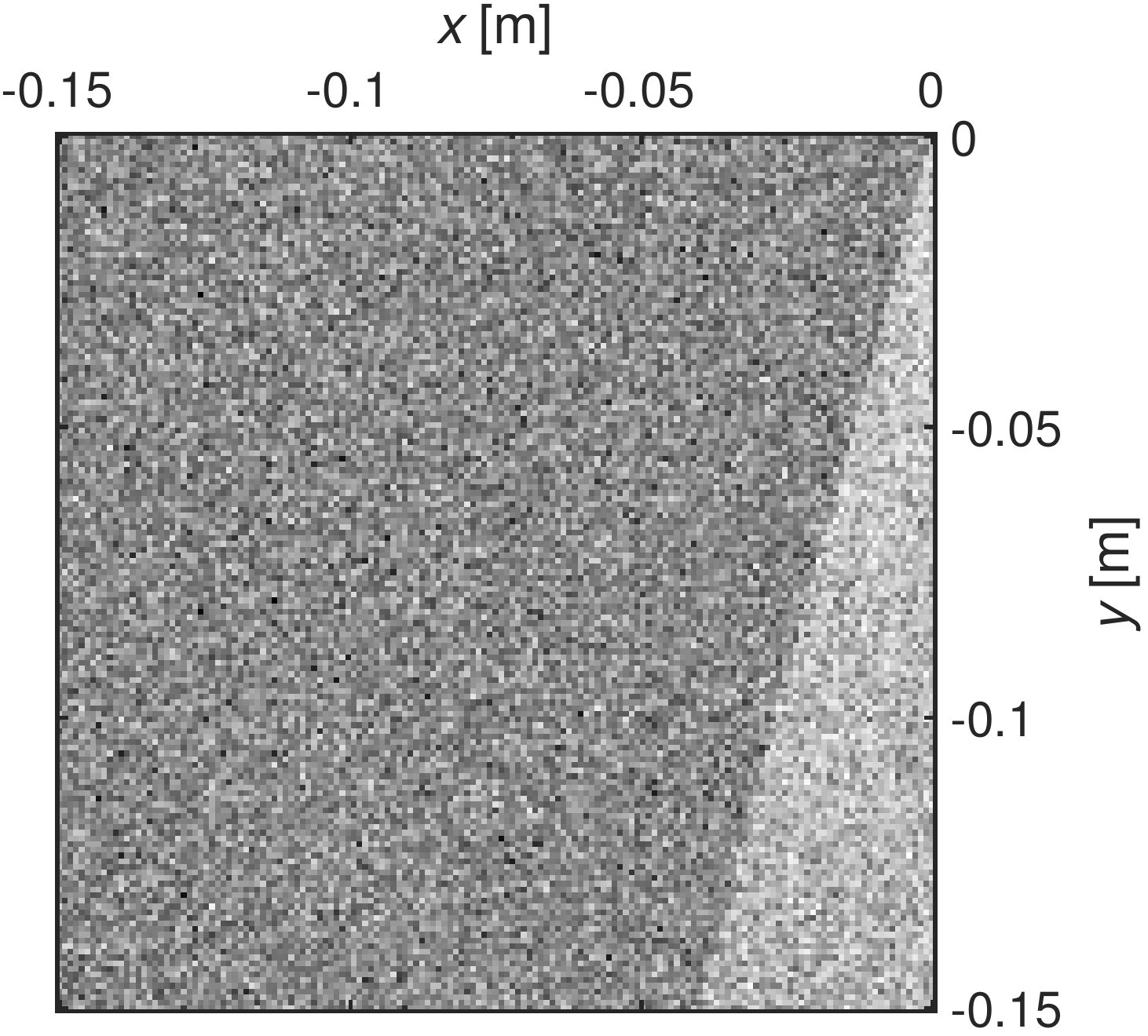}
    \end{minipage}%
    \hspace{4mm}
    \begin{minipage}{0.5\linewidth}
        \includegraphics[width=\linewidth]{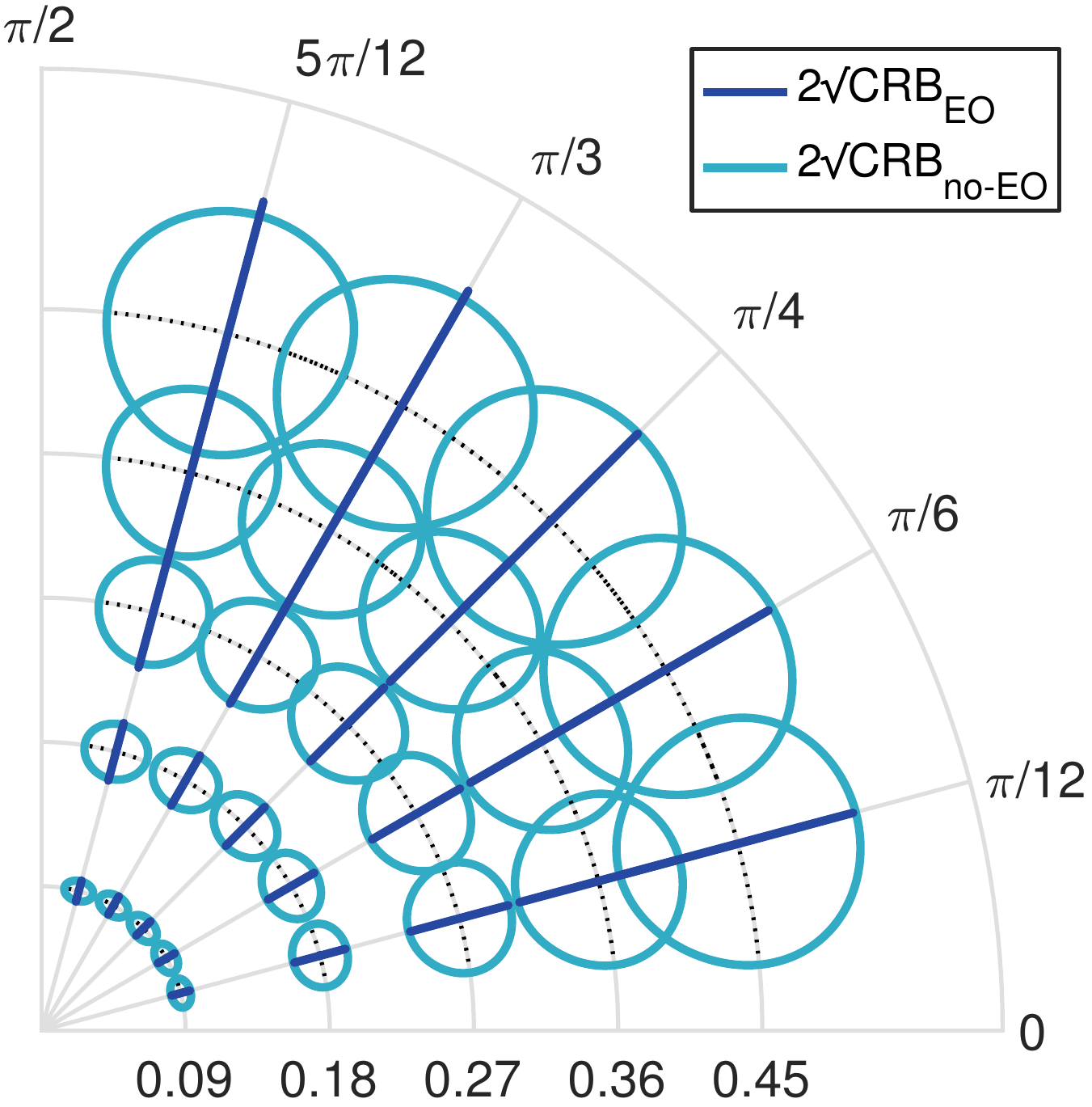}
    \end{minipage}
      \captionsetup{justification=centering}
      \caption{\footnotesize SNR = -5 dB}
      \label{fig:SNR-5dB}
    \end{subfigure}
    \\
    \begin{subfigure}{\linewidth}
    \begin{minipage}{.32\linewidth}
        \centering
        \includegraphics[width=\linewidth]{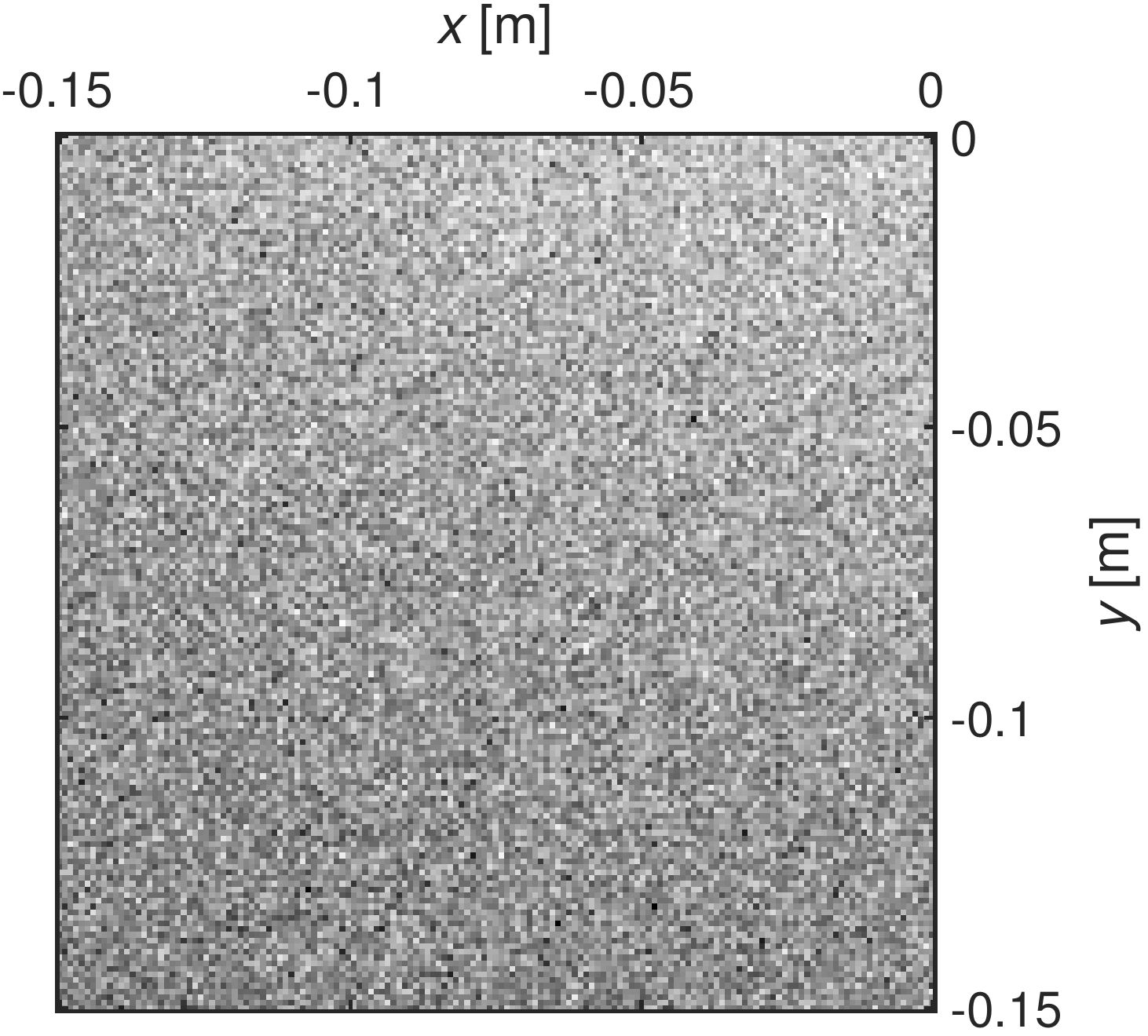}\\[0.5em]
        \includegraphics[width=\linewidth]{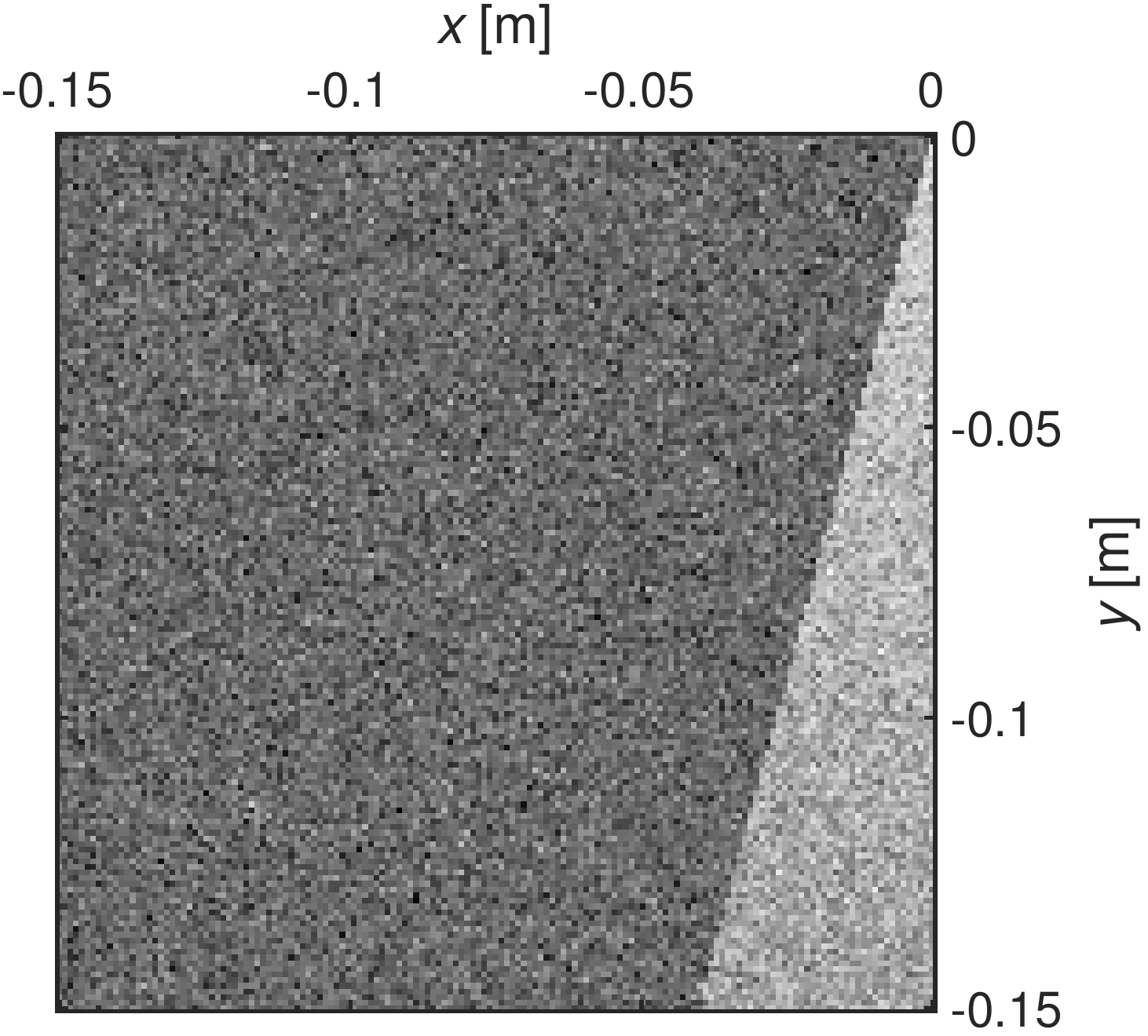}
    \end{minipage}%
    \hspace{4mm}
    \begin{minipage}{0.5\linewidth}
        \includegraphics[width=\linewidth]{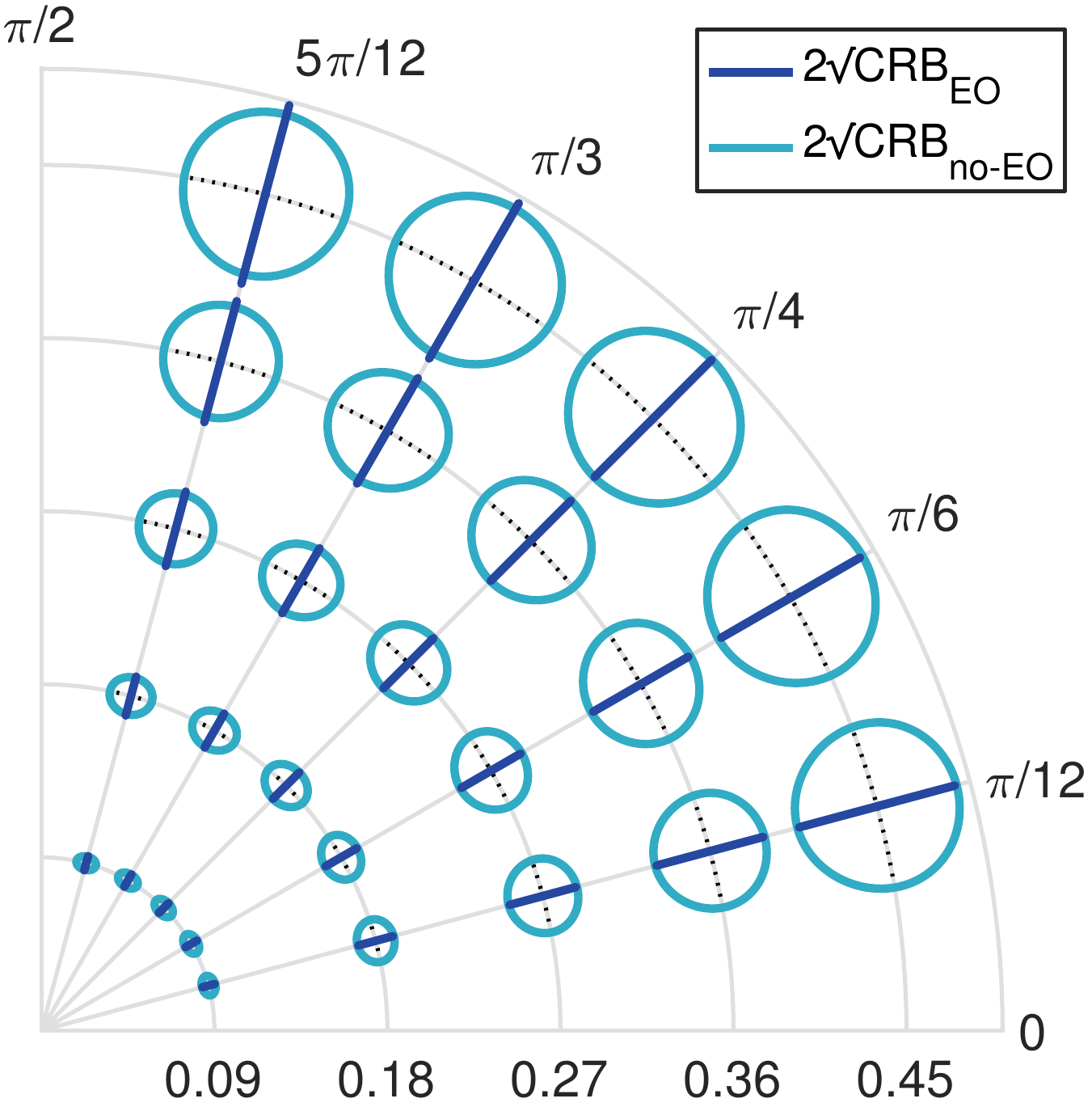}
    \end{minipage}
      \captionsetup{justification=centering}
      \caption{\footnotesize SNR = 0 dB}
      \label{fig:SNR0dB}
    \end{subfigure}
    \caption{$2{\times}\sqrt{\rm CRB}$ uncertainty regions (right) for various measurement SNR levels. Each uncertainty region is an ellipse (in polar coordinates) with minor and major axis length set to $4{\times}\sqrt{\rm CRB}$ for the corresponding dimension. Camera FOV = 0.15m $\times$ 0.15m, one typical realization of the camera measurement made at the corresponding SNR assuming no occluding wall (insets: top) and with an occluding wall (insets: bottom). The number of camera pixels $M=155^2$.}
    \label{fig:CRB_UncertaintyBubbles}
\end{figure}

Using the CRBs, one can compute theoretical spatial uncertainty regions for a hidden target.
These are regions within which the majority of a target's estimates are expected to fall.
Specific examples for a camera FOV of $0.15\,{\rm m} \times 0.15\,{\rm m}$ and $M = 155^2$~pixels, and at SNR levels resembling real experimental measurements are shown in \Cref{fig:CRB_UncertaintyBubbles}.
Assuming unbiased estimators that achieve the CRB, these bubbles depict regions within which three standard deviations of a target's estimate are expected to fall.
We observe that the uncertainty regions are very different with and without the occluding wall:
almost circular for the latter, while the angular uncertainties are virtually imperceptible for the former.
The presence of the corner collapses these bubbles into lines, with the length of each line representing the uncertainty in range, while angular uncertainties are almost completely removed.

\begin{figure*}
    \centering
    \begin{subfigure}{0.275\linewidth}
      \includegraphics[width=\linewidth]{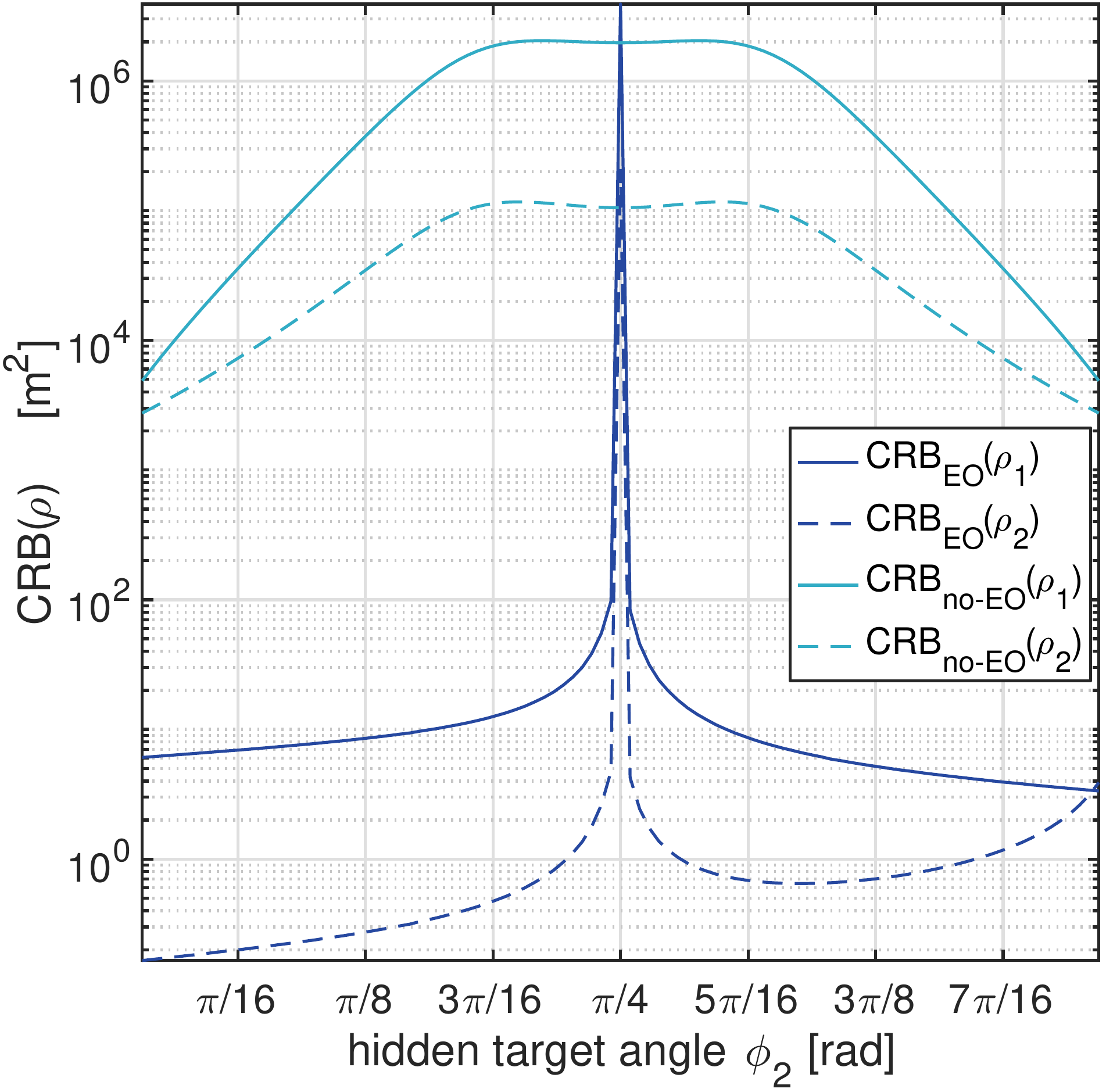}
      \captionsetup{justification=centering}
        \caption{\footnotesize
        }
      \label{fig:CRBrho_twotargets}
    \end{subfigure}
    \hspace{2em}
    \begin{subfigure}{0.275\linewidth}
      \includegraphics[width=\linewidth]{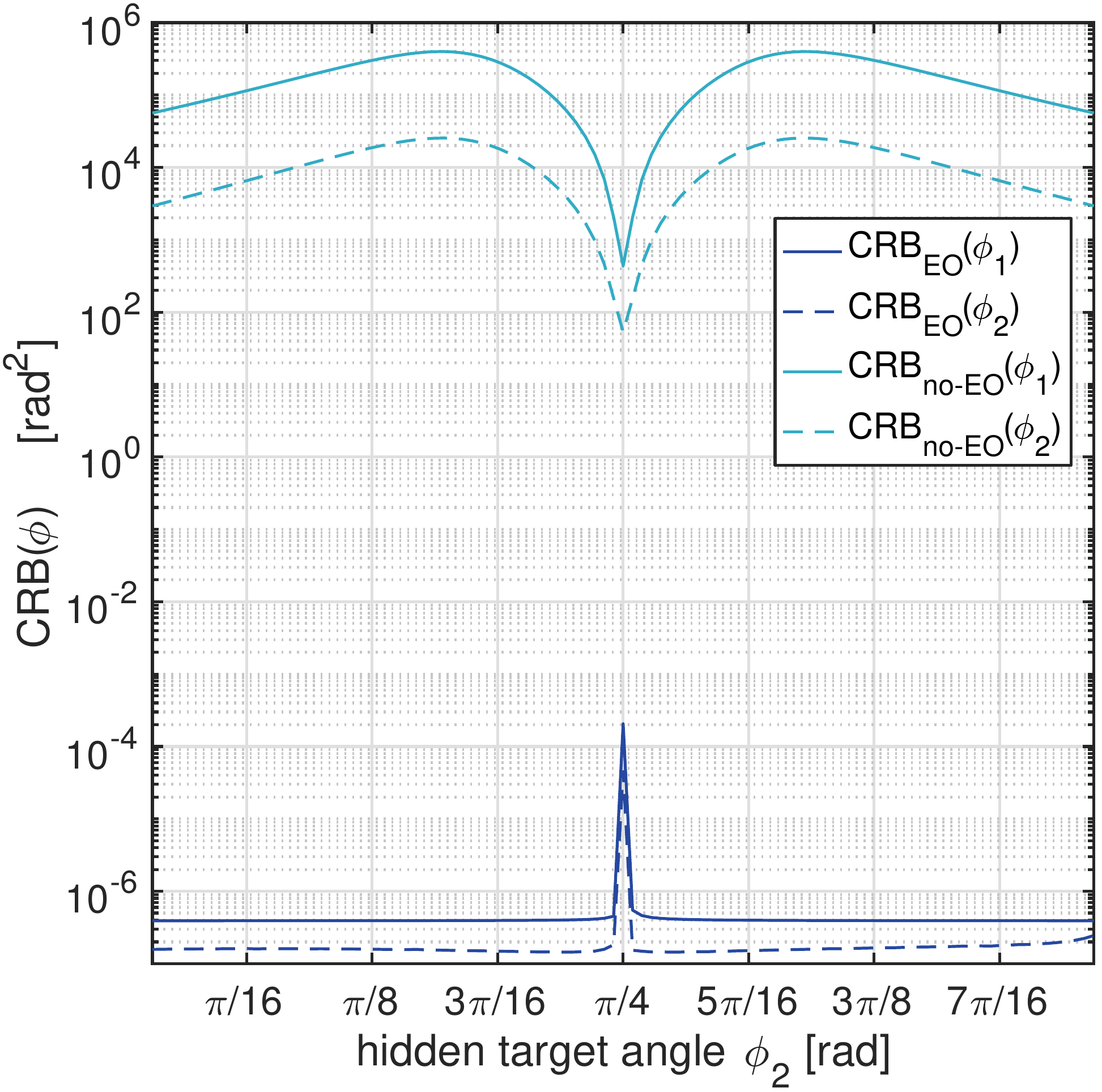}
      \captionsetup{justification=centering}
      \caption{\footnotesize
      }
      \label{fig:CRBphi_twotargets}
    \end{subfigure}
    \hspace{2em}
    \begin{subfigure}{0.275\linewidth}
      \includegraphics[width=\linewidth]{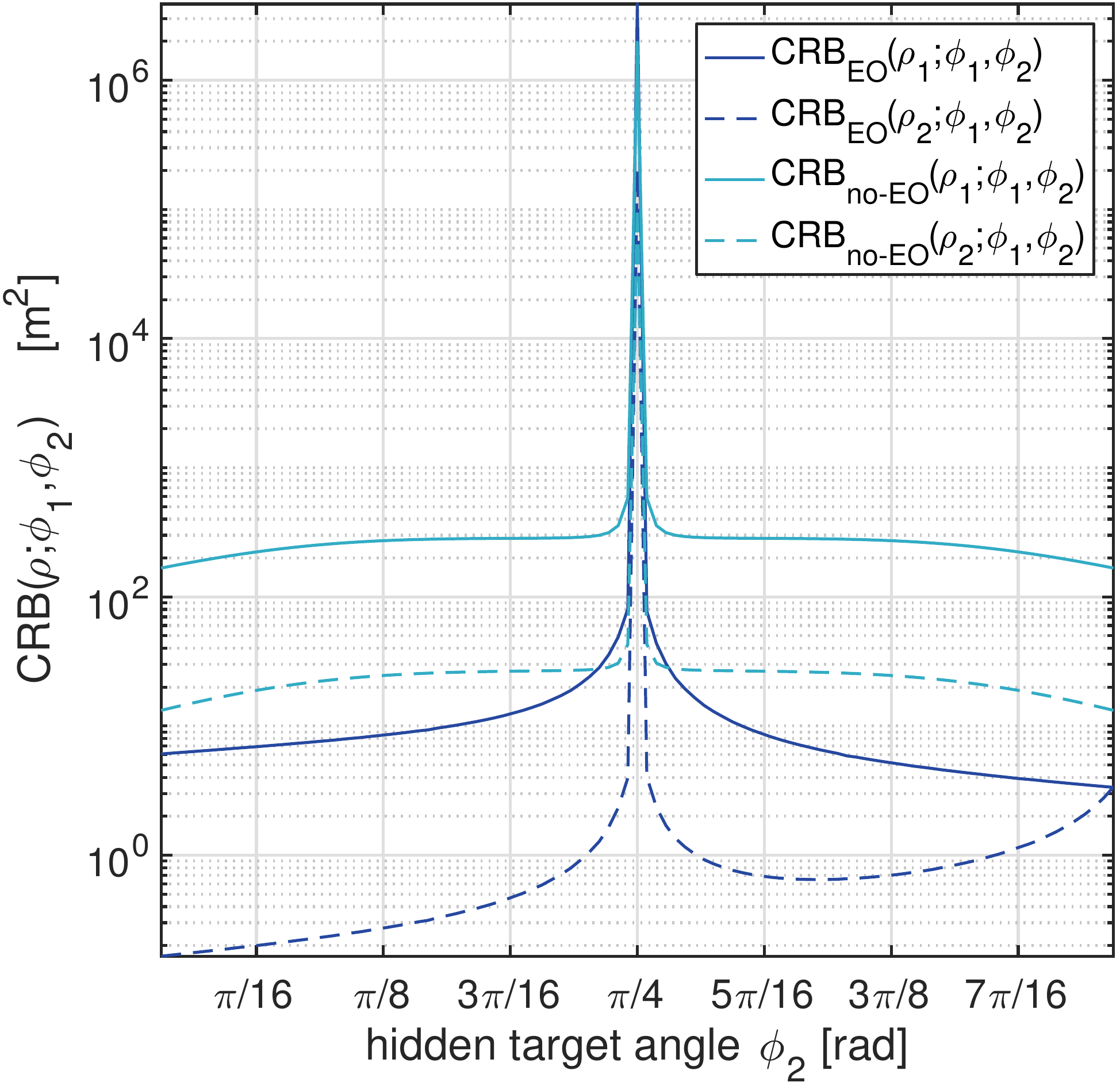}
      \captionsetup{justification=centering}
      \caption{\footnotesize
      }
      \label{fig:CRBrho_twotargets_knownphi}
    \end{subfigure}
    \caption{Variation of the CRBs for estimating two hidden targets for different target positions. The number of measurement pixels, $M=155^2$ and the measurement FOV is $0.2{\rm m} \times 0.2{\rm m}$, with fixed noise variance $\sigma^2 = 10$. Target 1 is fixed at  $(\rho_{\rm 1}, \phi_{\rm 1})= (1{\rm m}, \Frac{\pi}{4})$; target 2 is at range $\rho_2 = 2{\rm m}$ and moved in angle $\phi_2$.}
    \label{fig:CRB_twotargets}
\end{figure*}

\subsection{Multiple Hidden Targets}
\label{ssec:multipletargets}
Although our single point-target CRB analysis showed the incredible benefits of the occluding wall in $\phi_{\rm s}$ estimation, estimation of $\rhos$ was actually shown to be slightly more challenging, especially for hidden targets at greater angular depths. The benefit of the occluding wall in range estimation is realized when the hidden scene is more complicated. We extend our single point target CRB analysis to include a second hidden point target to demonstrate this effect.

In \Cref{fig:CRB_twotargets}, Target 1 is fixed at $(\rho_{\rm 1}, \phi_{1})= (1~{\rm m}, \Frac{\pi}{4}~{\rm rads})$ while Target 2 is held at $\rho_2 = 2~{\rm m}$ and moved in angle $\phi_2$.
The CRB for both parameters and targets are compared for scenarios with and without the corner in place.
\Cref{fig:CRBrho_twotargets} shows that ${\rm CRB}(\rho_1)$ and ${\rm CRB}(\rho_2)$ are, generally, over an order of magnitude smaller when the corner is in place, the only exception being when both targets are at or very near the same angle.
In this case, it understandably becomes difficult to isolate the two targets in range.
Just like the single-target scenario, $\CRBcc(\phi_1)$ and $\CRBcc(\phi_2)$ are seen to be many orders of magnitude smaller (than $\CRBnocc(\phi_1)$ and $\CRBnocc(\phi_2)$) in \Cref{fig:CRBphi_twotargets}.
This significantly improved angular resolution depends on the ability to separate angular derivatives due to each target, which becomes more challenging when they are very close to each other in angle, causing the peak at $\phi_2 =  \Frac{\pi}{4}$ in  \Cref{fig:CRBphi_twotargets}.
When the two targets are at or near the same angle, the no-corner case, which relies exclusively on radial falloff, shows improvement due to contributions from each target adding constructively in the measurement.
Though that improvement is marginal relative to improvement from having a corner.

Even when the angular location of both targets is given, ${\rm CRB}(\rho; \phi_1, \phi_2)$ is still substantially lower for the corner camera case, as shown in \Cref{fig:CRBrho_twotargets_knownphi}. This may be explained by the fact that light from the shallowest (in angle) target in the hidden scene affects a larger angular wedge in the measurement than the less shallow target. The difference between these two wedges is a swath of pixels affected only by the shallowest target, making range estimation for that target easier. In contrast, light from both targets without the corner strikes all pixels in the measurement, making separating each radial falloff pattern more challenging.

In addition, note that knowing the angular location of both targets has a negligible effect on ${\rm CRB}(\rho; \phi_1, \phi_2)$ when the wall is in place, likely due to the fact that angular uncertainty is already so low in that situation. Inspired by these observations, we introduce an algorithm that alternates between estimating $\phi$ and $\rho$ in \Cref{ssec:NonlinearModel}.

\section{Inverse Problems \& Algorithms}
\label{sec:Inverse}

In this section, we present two approaches to form a plan-view reconstruction from a photograph of the penumbra. The first method, described in \Cref{ssec:LinearModel}, discretizes the hidden scene into a polar grid of pixels; thus transforming our inverse problem into a linear problem of estimating the intensity of each \textit{polar pixel}. While this method is straightforward, we demonstrate improved reconstructions using a second approach that solves the inverse problem introduced in \ref{sec:discrete-model} by alternating between estimating angular and range information. First, by exploiting the high angular resolution provided by the corner an initial estimate of the scene is formed, as a function of angle. From this initial profile of the scene, the number of hidden targets is estimated. Finally, we alternate between estimating a single range for each target (i.e., \textit{learning} the true forward model), and updating the angular profile.

\subsection{Floor Albedo and Ambient Light}
\label{sec:albedoAmbient}
Jointly estimating $\fvec$ along with a 1D projection of the hidden scene has been studied in~\cite{Seidel2019_corner}, with the assumption that ambient---or visible side---light contribution to the measurements $\avec \approx c_1 \bm{1}$ is approximately constant over the camera FOV\@.
This work assumes uniform floor albedo $f(r,\theta)$ (i.e., $\fvec = \bm{1}$), though we remark that both inversion methods can be similarly extended to handle the case of unknown floor albedo $\fvec$.
This is by no means trivial and we leave it for a future work.
In addition, because ambient light contributions in the camera measurements is slowly varying, it can be approximately decomposed into a sum of light contributions from sources near the measurement surface, $\avec_{\rm NF}$, and those in the far-field, $\avec_{\rm FF}$.
The far-field contribution is roughly constant over the camera FOV, $\avec \approx c_1 \bm{1} + c_2 \avec_{\rm NF} $, where $c_1$ and $c_2$ are constants that lead to dimensionless pixel values.
The term $\avec_{\rm NF}$ can be measured, or computed from our knowledge of the position of the visible side, so that the only unknown needed to describe $\avec$ is $\cvec = [c_1, c_2]$.
In the presence of ambient light, the inverse problem becomes estimating $\left(\svech, \rhovech, \cvec\right)$ from measurements $\yvec$, under the model
\begin{equation}
    \yvec = \Amat \cvec + (\Vmat \odot \Dmat(\rhovech))\svech + \bm{\epsilon},
    \label{eq:DiscreteForwardModel_NL_bg}
\end{equation}
where $\Amat = \left[\bm{1}, \avec_{\rm NF} \right]$.

\subsection{A Linear Model and Inverse Algorithm}
\label{ssec:LinearModel}

\begin{figure}
    \centering
    \includegraphics[width=0.4\linewidth]{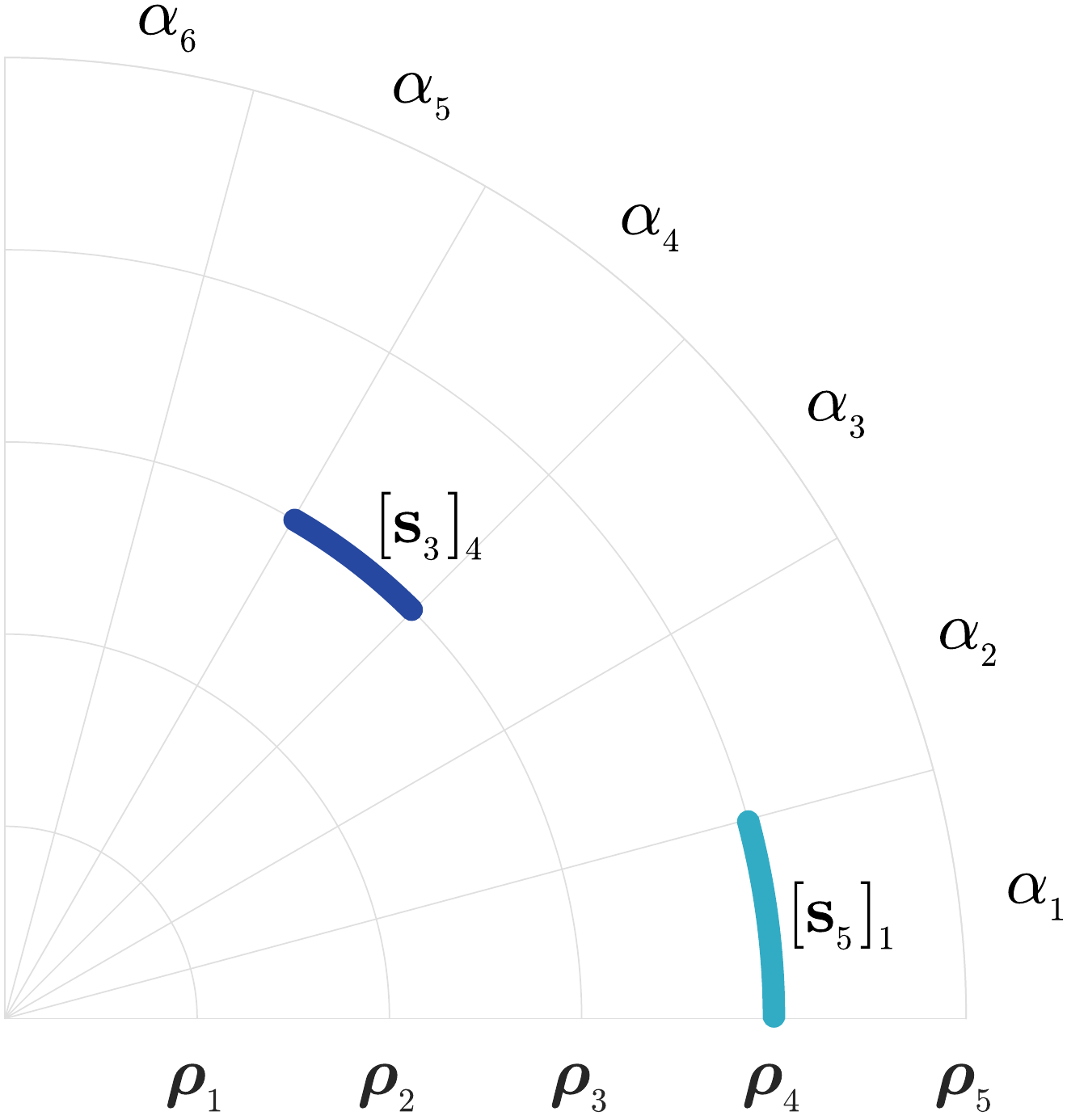}
    \caption{Polar partitioning of the hidden space to obtain polar pixels at six discrete angles and five discrete range values.
    The $n$-th column of each submatrix $\Dmatbar_l$, of $\Dmatbar$, describes propagation of light from hidden scene polar pixel $(\rho_l, \alpha_n)$, to the measurement plane.}
    \label{fig:ScenePolarDiscretization}
\end{figure}

Equation~\eqref{eq:DiscreteForwardModel_NL_bg} is linear in $\svech$ and nonlinear in $\rhovech$.
However, by discretizing the possible values of each element of $\rhoh(\alpha_n)$,
we can formulate a new system that is linear in all unknown parameters.
Specifically, let $\{\rho_1, \rho_2, \ldots, \rho_L\}$ be the set of allowed ranges. Then the Cartesian product $\{\rho_1, \rho_2, \ldots, \rho_L\} \times \{\alpha_1, \alpha_2, \ldots, \alpha_N\}$ gives a 2D polar partitioning of the hidden region, with each element $(\rho_\ell, \alpha_n)$ defining a hidden-scene \textit{polar pixel}. Shown in~\Cref{fig:ScenePolarDiscretization} is a (coarse) $5\times 6$ polar grid discretization of the hidden space.
Under this partitioning, the forward model~\eqref{eq:discreteforwardmodel_nonlinear} becomes
\begin{equation}
    \yvec = \Amat \cvec + \Dmatbar \svecbarh + \bm{\epsilon},
    \label{eq:discreteforwardmodel_linear}
\end{equation}
where $\Dmatbar = \left[ \Vmat\odot\Dmat(\rho_1 \bm{1}), \Vmat\odot\Dmat(\rho_2 \bm{1}), \ldots, \Vmat\odot\Dmat(\rho_L \bm{1}) \right]$, and
$\svecbarh = \mathrm{vec}\!\left([\svech{}{_{_1}}, \svech{}{_{_2}}, \ldots, \svech{}{_{_L}}]\right) \in \mathbb{R}_{+}^{NL}$
with $ [\svech{}{_{_\ell}}]_n$ representing the radiosity of pixel $(n,\ell)$ at range $\rho_\ell$ and angular bin $n$. 

Although \eqref{eq:discreteforwardmodel_linear} is linear in all unknown parameters and \eqref{eq:discreteforwardmodel_nonlinear} is not, there is an important difference.
Built into \eqref{eq:discreteforwardmodel_nonlinear} is the constraint that only a single hidden object per angle contributes to the measurement.
This constraint is based on the assumption that the scene is composed of opaque vertical facets, so light from objects that are behind other objects is blocked from reaching the corner.
In contrast, this constraint is \emph{not} built into \eqref{eq:discreteforwardmodel_linear}.
In this case, to model the fact that the vast majority of pixels in the hidden scene either do not contain a target or are occluded from the camera FOV by another visible to the camera FOV, we promote sparsity in our estimate of $\svecbarh$, resulting in the $\ell_1$-regularized problem
\begin{equation}
    [\hat{\svecbar}_{\rm h}, \hat{\cvec}]
    = \argmin_{\svecbarh,\cvec}
    \left[ \frac{1}{2} \left\| \yvec - \Amat \cvec - \Dmatbar \svecbarh \right\|_2^2 + \lambda \norm{\svecbarh}_1 \right]\!,
    \label{eq:OptimizationProblem_LinearIP_L1}
\end{equation}
where $\lambda$ is the regularization parameter.
The optimization problem \eqref{eq:OptimizationProblem_LinearIP_L1} is efficiently solved using the FISTA algorithm~\cite{Beck2009}.

\begin{figure}
    \centering
    \begin{subfigure}{0.53\linewidth}
      \includegraphics[width=\linewidth]{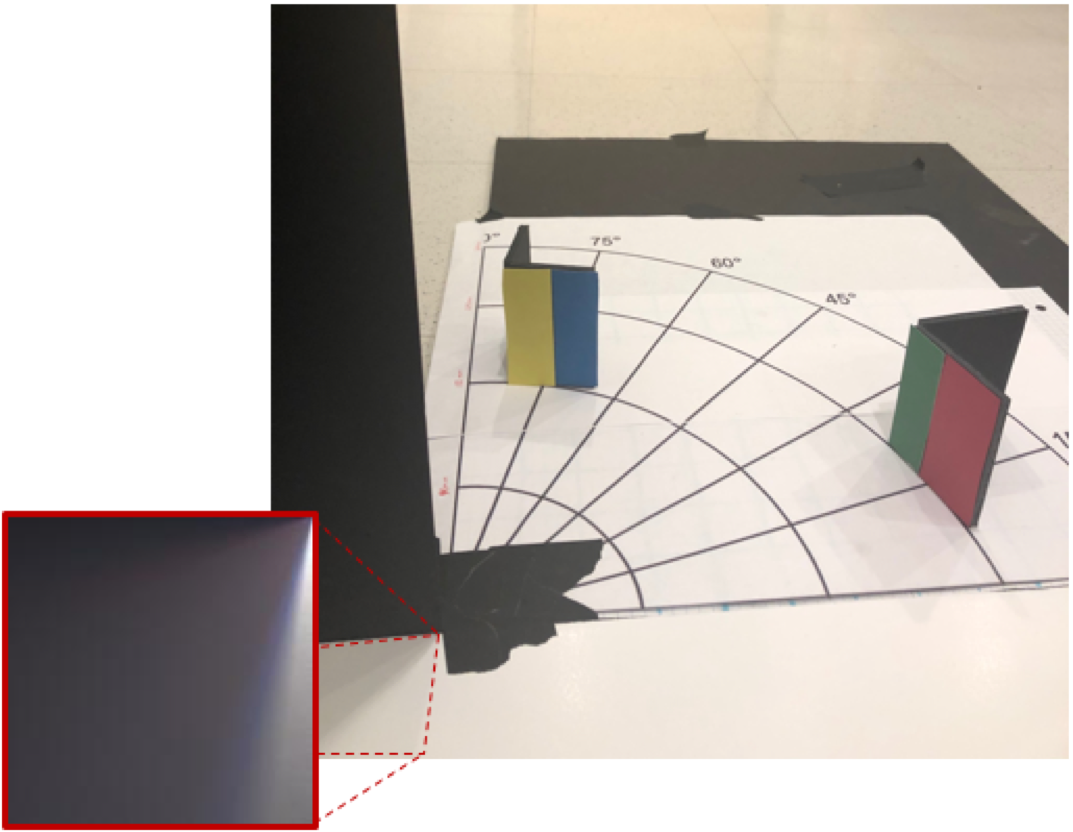}
      \captionsetup{justification=centering}
      \caption{\footnotesize RGB color measurement $\yvec$ and true hidden scene.}
      \label{fig:ColorStripes_measurement1}
    \end{subfigure}
    \\
    \begin{tabular}{cc}
    \begin{subfigure}{0.43\linewidth}
      \includegraphics[width=\linewidth]{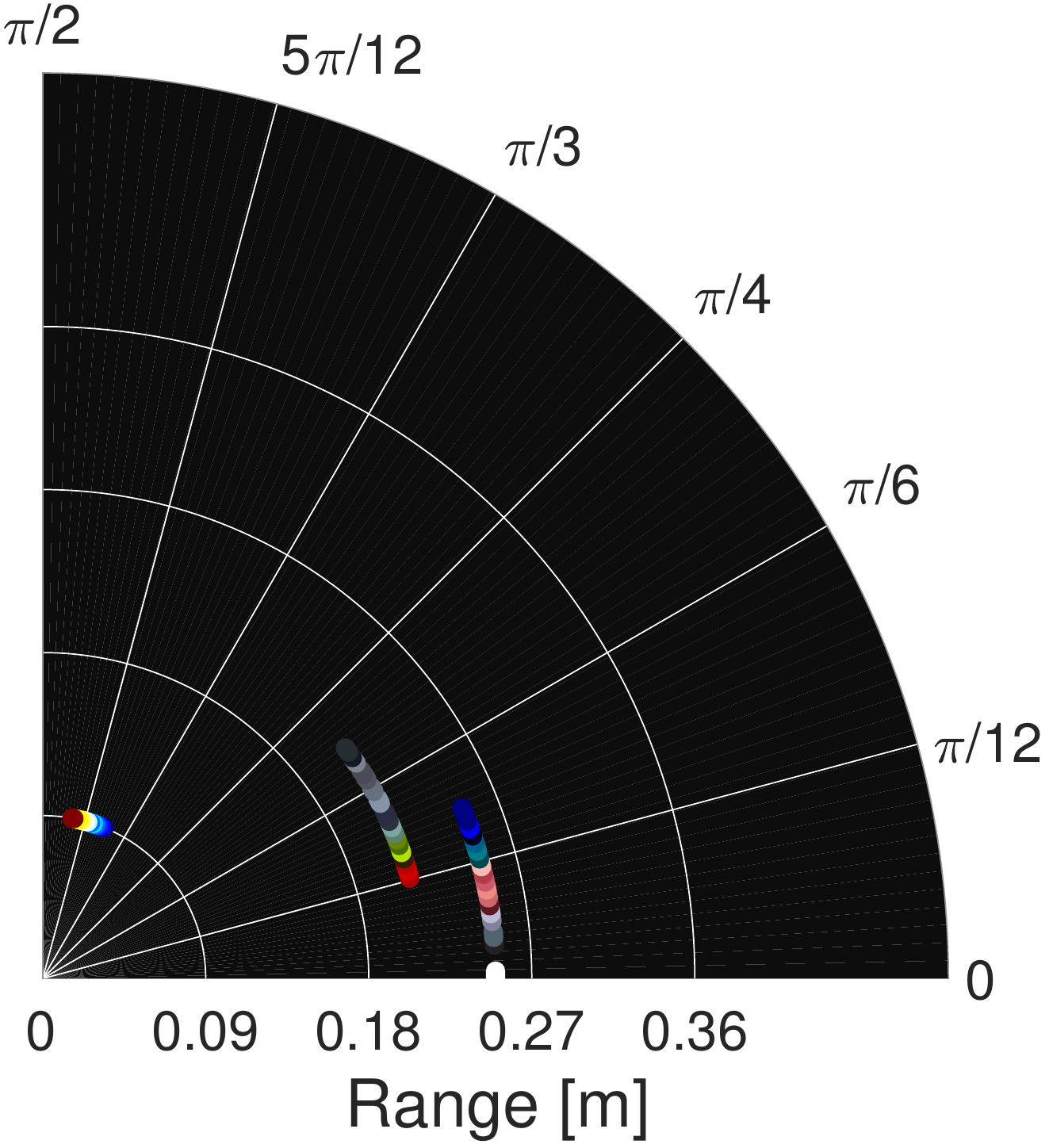}
       \captionsetup{justification=centering}
       \caption{\footnotesize $N=90$, $L=10$}
        \label{fig:LinearInversionRecons_90x10}
    \end{subfigure} 
    & 
    \begin{subfigure}{0.43\linewidth}
    \centering
      \includegraphics[width=\linewidth]{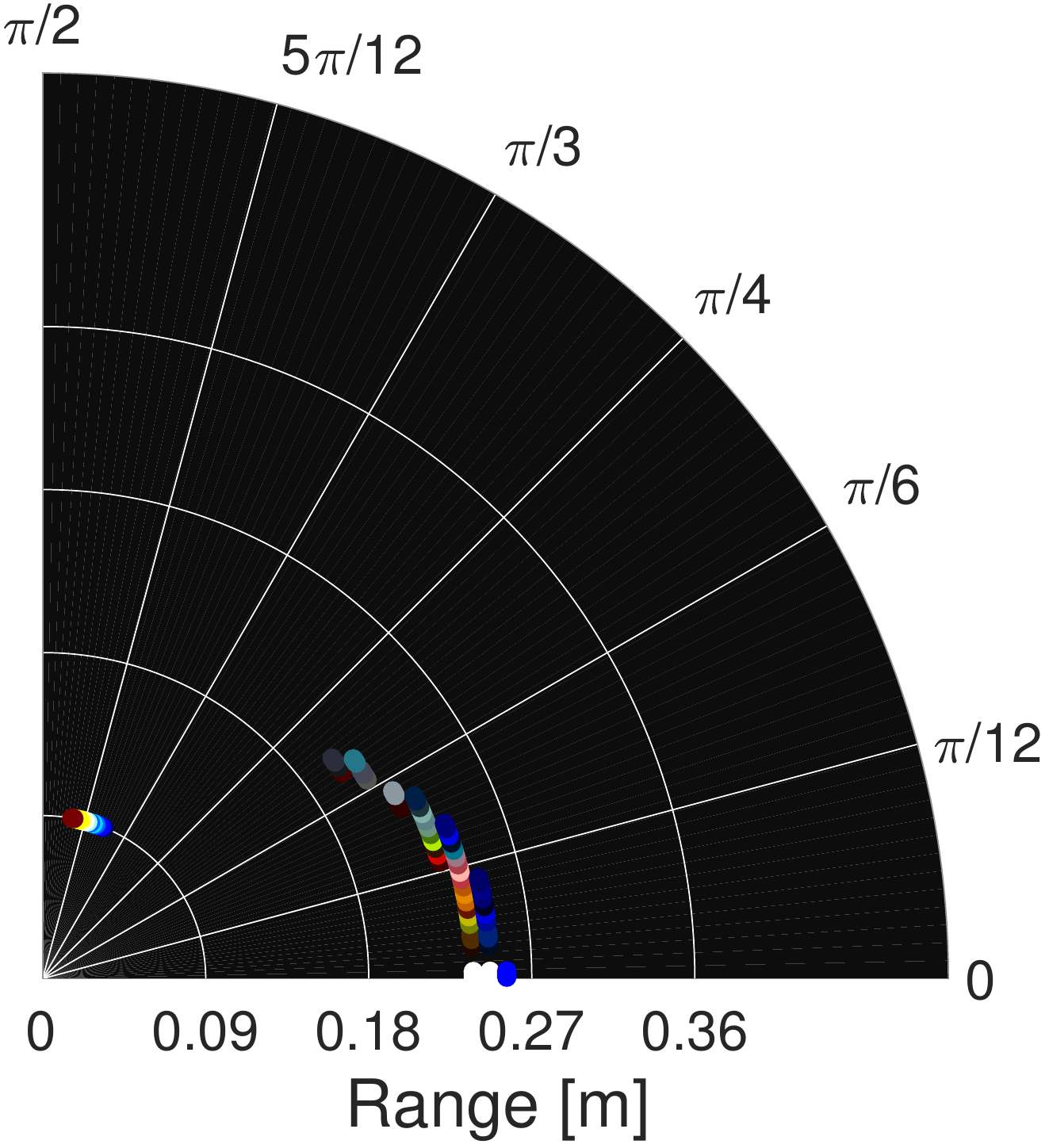}
      \captionsetup{justification=centering}
      \caption{\footnotesize $N=90$, $L=40$}
      \label{fig:LinearInversionRecons_90x40}
    \end{subfigure}
    \end{tabular}
    \caption{Demonstration of linear inversion algorithm for hidden scene and measurement shown in (a).
    The hidden region is discretized into $N$ angles and $L$ ranges.
    }
    \label{fig:LinearInversionRecons}
\end{figure}

We evaluate the linear model approach for the hidden scene and measurement in
\Cref{fig:ColorStripes_measurement1}.
Reconstructions at range resolutions of $L=10$ and $L=40$ are shown in \Cref{fig:LinearInversionRecons_90x10} and \Cref{fig:LinearInversionRecons_90x40}  for angular resolution $N=90$.
Both reconstructions exhibit two clusters of pixels with intensities larger than zero, corresponding to the two hidden objects in the scene.
While the relative order of the objects is correct, the yellow-blue stripe is estimated to be closer than its true location in both reconstructions.
Both targets are reconstructed with mostly correct color content, though several angular bins have different range estimates across the three different color channels causing some misalignment in the reconstructions.
Although both targets are at an approximately constant range across their angular extent, this is not the case in both reconstructions, particularly in the more coarse reconstruction of \Cref{fig:LinearInversionRecons_90x10}. Our nonlinear, more physically-inspired, model addresses some of these challenges.

\begin{myRem}
\label{rem:SGL}
Under the assumption of at most one target in any angular bin, each sub-vector $\svec_{{\rm h}_l}$ in $\svecbarh $ is either $1$-sparse or zero.
Combining this with the existence of only a few targets means that there is a small number of $1$-sparse groups in $\svecbarh$,
i.e., sparsity both within and across groups.
This could be incorporated by solving a Sparse-Group Lasso problem~\cite{Simon2013_SGL}:
\begin{equation}
    \argmin_{\svecbarh, \cvec}
    \left[
    \frac{1}{2} \left\| \yvec - \Amat \cvec - \Dmatbar \svecbarh \right\|_2^2 + \lambda_1 \sum_{l=1}^L \norm{\svec_{{\rm h}_l}}_2 + \lambda_2 \norm{\svecbarh}_1
    \right]\!.
    \label{eq:OptimizationProblem_LinearIP_SGL}
\end{equation}
Empirically, we found no compelling evidence that solving \eqref{eq:OptimizationProblem_LinearIP_SGL} is superior to solving \eqref{eq:OptimizationProblem_LinearIP_L1}.
Consequently, all results for the linear inverse problem~\eqref{eq:discreteforwardmodel_linear} are based on solving \eqref{eq:OptimizationProblem_LinearIP_L1}, separately, for each color channel.
\end{myRem}

\subsection{Nonlinear Modeling and Inversion}
\label{ssec:NonlinearModel}
In many practical scenarios, the hidden scene is composed of only a few hidden targets of interest, with each target having some angular extent and being roughly at a constant distance from the corner.
Solving \eqref{eq:OptimizationProblem_LinearIP_L1} with fine range and angular discretization is computationally expensive;
similarly, finely discretizing the angular dimension and estimating a unique range value $\rho(\alpha_n)$ for each hidden-scene angle $\alpha_n$, using \eqref{eq:DiscreteForwardModel_NL_bg}, is unnecessarily ambitious.
Alternatively, we can assume that there is an unknown number $\Nt \ll N$ of disjoint targets to be estimated, each with unknown range and radiosity.
Mathematically,
\begin{equation}
    \Shbar (\rho, \alpha) = \sum_{j=1}^{\Nt} s_j(\alpha) \delta(\rho - \bar{\rho}_{j}) u\!\left(\frac{\alpha - \bar{\alpha}_{j}}{\Delta_{j}}\right)\!,
    \label{eq:Shbar_model}
\end{equation}
with the $j$th hidden target having
angular position $\bar{\alpha}_{j}$,
angular extent $\Delta_{j}$,
range $\bar{\rho}_{j}$, and
radiosity $s_j(\alpha)$;
$u(\cdot)$ is the zero-centered unit rectangular function.

Angular bins containing no detected targets are attributed to background, and the minimal light coming from those regions will be assumed to be coming from very far away.
Under the model \eqref{eq:Shbar_model},
instead of having $N$ different range values, contiguous elements of $\svec$ will have the same range $\bar{\rho}_j$ if they contain the same target.
Letting $\rhobarbm = [\bar{\rho}_1, \bar{\rho}_2, \ldots, \bar{\rho}_{N_{\rm t}}]\transpose$:
\begin{equation}
    \yvec = \Amat \cvec + (\Vmat \odot \Dmat(\rhobarbm))\svec + \bm{\epsilon},
    \label{eq:discreteforwardmodel_nonlinear1}
\end{equation}
where for any $m=1,2,\ldots,M$ and $n=1,2,\ldots,N$,
\begin{equation}
[\Dmat(\rhobarbm)]_{m,n} = \frac{\bar{\rho}_j}{d^2(r_m,\theta_m,\bar{\rho}_j,\alpha_n)}
\label{eq:dmat_form}
\end{equation}
when $\alpha_n \in \left[\bar{\alpha}_j - \Frac{\delta_{\bar{\alpha}_j}}{2}, \bar{\alpha}_j + \Frac{\delta_{\bar{\alpha}_j}}{2}\right)$.
Note that $\svec$ represents the discretization of $s_j(\alpha) u\left(\frac{\alpha - \bar{\alpha}_j}{\delta_{\bar{\alpha}_j}}\right)$ over $\alpha \in [\Frac{\pi}{2},\pi]~{\rm rads}$\@.
We propose to estimate $\svec$, $\widetilde{\rhovec}$ and $\cvec$ by solving
\begin{equation}
\begin{split}
\underset{\svec,\rhobarbm,\cvec }{\text{min}}\:&
 \bigg(\underbrace{\frac{1}{2} \norm{\yvec -\Amat \cvec - (\Vmat \odot \Dmat(\rhobarbm))\svec}^2_2}_{\text{data fidelity}} \\
 &+\underbrace{\lambda_1 \norm{\Wmat\svec}_1
 + \lambda_2 \norm{\Bmat\svec}_2^2+\iota_{[0,\infty)^{N}}(\svec )}_{\text{regularizers for $\svec$}}
  + \underbrace{\iota_{[c,\infty)^{N_{\rm t}}}(\rhobarbm )}_{\text{regularizer for $\rhobarbm$}}\bigg),
 \label{eq:NonLinearCostFun}
\end{split}
\end{equation}
where $\Wmat$ is a wavelet transform matrix (we use the Daubechies wavelet of order 4), $\Bmat$ returns the difference between subsequent entries in $\svec$ that are attributed to hidden-scene background terms, $\lambda_1$ and $\lambda_2$ are tuning parameters, and $$ \iota_{\mathcal{C}}(\mathbf{x})=
\begin{cases}
0, &\text{if } \mathbf{x}\in \mathcal{C};\\
\infty, &\text{otherwise}
\end{cases}$$
is the indicator function for a set $\mathcal{C}$.
In \eqref{eq:NonLinearCostFun}, the regularizers for $\svec$ promote sparsity in the wavelet basis, smoothness in hidden-scene background contributions, and positivity in $\svec$, respectively.
The regularizer for range $\rhobarbm$ enforces range estimates to be at least $c>0$ (a small constant). This optimization problem is solved using an alternating approach described below.
\begin{enumerate}
\item  \textbf{Initialize} $\svec$ and $\cvec$ by solving
\begin{equation}
\begin{split}
    [\svec^{\rm 0},\cvec^{\rm 0}] &= \argmin_{\svec, \cvec} \frac{1}{2}\norm{\yvec - \Amat \cvec - (\Vmat \odot \Dmat(\rhobarbm_0))\svec}^2_2  \\
    &\qquad\qquad+ \lambda \norm{\Wmat\svec}_1 + \iota_{[0,\infty)^{N}}(\svec),
\end{split}
    \label{eq:alternating_initialization}
\end{equation}
with $\rhobarbm^0 = \rho_{\rm FF} \bm{1}$, initialized to represent a single target ($N_{\rm t} = 1$) in the far field ($\rho_{\rm FF} \gg 0 $). Our motivation to first estimate $\svec$ is because, given $\rhobarbm^0$, the resulting problem is well-conditioned (\Cref{sec:CRB}).
\item \textbf{Determine number of targets} $N_{\rm t}$
by comparing $[\svec^0]_{n}$ to the threshold
$\kappa_{n} = \Frac{\alpha}{(2\ell+1)}\sum_{i=n - \ell}^{n + \ell} \left[\svec^{0}\right]_i$,
where $\alpha\in\mathbb{R}_{+}$ and (odd) filter length $(2\ell+1) \in\mathbb{Z}_{+}$ are tuneable parameters.
It is assumed that $[\svec^{t-1}]_n = 0$ for
$n \notin \{1,2,\ldots,N\}$.
Consecutive threshold crossings in $\svec^0$ represent the edges of a single target.

\item \textbf{Update} $\rhobarbm^t$ by
\begin{equation}
\begin{split}
[\rhobarbm^t, \zvec^t] = &\argmin_{\rhobarbm, \zvec} \frac{1}{2}\norm{\yvec - \Amat \cvec^{t{-}1} {-}(\Vmat\!\odot\!\Dmat(\rhobarbm))\svec^{t{-}1}}^2_2  \\
                    &+\iota_{[c,\infty)^{N_{\rm t}}}(\rhobarbm)+\iota_{[0,\infty)^{N_{\rm t}}}(\zvec),
\end{split}
\label{eq:rho_update}
\end{equation}
where $\zvec=[z_1, z_2, \ldots,z_{N_{\rm t}}]\transpose \in \mathbb{R}_{+}^{N_{\rm t}}$ is 
such that for any $n=1,2,\ldots,N$,
\begin{equation*}
    [\Dmat(\rhobarbm)]_{m,n} = \frac{z_j \tilde{\rho}_j} {d^2\left(r_m,\theta_m,\tilde{\rho}_j,\alpha_n\right)}
\end{equation*}
when $\alpha_n \in \left[\tilde{\alpha}_j - \Frac{\delta_{\tilde{\alpha}_j}}{2}, \tilde{\alpha}_j + \Frac{\delta_{\tilde{\alpha}_j}}{2}\right)$.
The introduction of $\zvec$ couples the minimization problems \eqref{eq:rho_update} and \eqref{eq:s_update}, permitting radiosities $\svec^{t-1}$ to be scaled, appropriately, as $\rhobarbm$ is updated.
\item \textbf{Update} $\svec$ and $\cvec$ by solving
\begin{equation}
\begin{aligned}
[\svec^{t},\cvec^{t}] &= \argmin_{\svec,\cvec} \frac{1}{2}\norm{\yvec - \Amat\cvec - (\Vmat\!\odot\!\Dmat(\rhobarbm^{t}))\svec}^2_2  \\
&+ \lambda_1 \norm{\Wmat\svec}_1 + \lambda_2 \norm{\Bmat\svec}_2^2 + \iota_{[0,\infty)^{N}}(\svec).
\end{aligned}
\label{eq:s_update}
\end{equation}
\item \textbf{Increment iteration} counter $t$ by one.
\item \textbf{Repeat} steps 3, 4 and 5 until convergence.
\item \textbf{Return} $\widehat{\rhobarbm} \leftarrow \rhobarbm^t$, and $\widehat{\svec} \leftarrow \svec^t$.
\end{enumerate}
\begin{figure}
    \centering
    \begin{tabular}{c@{}c@{}c}
    \begin{subfigure}{0.25\linewidth}
    \vspace{8mm}
      \includegraphics[width=\linewidth]{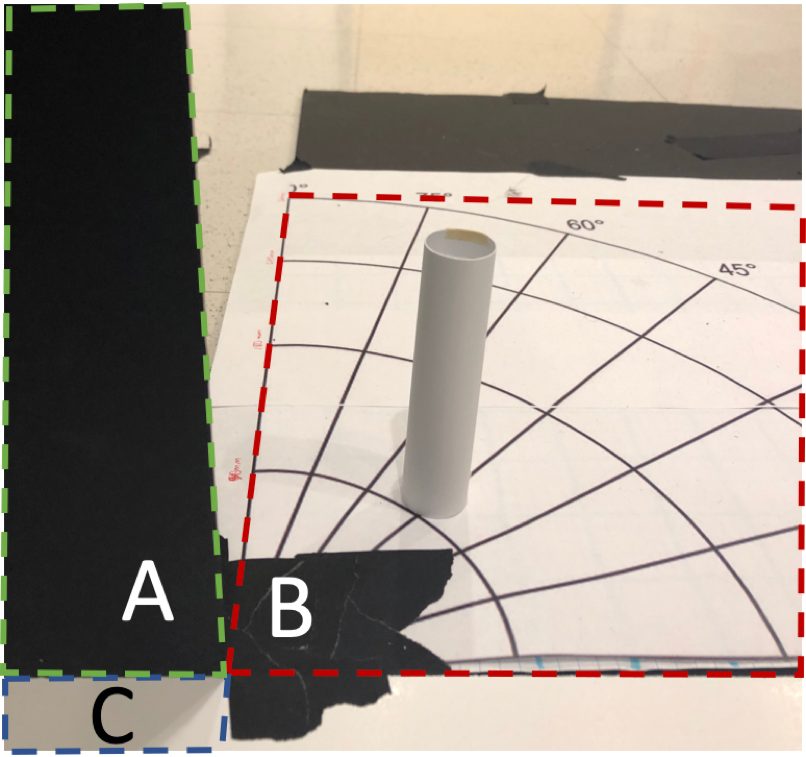}
       \captionsetup{justification=centering}
       \vspace{1mm}
       \caption{\footnotesize  Photograph of hidden scene }
        \label{fig:demoPhoto}
    \end{subfigure} 
    & 
    \begin{subfigure}{0.33\linewidth}
    \centering
     \includegraphics[width=\linewidth]{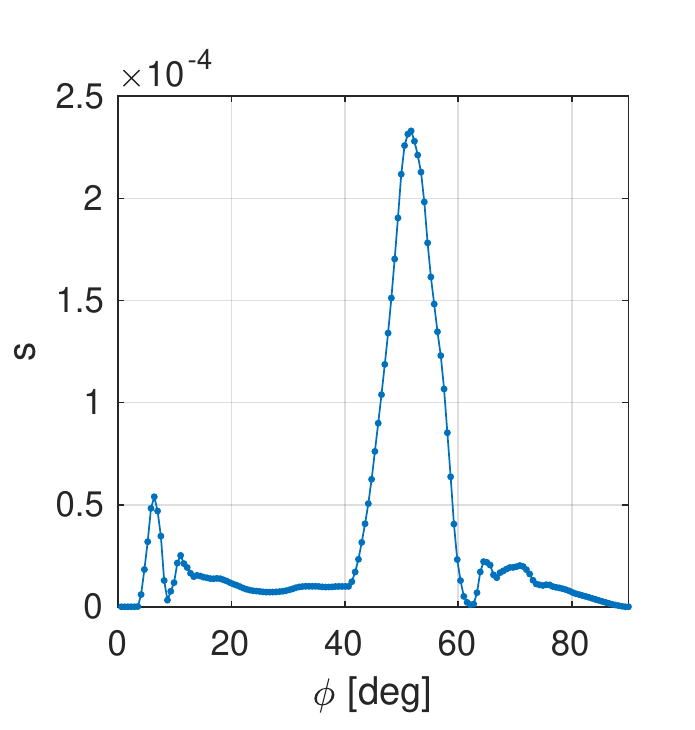}
      \captionsetup{justification=centering}
      \caption{\footnotesize Initial estimate ${\svec_0}$}
      \label{fig:SingleTarget_Example_InitialEst}
    \end{subfigure}
    &
    \begin{subfigure}{0.33\linewidth}
    \centering
      \includegraphics[width=\linewidth]{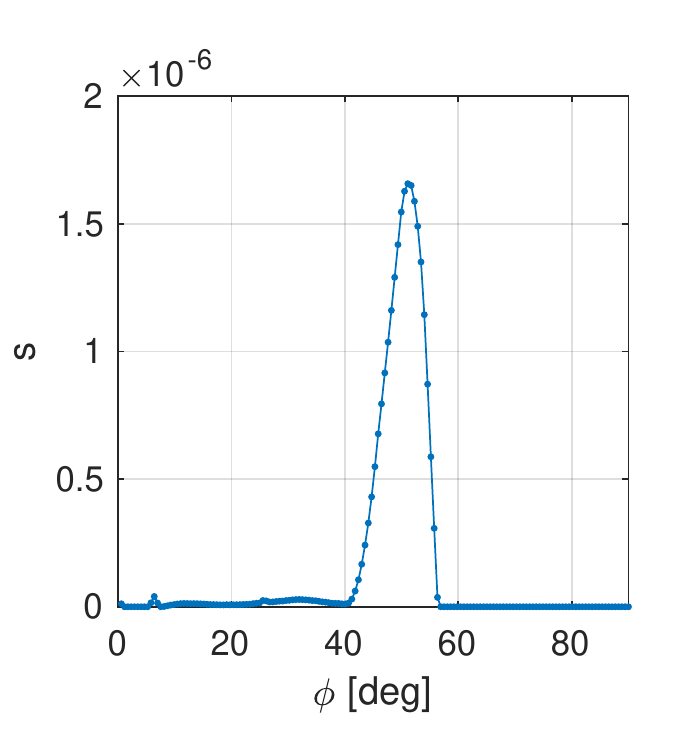}
      \captionsetup{justification=centering}
      \caption{\footnotesize Final estimate $\hat{\svec}$ }
      \label{fig:SingleTarget_Example_FinalEst}
    \end{subfigure}
    \\
   \begin{subfigure}{0.31\linewidth}
   \vspace{-6mm}
      \includegraphics[width=\linewidth]{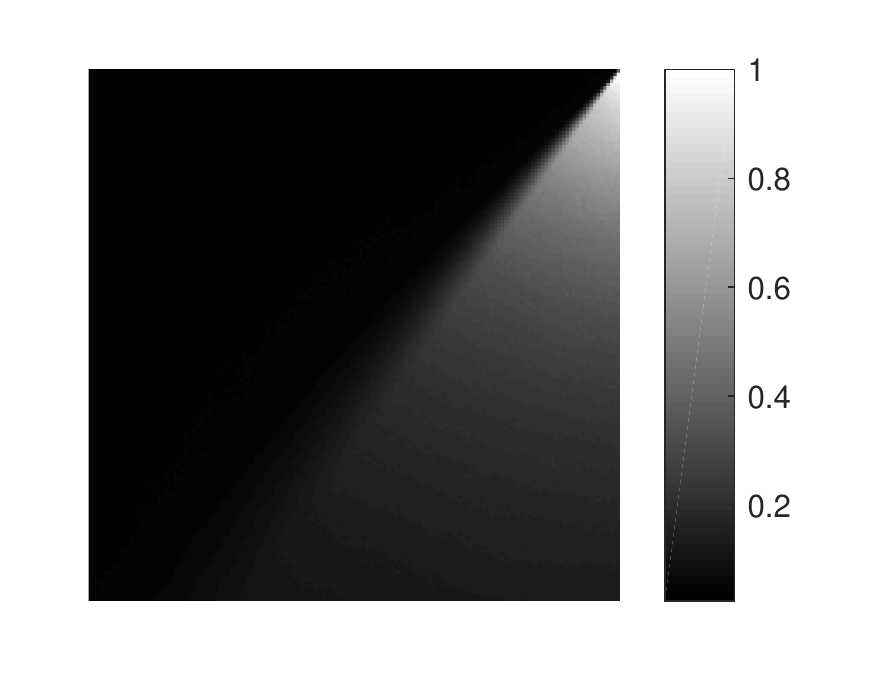}\vspace{+1mm}
      \captionsetup{justification=centering}
      \caption{\footnotesize Measurement $\yvec$}
      \label{fig:SingleTarget_Example_Meas}
    \end{subfigure}
    &
    \begin{subfigure}{0.31\linewidth}
      \includegraphics[width=\linewidth]{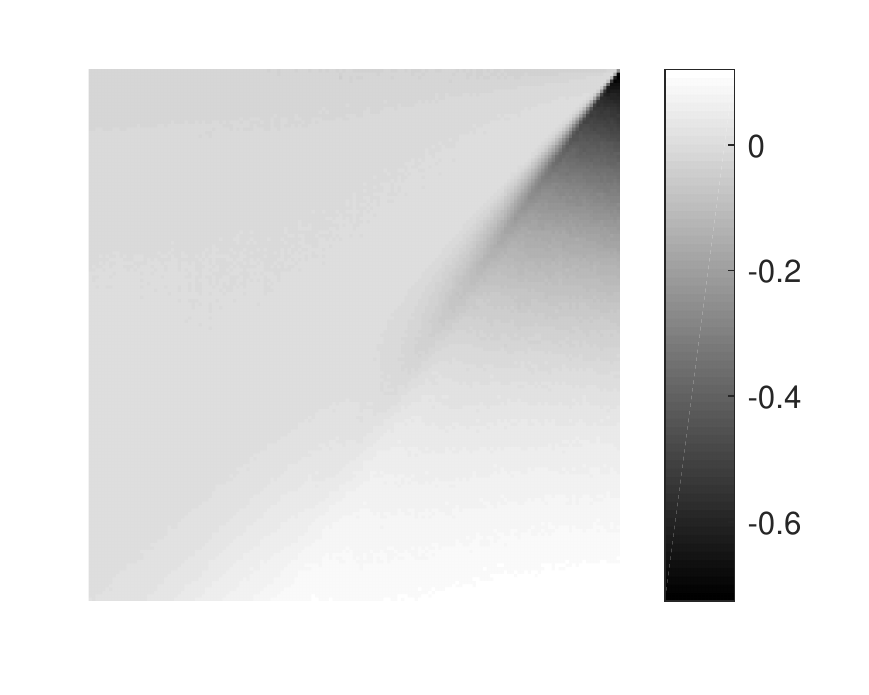}
      \captionsetup{justification=centering}
      \caption{\footnotesize Initial residual: $\yvec - \Amat\cvec^0 - \left(\Vmat \! \odot \! \Dmat(\rhobarbm^0)\right)\svec^0$}
      \label{fig:SingleTarget_Example_Residual}
    \end{subfigure}
    &
    \begin{subfigure}{0.31\linewidth}
    \vspace{-3.5mm}
      \includegraphics[width=\linewidth]{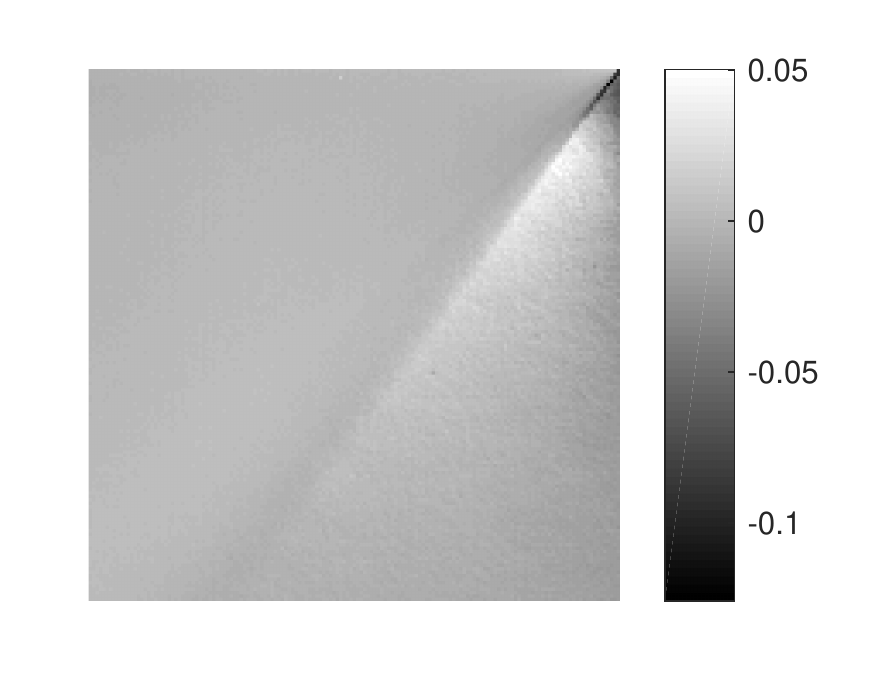}
      \captionsetup{justification=centering}
      \caption{\footnotesize Final residual: $\yvec - \Amat{\hat{\cvec}}-(\Vmat \odot \Dmat(\widehat{\rhobarbm}))\hat{\svec}$}
      \label{fig:singleTarget_Example_FinalResid}
    \end{subfigure}
    \end{tabular}
    \caption{Demonstration of model mismatch with scene initialization. The hidden scene was a narrow white cylinder with a diameter of 2.5cm, 18cm away from the corner, at $\phi = 45^{\rm o}$, as shown in (a).  When all light is assumed to originate in the far field, the initial estimate of $\svec_{\rm 0}$ (b) does not describe the radial falloff that is present in the measurement $\yvec$ (d), as shown in the initial residual (e). After the algorithm converges, the range estimate is updated allowing for a more accurate estimate of $\svec$ as shown in (c) with a much smaller residual (f).
    }
    \label{fig:singleTarget_Example}
\end{figure}
Steps 1, 3, and 4 are solved using projected gradient methods~\cite{Beck2009}.
\Cref{fig:singleTarget_Example} illustrates several algorithm steps for a scene containing a single hidden cylinder, shown in \Cref{fig:demoPhoto}, resulting in measurement $\yvec$, shown in \Cref{fig:SingleTarget_Example_Meas}.
The final estimate of $\svec$ (\Cref{fig:SingleTarget_Example_FinalEst}) does not contain the artifacts seen in the initial estimate $\svec_0$ (\Cref{fig:SingleTarget_Example_InitialEst}) because, instead of assuming the hidden scene is in the far field, the model has been updated to include the effects of radial falloff due to a target at estimated distance $\widehat{\rhobarbm}$.
In fact, the residual due to the initial far field assumption (\Cref{fig:SingleTarget_Example_Residual}) clearly contains unmodeled radial falloff, whereas the final residual (\Cref{fig:singleTarget_Example_FinalResid}) exhibits a much better overall fit.

\subsection{Nonlinear RGB Model Inversion}
\label{ssec:NonlinearRGB}
The algorithm described in \Cref{ssec:NonlinearModel}, which operates on a single measurement channel, may be adapted to operate on color (RGB) data. In this case, the camera measures $\yvec_{\rm R}$, $\yvec_{\rm G}$, and $\yvec_{\rm B}$ corresponding to each color channel.
Although our goal is still to estimate range values $\rhobarbm\in\mathbb{R}_{+}^{N_{\rm t}}$, we now seek radiosity estimates  $\widehat{\svec}_{\rm R}$, $\widehat{\svec}_{\rm G}$, and $\widehat{\svec}_{\rm B}$, as well as estimates of ambient light $\widehat{\cvec}_{\rm R}$, $\widehat{\cvec}_{\rm G}$, and $\widehat{\cvec}_{\rm B}$.
These estimates are obtained by solving \eqref{eq:NonLinearCostFun} with substitutions
\begin{align*}
  \yvec &\rightarrow \widetilde{\yvec}
                     = {\rm vec}\left(\left[\yvec_{\rm R}, \yvec_{\rm G}, \yvec_{\rm B}\right]\right)
                     \in \mathbb{R}^{3M}, \\
  \Amat &\rightarrow \widetilde{\Amat}
                     = \diag \left([\bm{1}, \avec_{\rm R}], [\bm{1}, \avec_{\rm G}], [\bm{1}, \avec_{\rm B}] \right)
                     \in \mathbb{R}^{3M \times 6}, \\
  \Vmat\odot\left(\Dmat(\rhobarbm)\right)
        &\rightarrow \widetilde{\Dmat}(\rhobarbm)
                     = \left(\Vmat\odot\Dmat(\rhobarbm)\right)\otimes \Imat
                     \in \mathbb{R}^{3M \times 3N}, \\
  \Bmat
        &\rightarrow \widetilde{\Bmat}
                     = \Bmat\otimes\Imat
                     \in \mathbb{R}^{3(N-1) \times 3N},
\end{align*}
where $\Imat$ is the $3\times3$ identity matrix.
The optimization becomes
\begin{equation}
\begin{split}
    \min_{\rhobarbm,\widetilde{\svec},\widetilde{\cvec}}  \, &\biggr( \frac{1}{2} \norm{\widetilde{\yvec} -\widetilde{\Amat}\widetilde{\cvec}-\widetilde{\Dmat}(\rhobarbm)\widetilde{\svec}}^2_2 + \lambda_2 \norm{\widetilde{\Bmat}\widetilde{\svec}}_2^2
    \\&
    + \lambda\norm{\Wmat\widetilde{\svec}}_1 + \iota_{[0,\infty)^{3N}}(\widetilde{\svec}) + \iota_{[c,\infty)^{N_{\rm t}}}(\rhobarbm) \biggr),
\end{split}
\label{eq:NonlinearMinimizationProblem_RGB}
\end{equation}
which, like before, is solved using an alternating approach, performing initial thresholding, or target counting, on
$\bar{\svec} = \frac{1}{3}(\svec_{\rm R}^0+\svec_{\rm G}^0+\svec_{\rm B}^0)$,
with \eqref{eq:rho_update} modified to update
$\widetilde{\zvec} = [\zvec_{\rm R}; \zvec_{\rm G}; \zvec_{\rm B}]\in\mathbb{R}^{3N_{\rm t}}$ instead of $\zvec$:
\begin{equation}
\begin{split}
[\rhobarbm^t,\widetilde{\zvec}^t] &= \argmin_{\rhobarbm,\widetilde{\zvec}} \frac{1}{2}\norm{\widetilde{\yvec}-\Amat\widetilde{\cvec}^{t-1}-(\widetilde{\Dmat}(\rhobarbm))\widetilde{\svec}^{t-1}}^2_2  \\
&\qquad\qquad+  \iota_{[c,\infty)^{N_{\rm t}}}(\rhobarbm)+\iota_{[0,\infty)^{3N_{\rm t}}}(\widetilde{\zvec}).
\end{split}
\label{eq:rho_update_RGB}
\end{equation}
The scene $\tilde{\svec}$ is updated by solving \eqref{eq:s_update} in parallel for each color channel.
This concatenation of the color channel measurements enforces consensus among channels in the range estimate and angular extent of a given hidden target,
thus avoiding the spurious range estimates observed in
the $\ell_1$-regularized solutions of the linear inverse problem formulation
 (\Cref{fig:LinearInversionRecons}).

\section{Experimental Evaluation using Real Data}
\label{sec:experimental_results}

Performances of the algorithms presented in \Cref{ssec:NonlinearModel,ssec:NonlinearRGB} were evaluated in a variety of conditions,
using the scaled-down laboratory setup shown in annotated photograph \Cref{fig:demoPhoto}.
A tripod-mounted \emph{FLIR Grasshopper3} camera model GS3-U3-41S4C-C equipped with a Tamron M118FM16 lens was used to photograph the floor (C) on the visible side of occluding wall (A).
A tuneable light source, positioned behind the occluding wall, was used to illuminate the hidden scene region (B).
In this work, we reconstruct a region that extends $\Frac{\pi}{2}$ radians into the hidden scene.
In principle, the full $\pi$ radians of hidden scene may be reconstructed by extending the photograph region (C) to the right.

\subsection{Empirical Performance Evaluation: Single Target}
\label{ssec:EmpiricalPerf}
In order to evaluate performance, a single white cylindrical target, shown in \Cref{fig:demoPhoto}, was placed at different positions $(\rho_1,\phi_1)$ in range and angle. For each position, with a camera FOV of $0.16~ {\rm m} \times 0.16~{\rm m}$, 150 snapshots of the visible floor were taken. By combining 1, 2, 4, 8, and 20 randomly selected snapshots (without replacement), we emulated decreasing measurement noise levels. Estimates' bias and variance were computed using the recovery results from 60 repetitions of each configuration.
In each trial, scalar range parameter estimates $\widehat{\varphi}_1$ and angular profiles $\widehat{\svec}$ are recovered for the target, as shown in \Cref{fig:SingleTarget_Example_FinalEst}.
We use the peak value of $\widehat{\svec}$ (after up-sampling) as a proxy for $\widehat{\varphi}_1$, in order to compute its bias and variance.
We take the measured center, in angle, of the target as the true ${\varphi}_1$. While it is expected that they are close, this measured center of the cylinder may not exactly match the brightest illuminated region of the cylinder.

\begin{figure}
    \centering
    \begin{subfigure}{0.43\linewidth}
      \includegraphics[width=\linewidth]{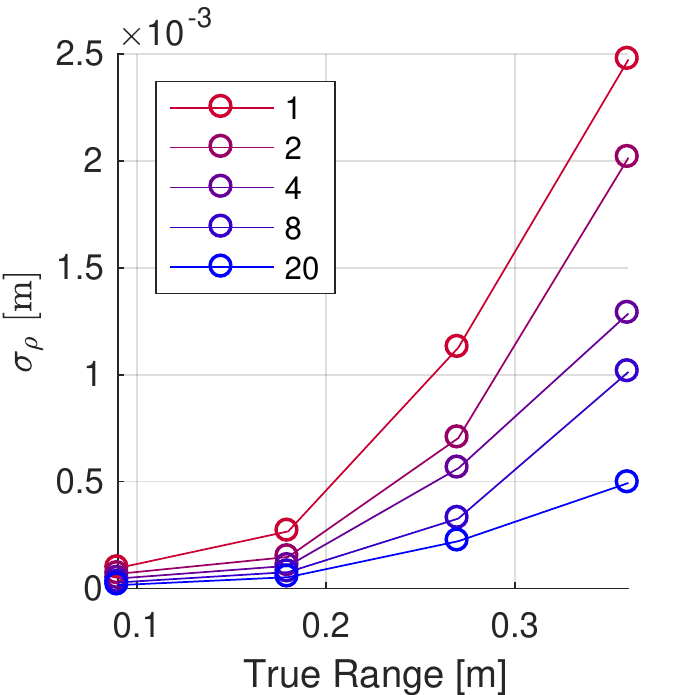}
      \captionsetup{justification=centering}
      \caption{\footnotesize  $\sigma_{\rho}$ vs. $\rho$ }
      \label{fig:RangeEstStdDev_vs_range}
    \end{subfigure}
    \begin{subfigure}{0.43\linewidth}
      \includegraphics[width=\linewidth]{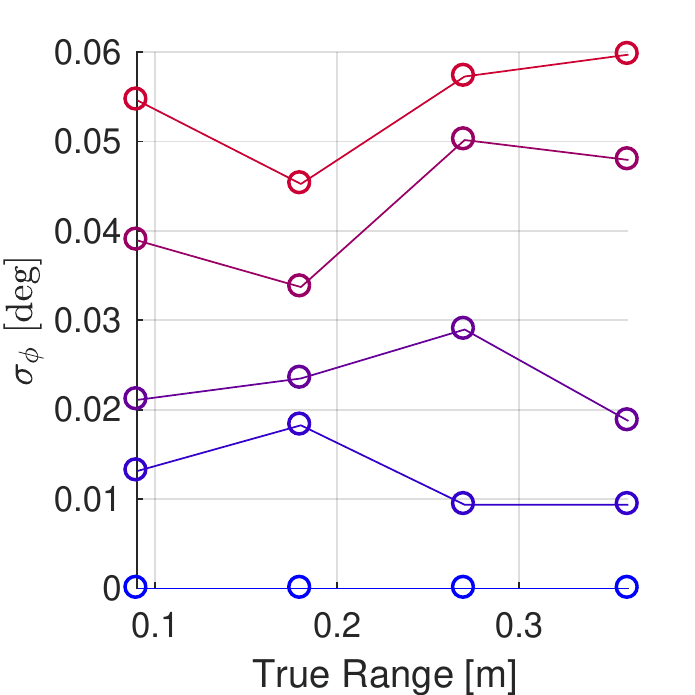}
      \captionsetup{justification=centering}
      \caption{\footnotesize $\sigma_{\varphi}$ vs $\rho$ }
      \label{fig:AngleEstStdDev_vs_range}
    \end{subfigure}\\
    \begin{subfigure}{0.43\linewidth}
      \includegraphics[width=\linewidth]{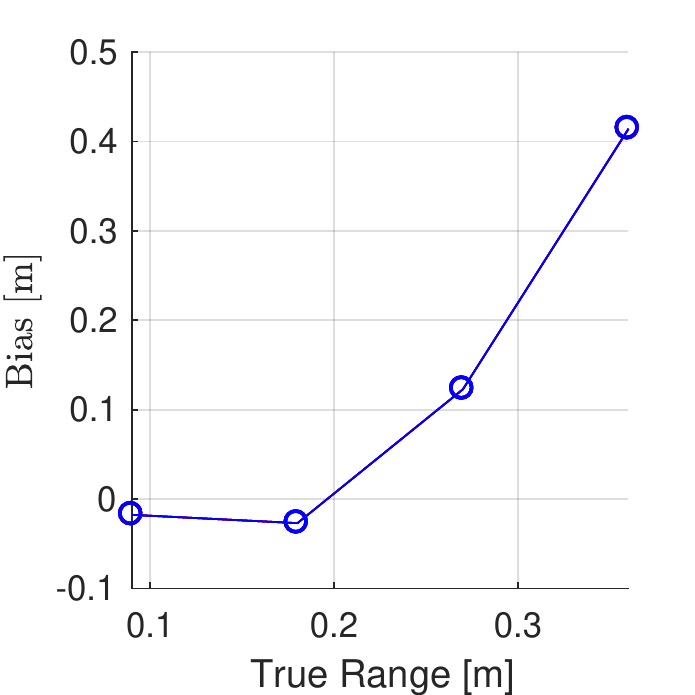}
      \captionsetup{justification=centering}
      \caption{\footnotesize ${\rm Bias}_\rho$ vs. $\rho$}
      \label{fig:RangeEstRMSE_vs_range}
    \end{subfigure}
    \begin{subfigure}{0.43\linewidth}
      \includegraphics[width=\linewidth]{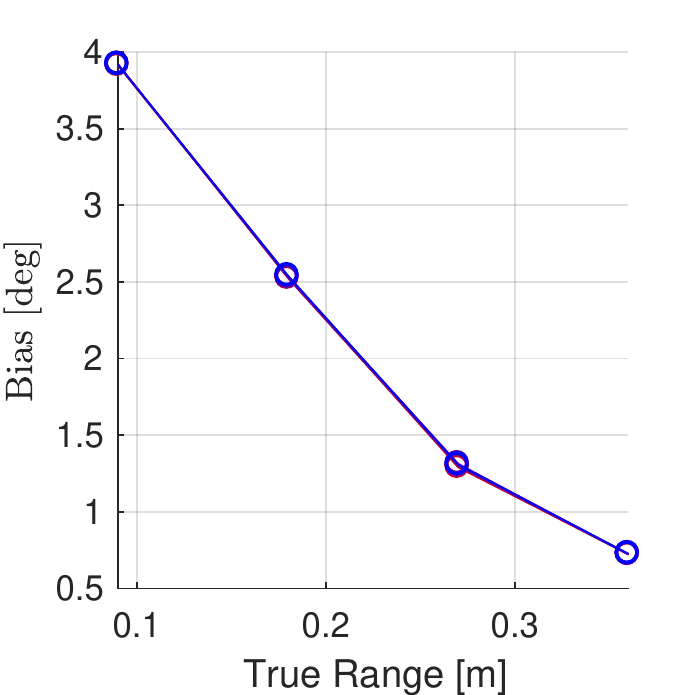}
      \captionsetup{justification=centering}
      \caption{\footnotesize ${\rm Bias}_\varphi$ vs. $\rho$  }
      \label{fig:AngleEstRMSE_vs_range}
    \end{subfigure}
    \caption{Evaluation of algorithm performance for a single target at four different ranges, in five different noise conditions, placed at $\varphi = 45^{\circ}$. The standard deviation of the range estimate (a), $\sigma_{\rho}$, increases with increasing range $\rho$, and the standard deviation of the angular estimate (b), $\sigma_{\varphi}$ is small at all ranges. Bias for the range estimate (c) increases with increasing range, while the bias for the angular estimate remains small at all ranges.
    }
    \label{fig:VaryingTargetRange}
\end{figure}

\begin{figure}
    \centering
    \begin{subfigure}{0.43\linewidth}
      \includegraphics[width=\linewidth]{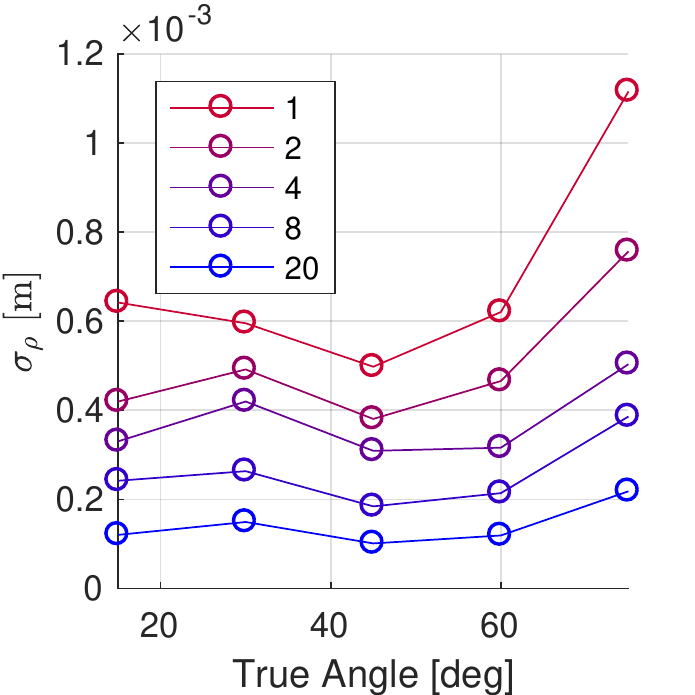}
      \captionsetup{justification=centering}
      \caption{\footnotesize $\sigma_\rho$ vs. $\phi$}
      \label{fig:RangeEstStdDev_vs_angle}
    \end{subfigure}
    \begin{subfigure}{0.43\linewidth}
      \includegraphics[width=\linewidth]{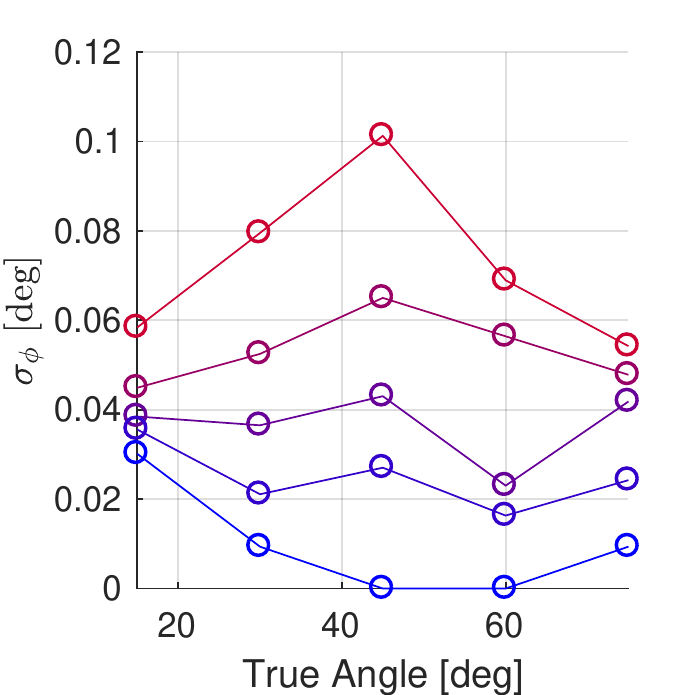}
      \captionsetup{justification=centering}
      \caption{\footnotesize $\sigma_\phi$ vs. $\phi$ }
      \label{fig:AngleEstStdDev_vs_angle}
    \end{subfigure}\\
    \begin{subfigure}{0.43\linewidth}
      \includegraphics[width=\linewidth]{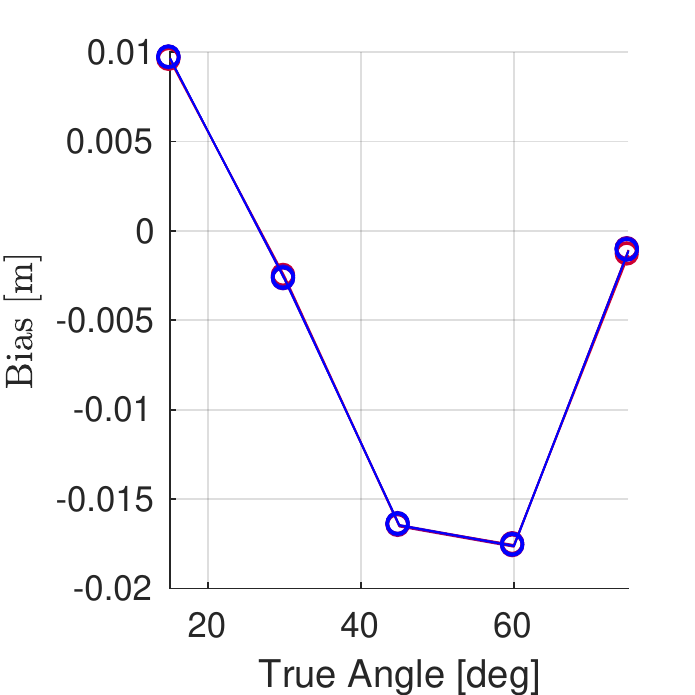}
      \captionsetup{justification=centering}
      \caption{\footnotesize ${\rm Bias}_{\rho}$ vs. $\phi$}
      \label{fig:RangeEstRMSE_vs_angle}
    \end{subfigure}
    \begin{subfigure}{0.43\linewidth}
      \includegraphics[width=\linewidth]{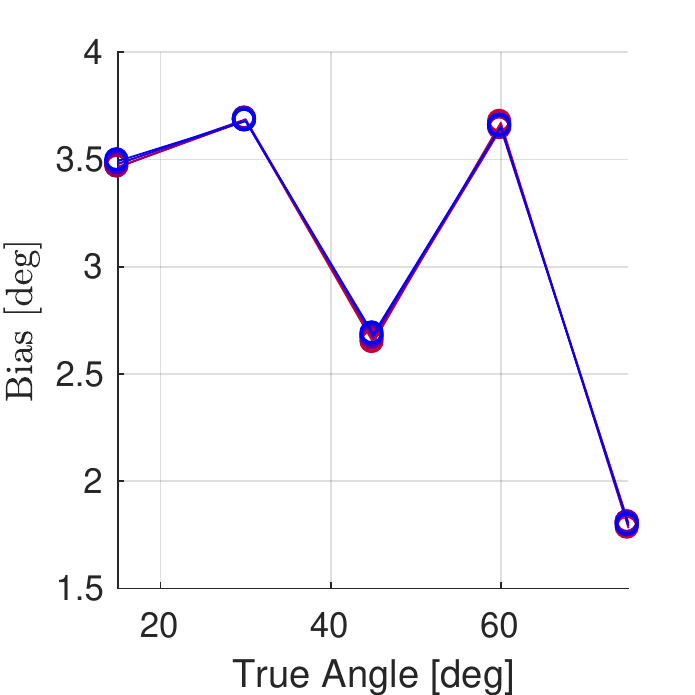}
      \captionsetup{justification=centering}
      \caption{\footnotesize  ${\rm Bias}_{\phi}$ vs. $\phi$ }
      \label{fig:AngleEstRMSE_vs_angle}
    \end{subfigure}
    \caption{Evaluation of algorithm performance for a single target at five different angles, in five different noise conditions, placed at $\rho = 0.18~{\rm m}$.
    The standard deviation of the range estimate (a), $\sigma_{\rho}$, is greatest when $\varphi = 75^{\circ}$, when the fewest pixels on the floor are exposed to penumbra. The standard deviation of the angular estimate (b), $\sigma_{\phi}$ is small at all angles.
    }
    \label{fig:VaryingTargetAngle}
\end{figure}

\subsubsection{Varying Range}
\Cref{fig:VaryingTargetRange} shows estimate bias and standard deviation computed for $\varphi_1 = 45^{\circ}$ and ranges $\rho_1 = 0.09~{\rm mm}$, $0.18~{\rm mm}$, $0.27~{\rm mm}$, and $0.36~{\rm mm}$.
As shown in \Cref{fig:RangeEstStdDev_vs_range}, range estimate standard deviation increases in noisier conditions (i.e., fewer combined frames) and at greater ranges.
\Cref{fig:AngleEstStdDev_vs_range} shows that, as predicted by the CRB analyses in~\Cref{sec:CRB}, the standard deviation of estimate $\widehat{\varphi}_1$ remains small at every position in range.
 
\Cref{fig:RangeEstRMSE_vs_range} and \Cref{fig:AngleEstRMSE_vs_range} show the bias for range and angle estimates respectively, at the four ranges in the same five noise conditions. For both range and angle estimates, bias is constant at a given range, regardless of the noise level. In both cases, the bias is orders of magnitude larger than the corresponding standard deviation. For the range estimate, we attribute this bias to model mismatch due to unmodelled reflections, nonzero target height, and edge imperfections. As shown in \Cref{fig:AngleEstRMSE_vs_range}, angular bias is much smaller, and may correctly reflect the fact that the brightest part of the cylinder changes in angle, as the cylinder moves with respect to the fixed hidden scene illumination.

\subsubsection{Varying Angle}
\Cref{fig:VaryingTargetAngle} shows estimate standard deviation and bias for fixed range  $\rho_1 = 0.18{\rm m}$ a set of angles $\phi_1 = 15^{\rm o}$, $30^{\rm o}$, $45^{\rm o}$, $60^{\rm o}$, and $75^{\rm o}$. We similarly observe lower estimate standard deviation in less noisy conditions, greater standard deviation in range than angle, and substantially higher bias than standard deviation for both range and angle.

\begin{figure*}
    \centering
    \begin{subfigure}{0.15\textwidth}
      \includegraphics[width=\linewidth]{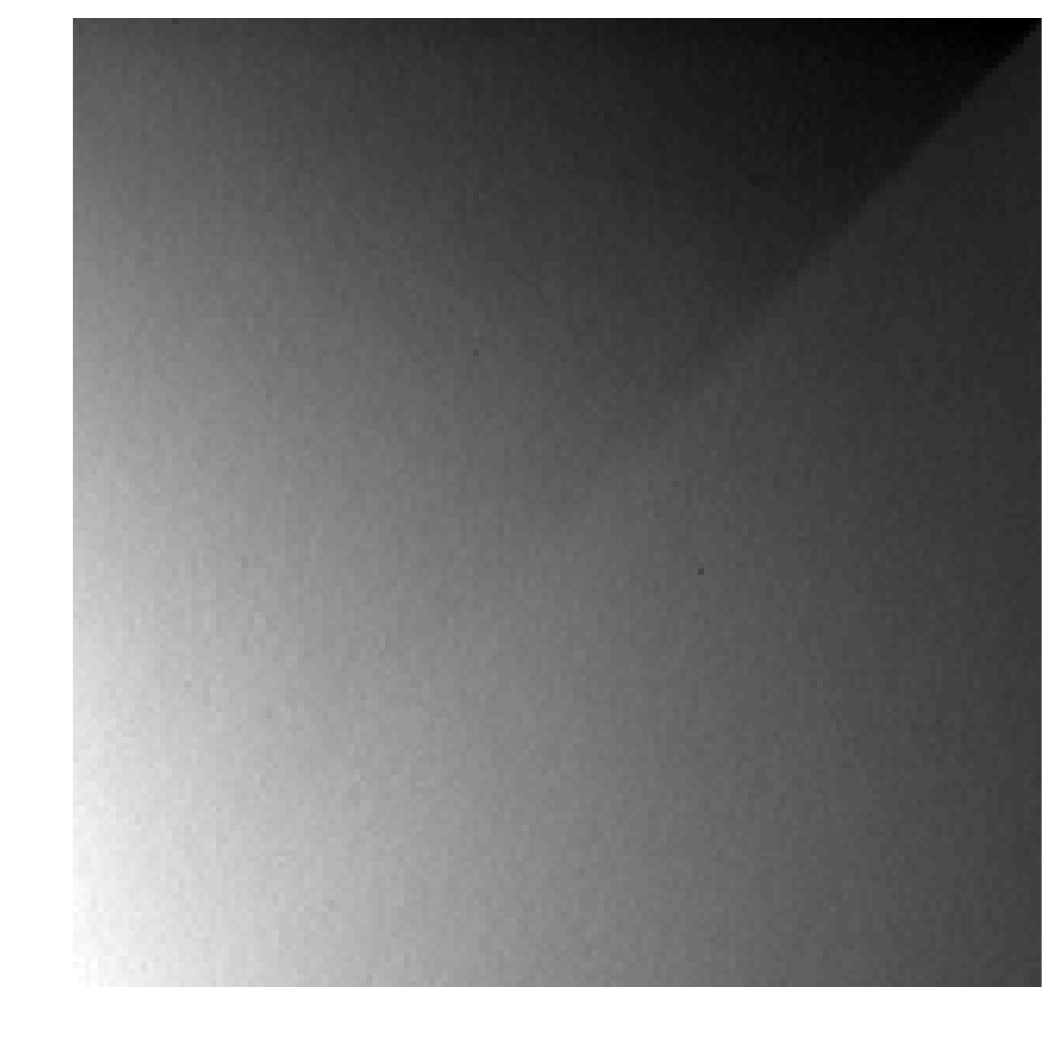}
      \captionsetup{justification=centering}
      \caption{\footnotesize $\yvec_1$}
      \label{fig:meas100}
    \end{subfigure}
    \begin{subfigure}{0.15\textwidth}
      \includegraphics[width=\linewidth]{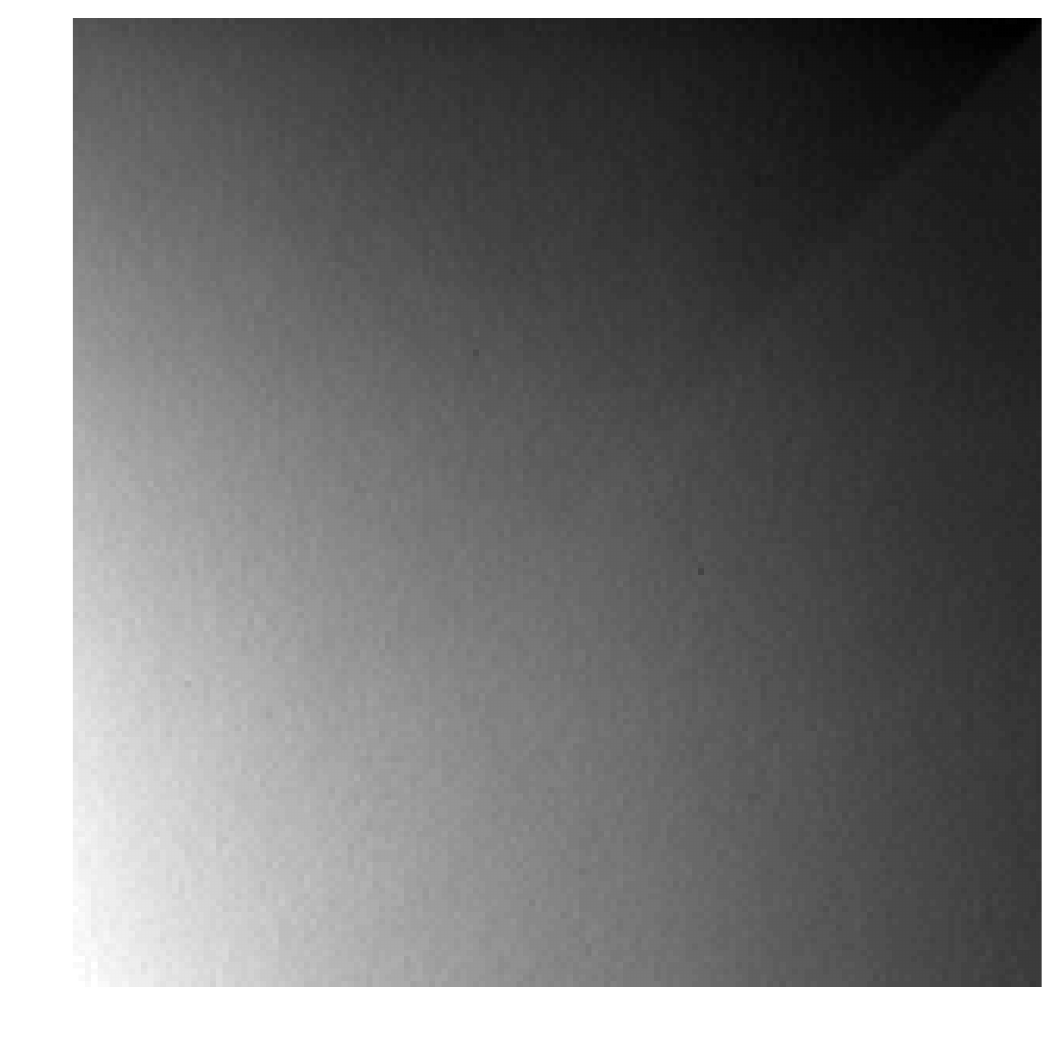}
      \captionsetup{justification=centering}
      \caption{\footnotesize $\yvec_2$}
      \label{fig:meas50}
    \end{subfigure}  
    \begin{subfigure}{0.15\textwidth}
      \includegraphics[width=\linewidth]{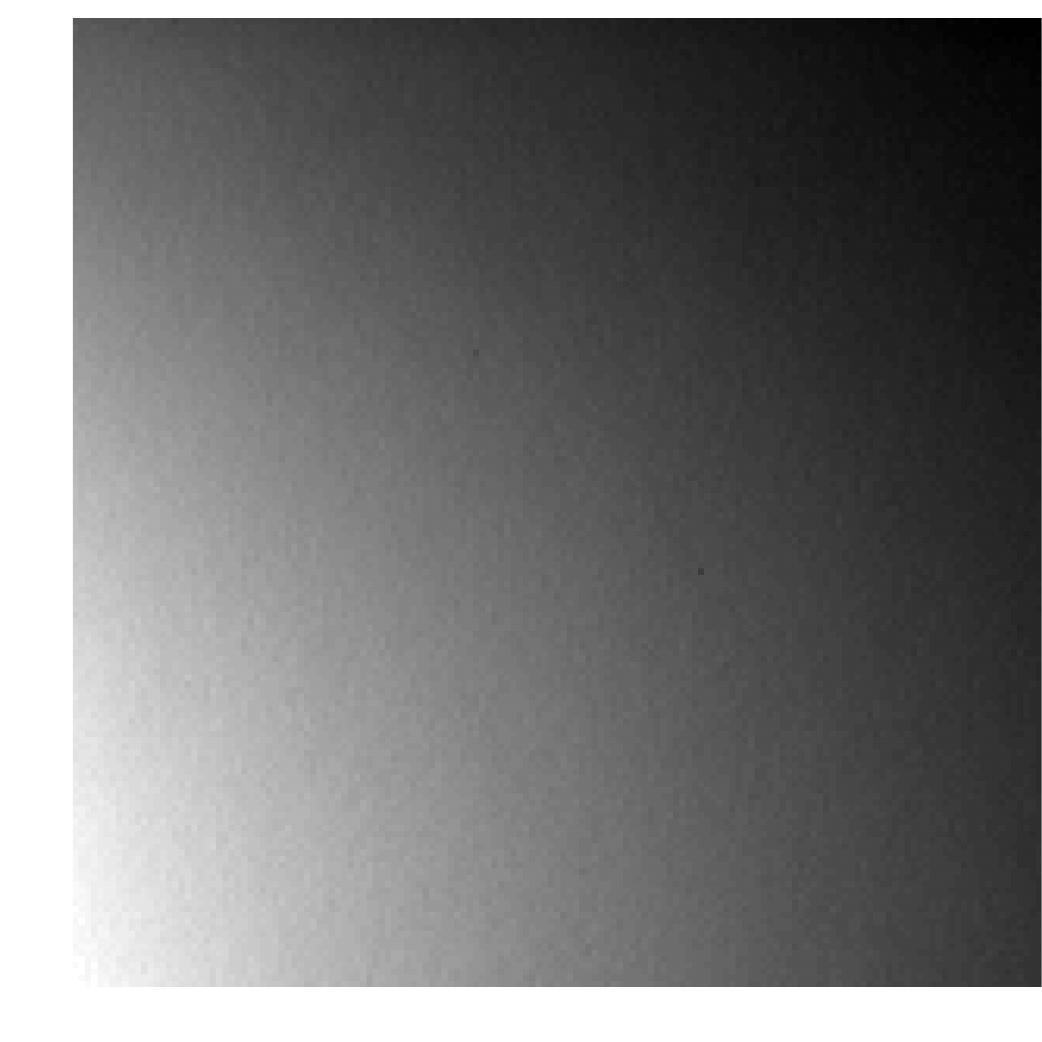}
      \captionsetup{justification=centering}
      \caption{\footnotesize $\yvec_3$}
      \label{fig:meas10}
    \end{subfigure}   
    \begin{subfigure}{0.15\textwidth}
      \includegraphics[width=\linewidth]{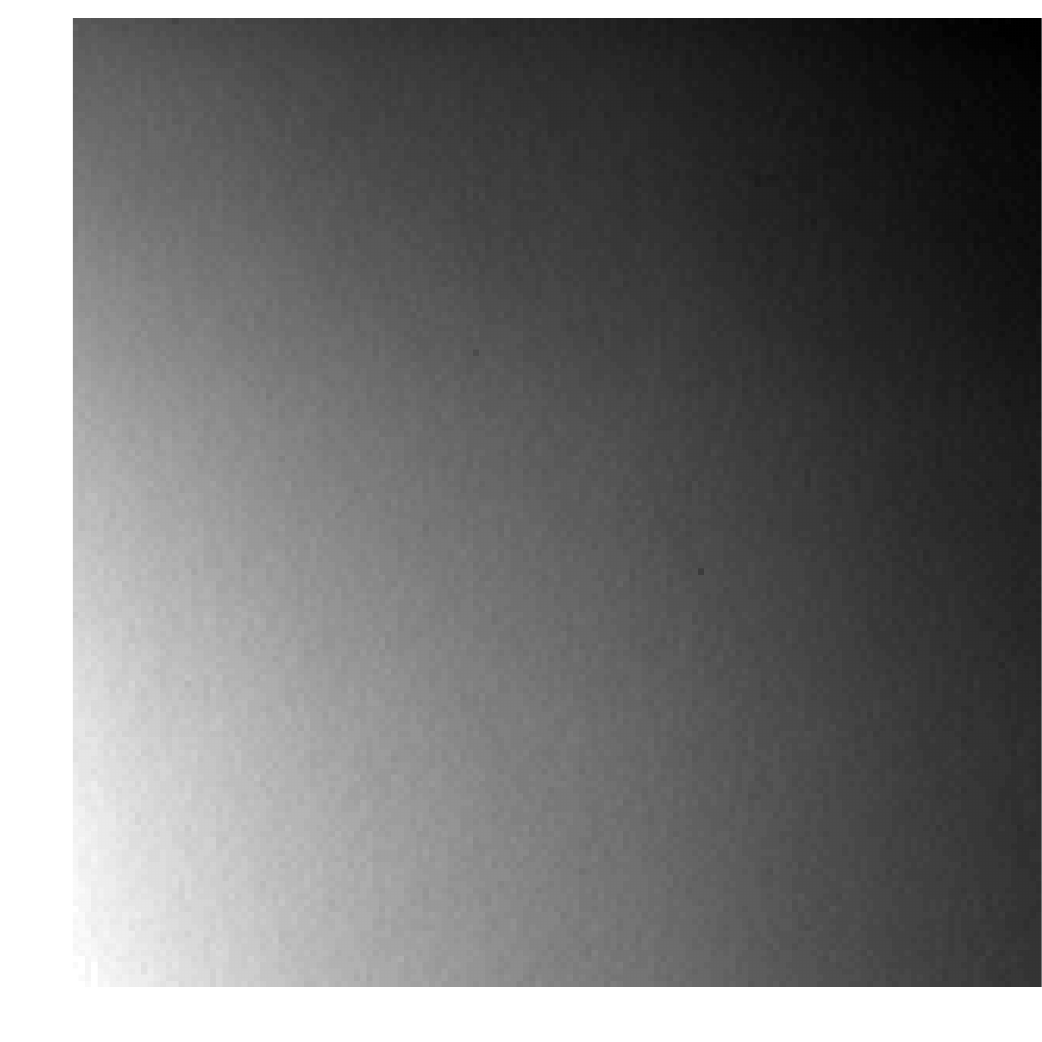}
      \captionsetup{justification=centering}
      \caption{\footnotesize $\yvec_4$}
      \label{fig:meas05}
    \end{subfigure}\\
    \begin{subfigure}{0.2\textwidth}
      \includegraphics[width=\linewidth]{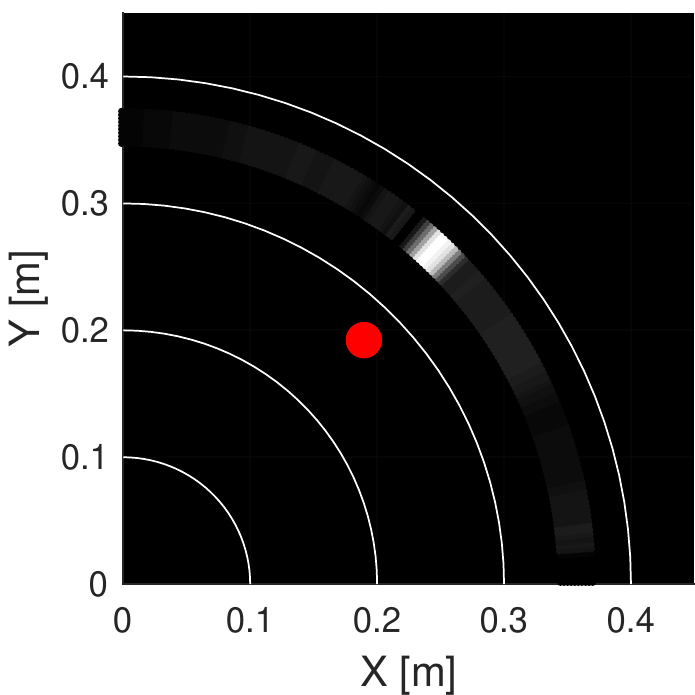}
      \captionsetup{justification=centering}
      \caption{\footnotesize ($\hat{\rhovec}_1$, $\hat{\bar{\svec}}_1$) }
      \label{fig:recon100}
    \end{subfigure}
    \begin{subfigure}{0.2\textwidth}
      \includegraphics[width=\linewidth]{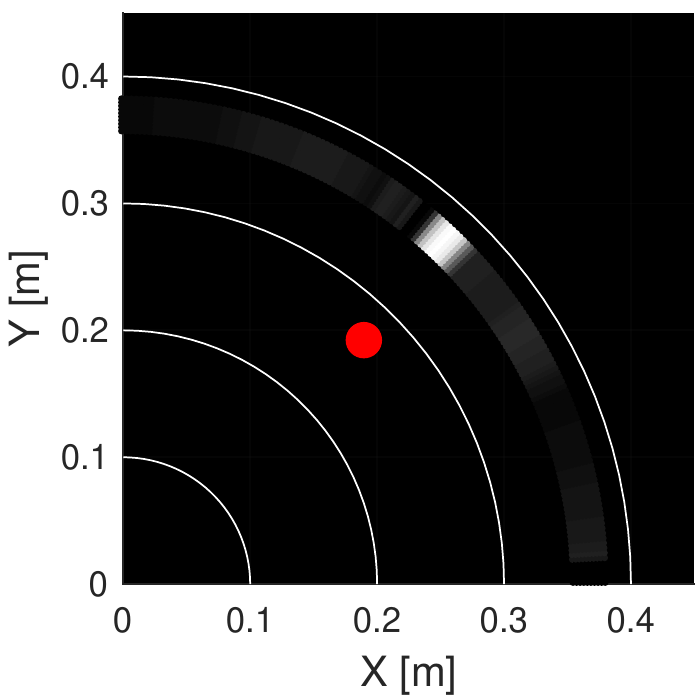}
      \captionsetup{justification=centering}
      \caption{\footnotesize  ($\hat{\rhovec}_2$, $\hat{\bar{\svec}}_2$)}
      \label{fig:recon50}
    \end{subfigure}
    \begin{subfigure}{0.2\textwidth}
      \includegraphics[width=\linewidth]{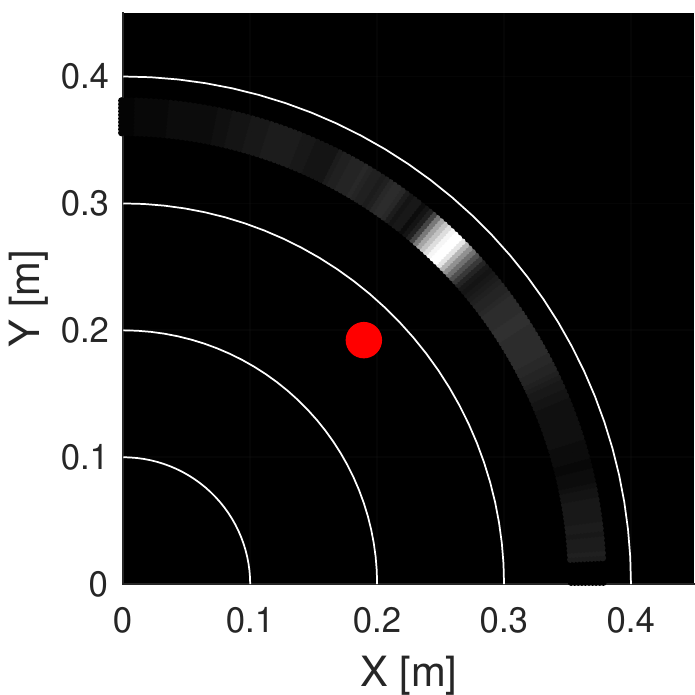}
      \captionsetup{justification=centering}
      \caption{\footnotesize ($\hat{\rhovec}_3$, $\hat{\bar{\svec}}_3$)}
      \label{fig:recon10}
    \end{subfigure}
    \begin{subfigure}{0.2\textwidth}
      \includegraphics[width=\linewidth]{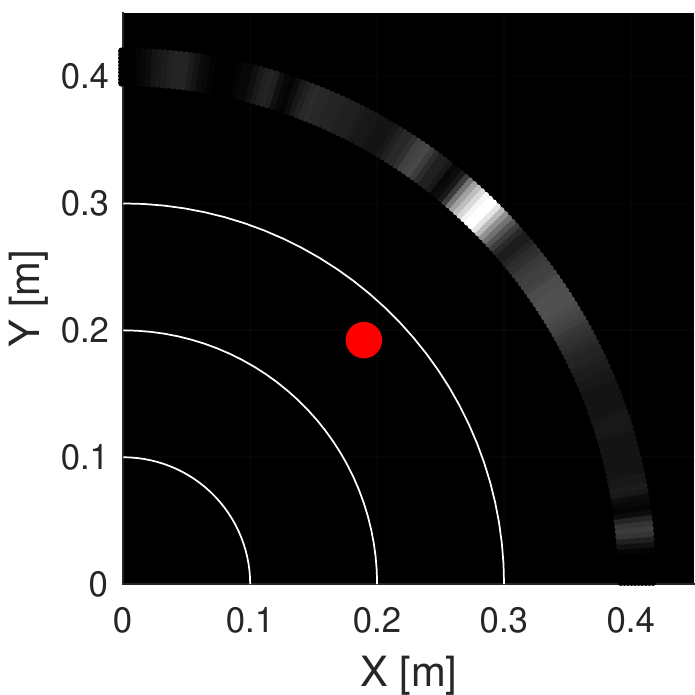}
      \captionsetup{justification=centering}
      \caption{\footnotesize ($\hat{\rhovec}_4$, $\hat{\bar{\svec}}_4$)}
      \label{fig:recon05}
    \end{subfigure}
        \caption{Demonstration of performance degradation as the penumbra becomes fainter. The true location of the hidden object, a white cylinder, is shown in red. When the penumbra is brighter, (a) and (e), the estimated range $\hat{\rhovec}$ is closer to the truth, and $\hat{\bar{\svec}}$ contains a sharp peak at the true angular location of the hidden object. When the penumbra is more faint, (d) and (h), the estimated range $\hat{\rhovec}$ is further from the truth and the peak in $\hat{\bar{\svec}}$ less sharp.        }
    \label{fig:VaryingAmbientLight}
\end{figure*}

\subsubsection{Varying Ambient Light}
Measurements were also taken of this same target in a fixed position, with different levels of ambient light.
A constant light source on the visible side introduced ambient light while a light source on the hidden side was tuned to vary penumbra brightness.
\Cref{fig:meas100}, \Cref{fig:meas50}, \Cref{fig:meas10}, and \Cref{fig:meas05} show measurements as the penumbra becomes faint to the point of not being visible to the naked eye;
\Cref{fig:recon100}, \Cref{fig:recon50}, \Cref{fig:recon10}, and \Cref{fig:recon05} show the corresponding reconstructions with the true target location marked by a red dot.
All four reconstructions correctly pick out the target in angle demonstrating robustness to a surprising amount of ambient light, although the higher SNR case is both sharper in angle and more accurate in range estimation.

\subsection{Color Reconstructions}
The RGB nonlinear inversion algorithm was tested on scenes with colored objects in several different configurations. Testing was also performed on a multi-object, colored scene in the presence of increasingly bright ambient light to demonstrate algorithm robustness to low SNR conditions.

\subsubsection{Multiple Targets}
\begin{figure}
    \centering
    \begin{subfigure}{0.53\linewidth}
      \includegraphics[width=\linewidth]{Figures/barcode/barcode_inset.png}
      \captionsetup{justification=centering}
      \caption{\footnotesize RGB color measurement $\yvec_1$ and true hidden scene.}
      \label{fig:ColorStripes_measurement}
    \end{subfigure}
    \begin{subfigure}{0.37\linewidth}
      \includegraphics[width=\linewidth]{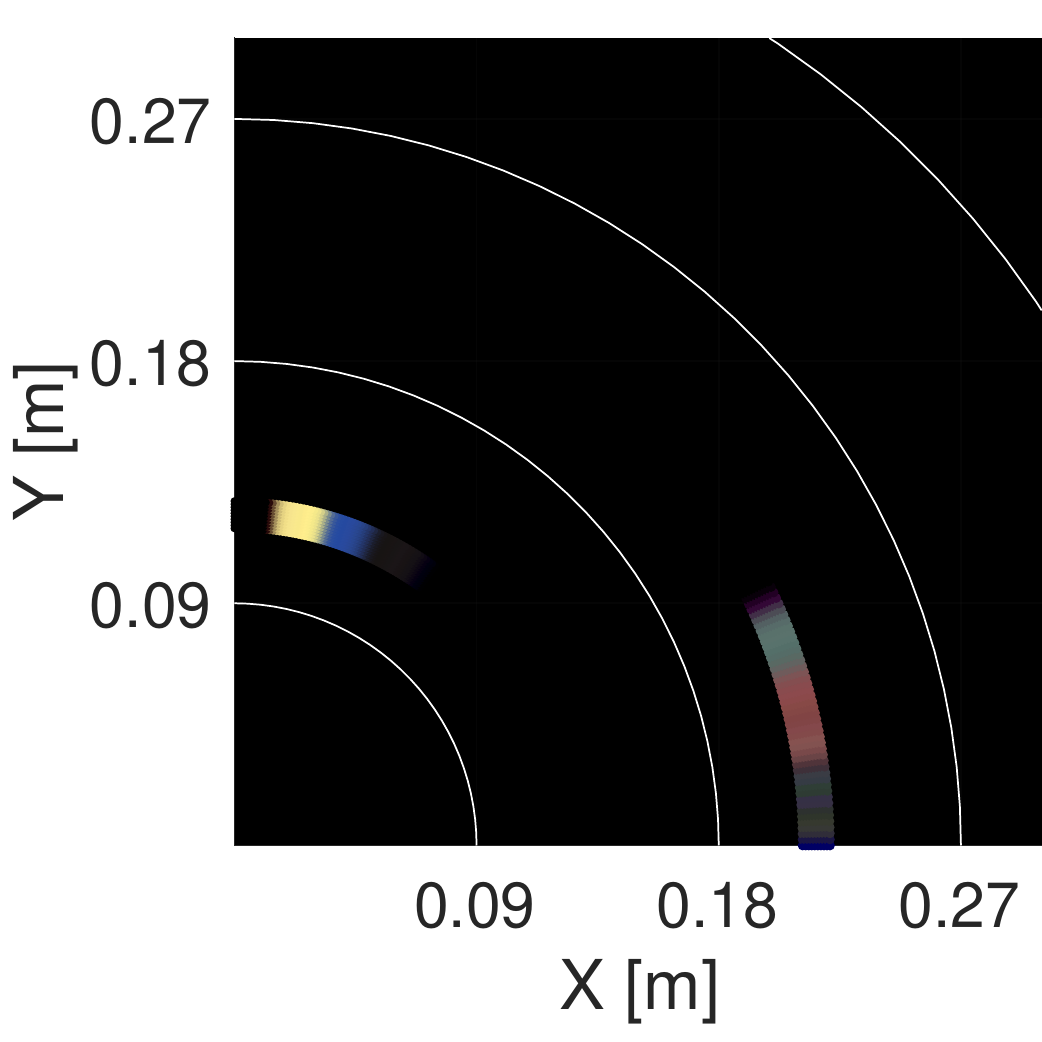}
      \captionsetup{justification=centering}
      \caption{\footnotesize Reconstruction ($\hat{\rhovec}_1$, $\hat{{\svec}}_1$) }
      \label{fig:ColorStripes_recon}
    \end{subfigure}\\
    \begin{subfigure}{0.53\linewidth}
      \includegraphics[width=\linewidth]{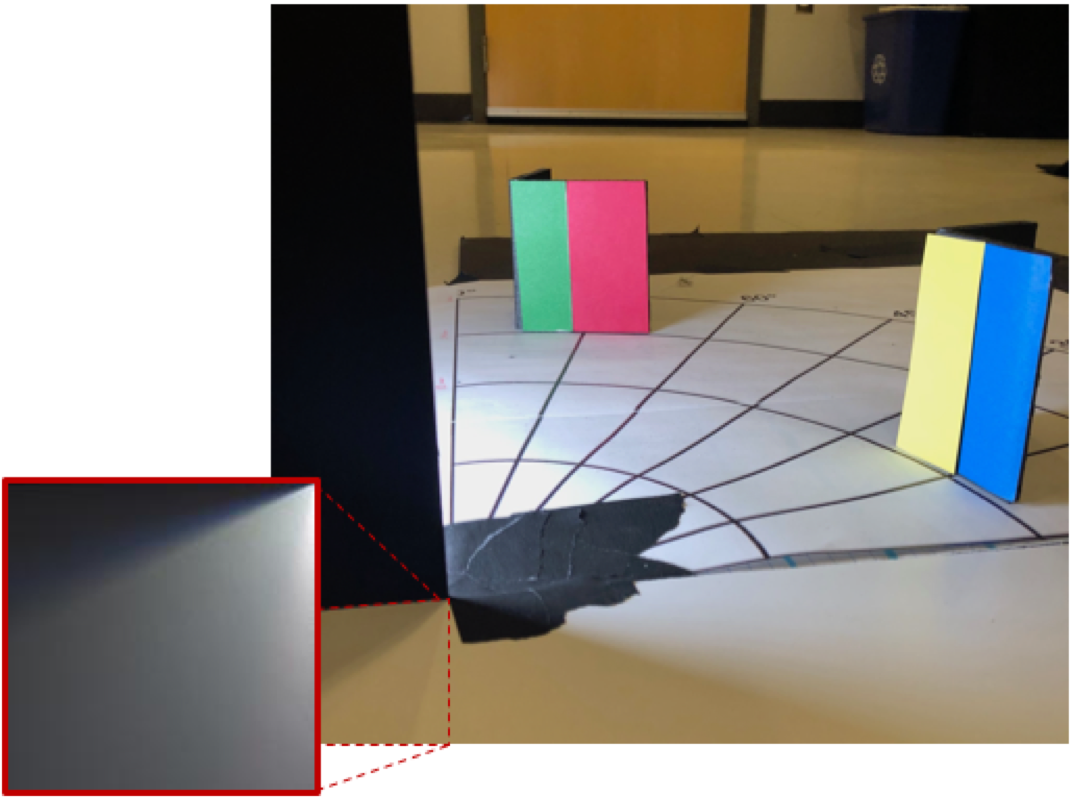}
      \captionsetup{justification=centering}
      \caption{\footnotesize RGB color measurement $\yvec_2$ and true hidden scene.}
      \label{fig:ReverseColorStripes_measurement}
    \end{subfigure}
    \begin{subfigure}{0.37\linewidth}
      \includegraphics[width=\linewidth]{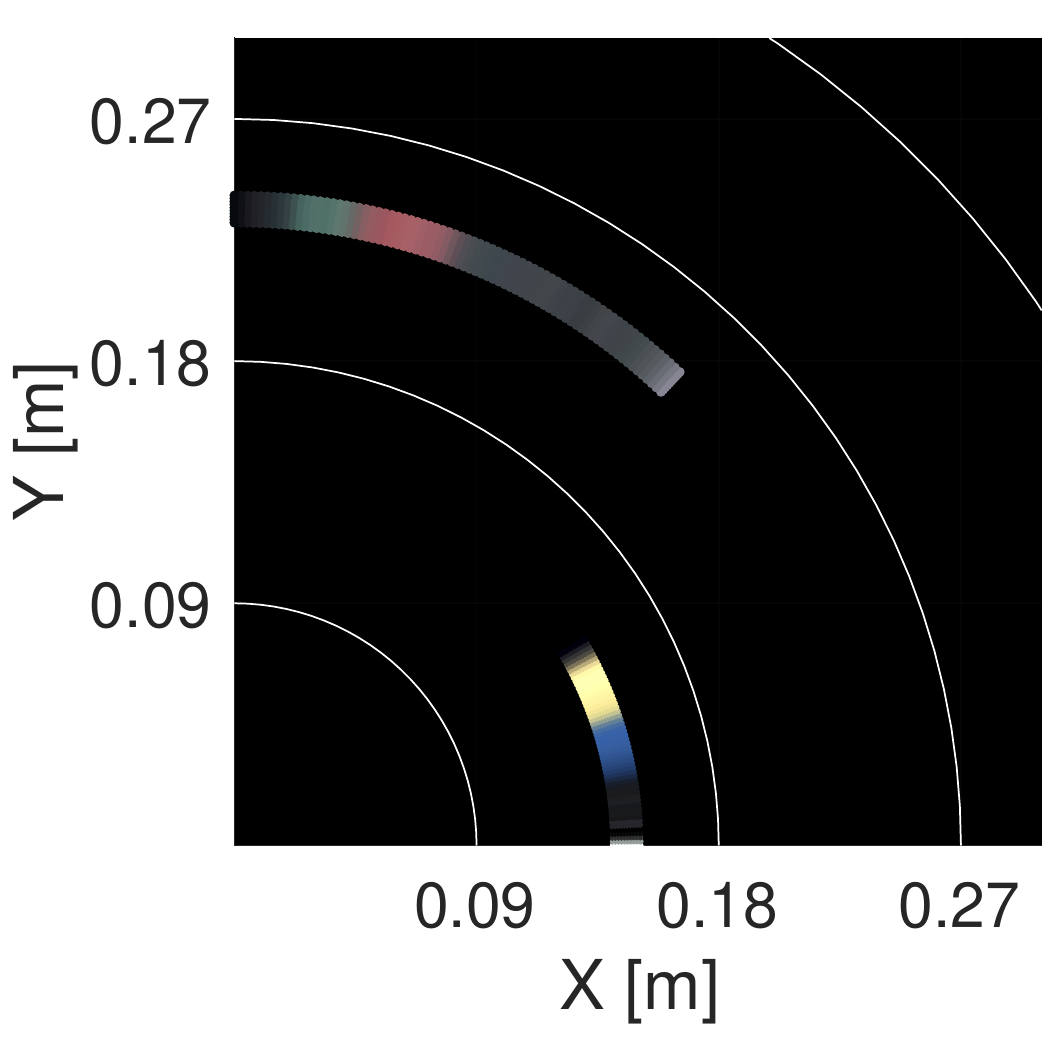}
      \captionsetup{justification=centering}
      \caption{\footnotesize Reconstruction ($\hat{\rhovec}_2$, $\hat{{\svec}}_2$) }
      \label{fig:ReverseColorStripes_recon}
    \end{subfigure}\\
    \begin{subfigure}{0.53\linewidth}
      \includegraphics[width=\linewidth]{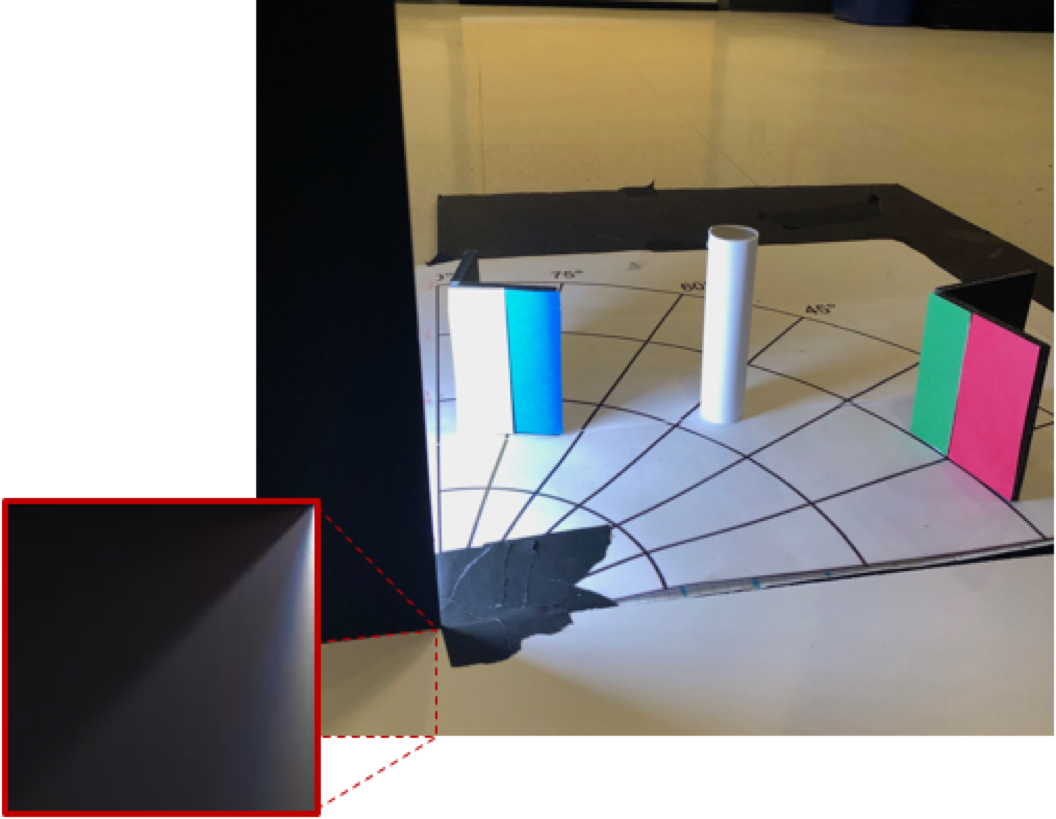}
      \captionsetup{justification=centering}
      \caption{\footnotesize RGB color measurement $\yvec_3$ and true hidden scene.}
      \label{fig:threeTarg_measurement}
    \end{subfigure}
    \begin{subfigure}{0.37\linewidth}
      \includegraphics[width=\linewidth]{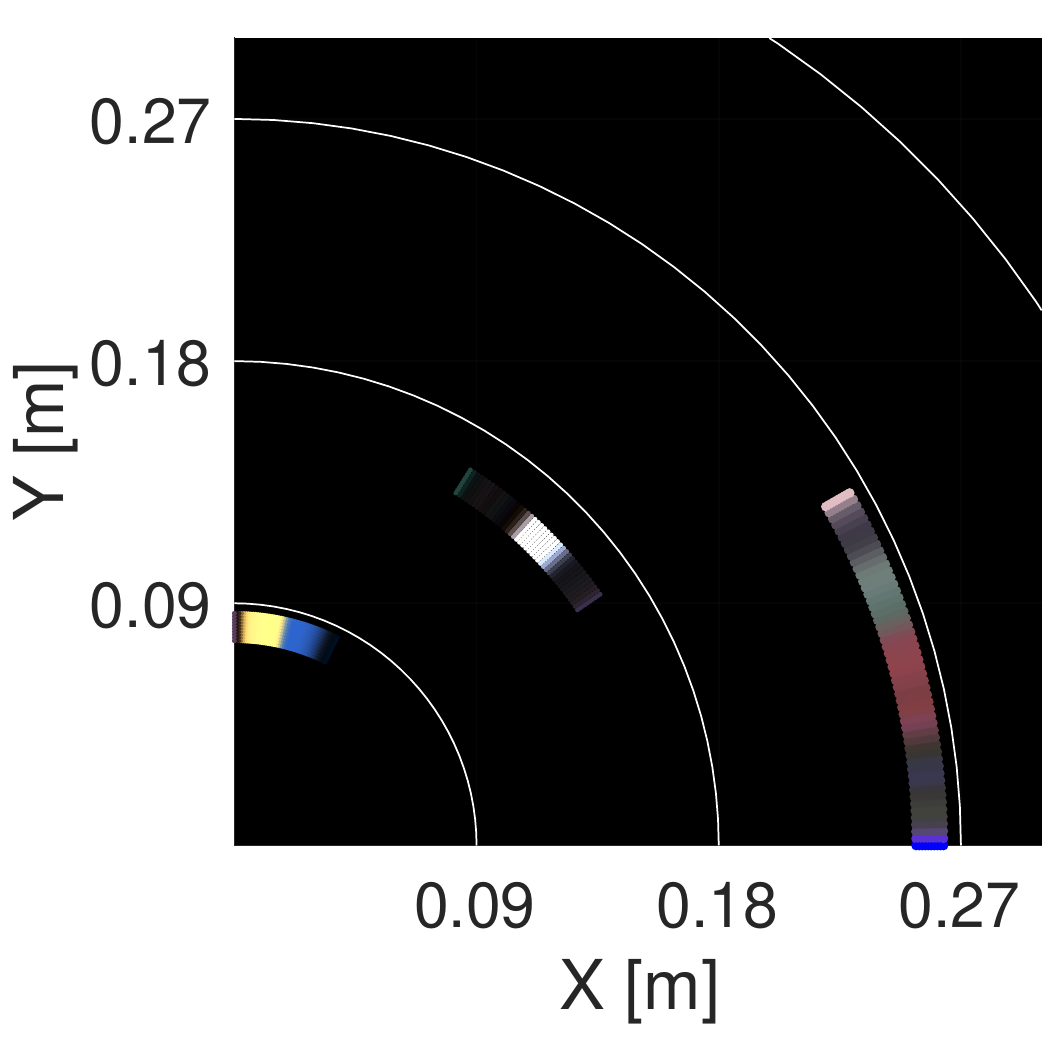}
      \captionsetup{justification=centering}
      \caption{\footnotesize Reconstruction ($\hat{\rhovec}_3$, $\hat{{\svec}}_3$) }
      \label{fig:threeTarg_recon}
    \end{subfigure}
    \caption{Demonstration of color reconstruction with three different scenes containing multiple hidden objects. The widths of yellow, blue, white, green, and red objects are $2.9~{\rm cm}$, $2.7~{\rm cm}$, $2.5~{\rm cm}$, $2.9~{\rm cm}$, and $4.2~{\rm cm}$ respectively.
    The black arcs on the floor in
    (a), (c), and (e)
    correspond to the ranges marked in white in
    (b), (d), and (f).
    }
    \label{fig:ColorStripes}
\end{figure}

Measurement and hidden scene pairs are shown in \Cref{fig:ColorStripes_measurement}  and \Cref{fig:ReverseColorStripes_measurement}, where the same two colored objects have been placed in reverse positions. Reconstructions \Cref{fig:ColorStripes_recon} and \Cref{fig:ReverseColorStripes_recon} show that in both scenarios, the two targets are accurately found in angle, and placed in range correctly with respect to each other. High angular resolution is demonstrated in both reconstructions, with the red-green object correctly portrayed to have a slightly wider red section, just like the yellow-blue object has a slightly wider yellow section. The scenario and measurement shown in \Cref{fig:threeTarg_measurement} tests our algorithm on a scene that includes three targets instead of two. Still, the reconstruction shown in \Cref{fig:threeTarg_recon} accurately picks out all three targets in angle and places them at ranges that are correct with respect to each other.

\subsubsection{Varying Ambient Light}
\begin{figure*}
    \centering
     \hspace{33mm}%
    \begin{subfigure}{0.14\textwidth}
      \includegraphics[width=\linewidth]{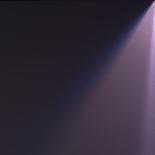}
      \captionsetup{justification=centering}
      \caption{\footnotesize $\yvec_1$}
      \label{fig:meas100Color}
    \end{subfigure} \hspace{9mm}%
    \begin{subfigure}{0.14\textwidth}
      \includegraphics[width=\linewidth]{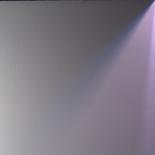}
      \captionsetup{justification=centering}
      \caption{\footnotesize $\yvec_2$}
      \label{fig:meas10Color}
    \end{subfigure}  \hspace{9mm}%
    \begin{subfigure}{0.14\textwidth}
      \includegraphics[width=\linewidth]{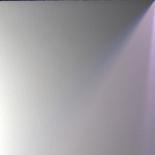}
      \captionsetup{justification=centering}
      \caption{\footnotesize $\yvec_3$}
      \label{fig:meas5Color}
    \end{subfigure}   \hspace{9mm}%
    \begin{subfigure}{0.14\textwidth}
      \includegraphics[width=\linewidth]{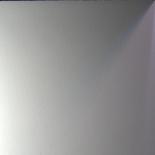}
      \captionsetup{justification=centering}
      \caption{\footnotesize $\yvec_4$}
      \label{fig:meas2_5Color}
    \end{subfigure}\\
        \begin{subfigure}{0.14\textwidth}
      \includegraphics[width=\linewidth]{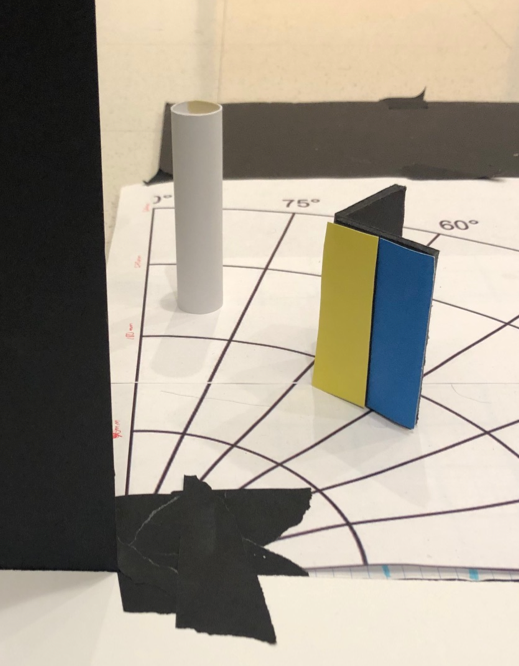}
      \captionsetup{justification=centering}
      \caption{\footnotesize Line of sight photograph of hidden scene }
      \label{fig:colorAmbientSetup}
    \end{subfigure}
    \begin{subfigure}{0.19\textwidth}
      \includegraphics[width=\linewidth]{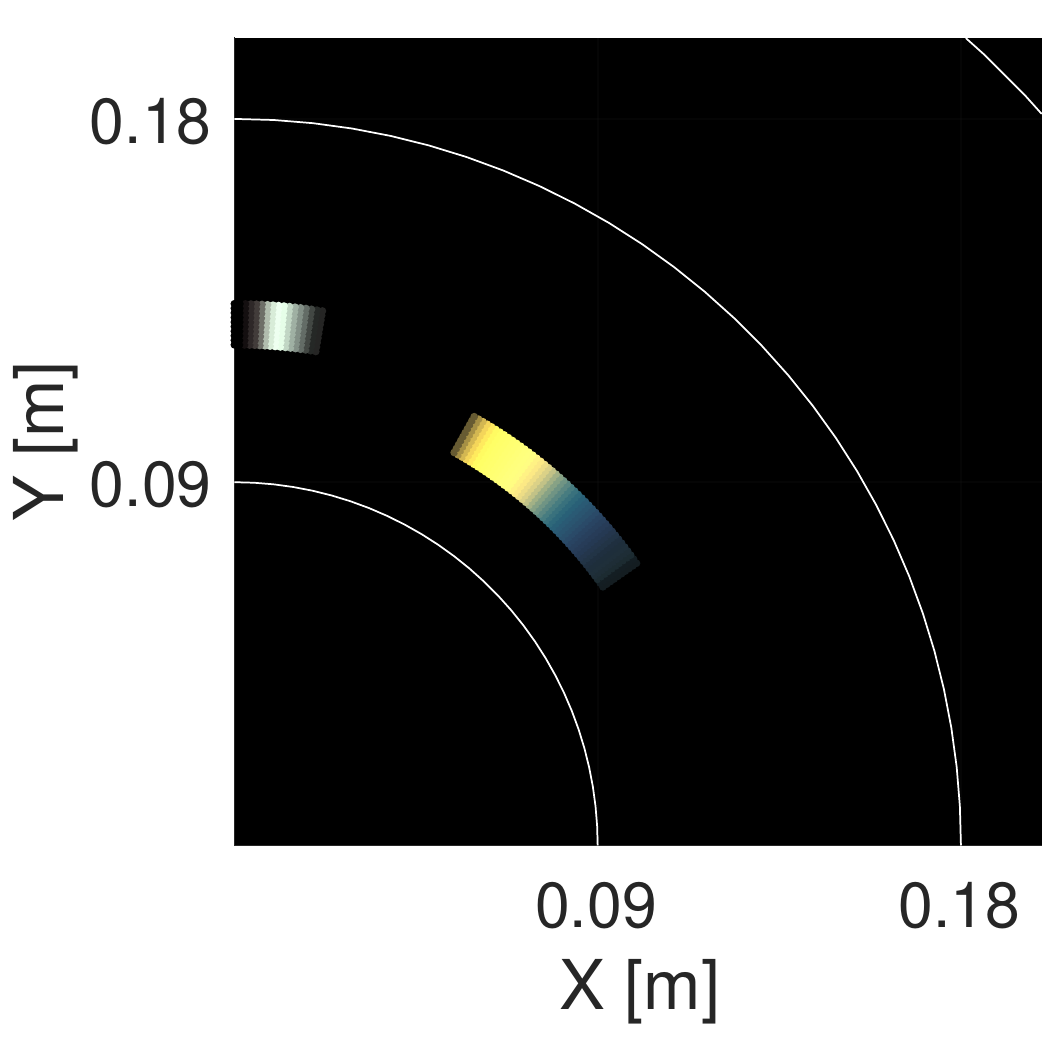}
      \captionsetup{justification=centering}
      \caption{\footnotesize ($\hat{\rhovec}_1$, $\hat{\bar{\svec}}_1$) }
      \label{fig:recon100Color}
    \end{subfigure}
    \begin{subfigure}{0.19\textwidth}
      \includegraphics[width=\linewidth]{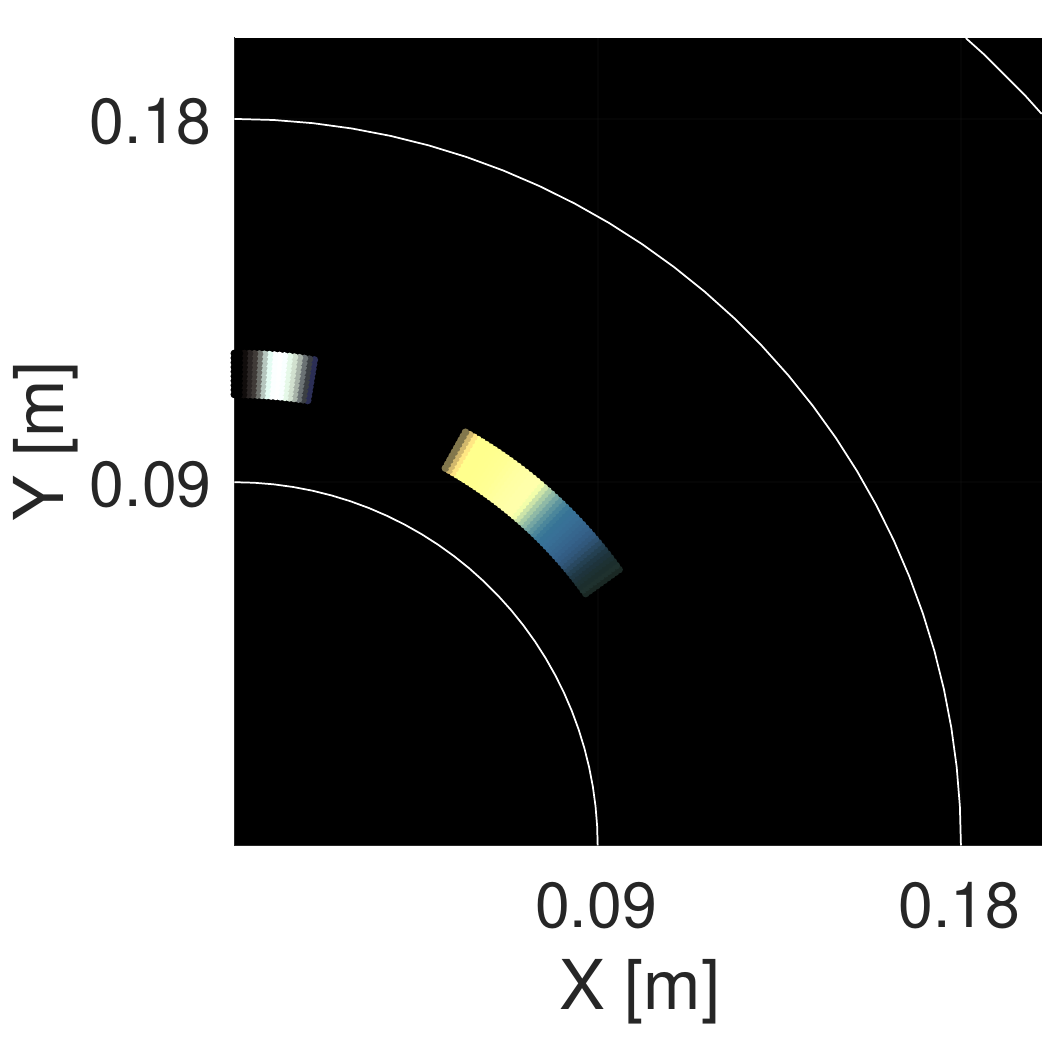}
      \captionsetup{justification=centering}
      \caption{\footnotesize  ($\hat{\rhovec}_2$, $\hat{\bar{\svec}}_2$)}
      \label{fig:recon10Color}
    \end{subfigure}
    \begin{subfigure}{0.19\textwidth}
      \includegraphics[width=\linewidth]{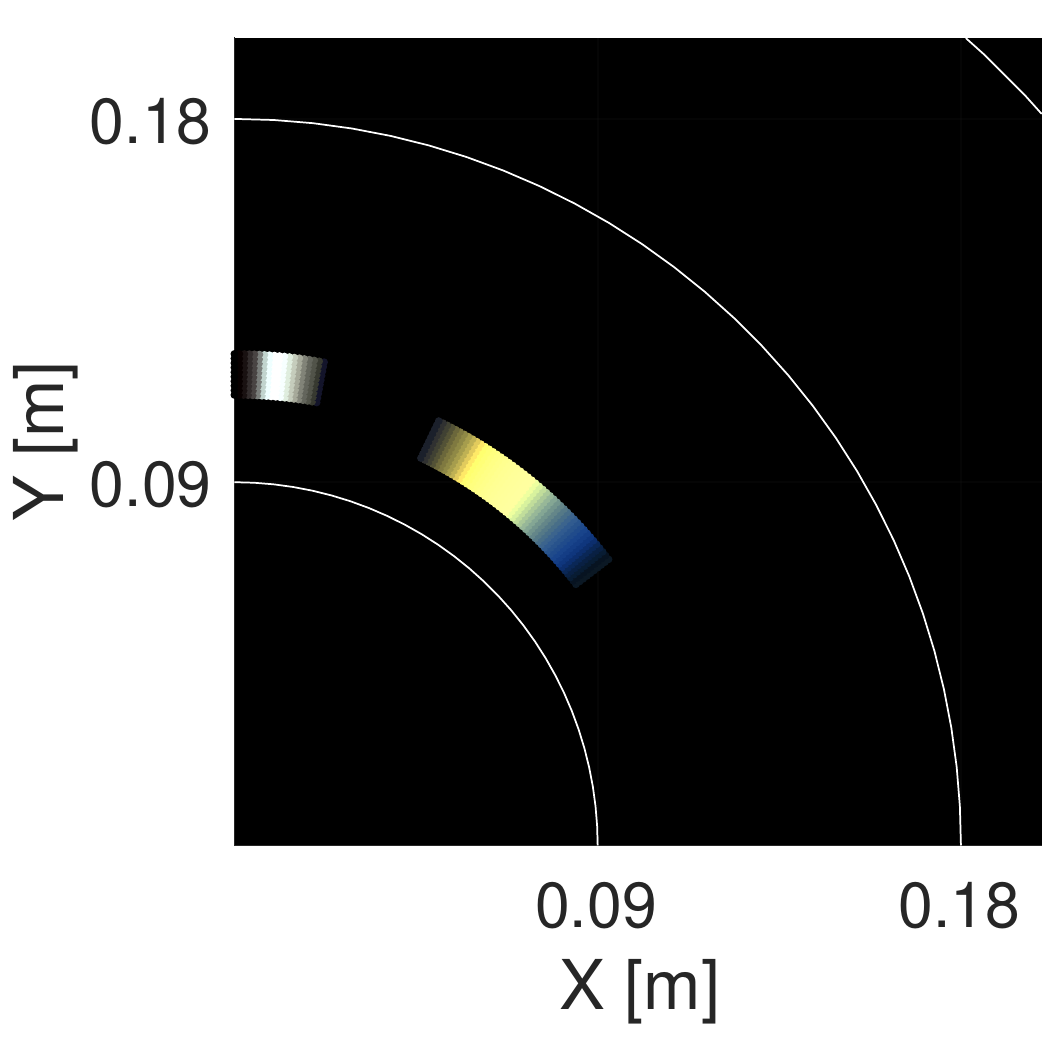}
      \captionsetup{justification=centering}
      \caption{\footnotesize ($\hat{\rhovec}_3$, $\hat{\bar{\svec}}_3$)}
      \label{fig:recon5Color}
    \end{subfigure}
    \begin{subfigure}{0.19\textwidth}
      \includegraphics[width=\linewidth]{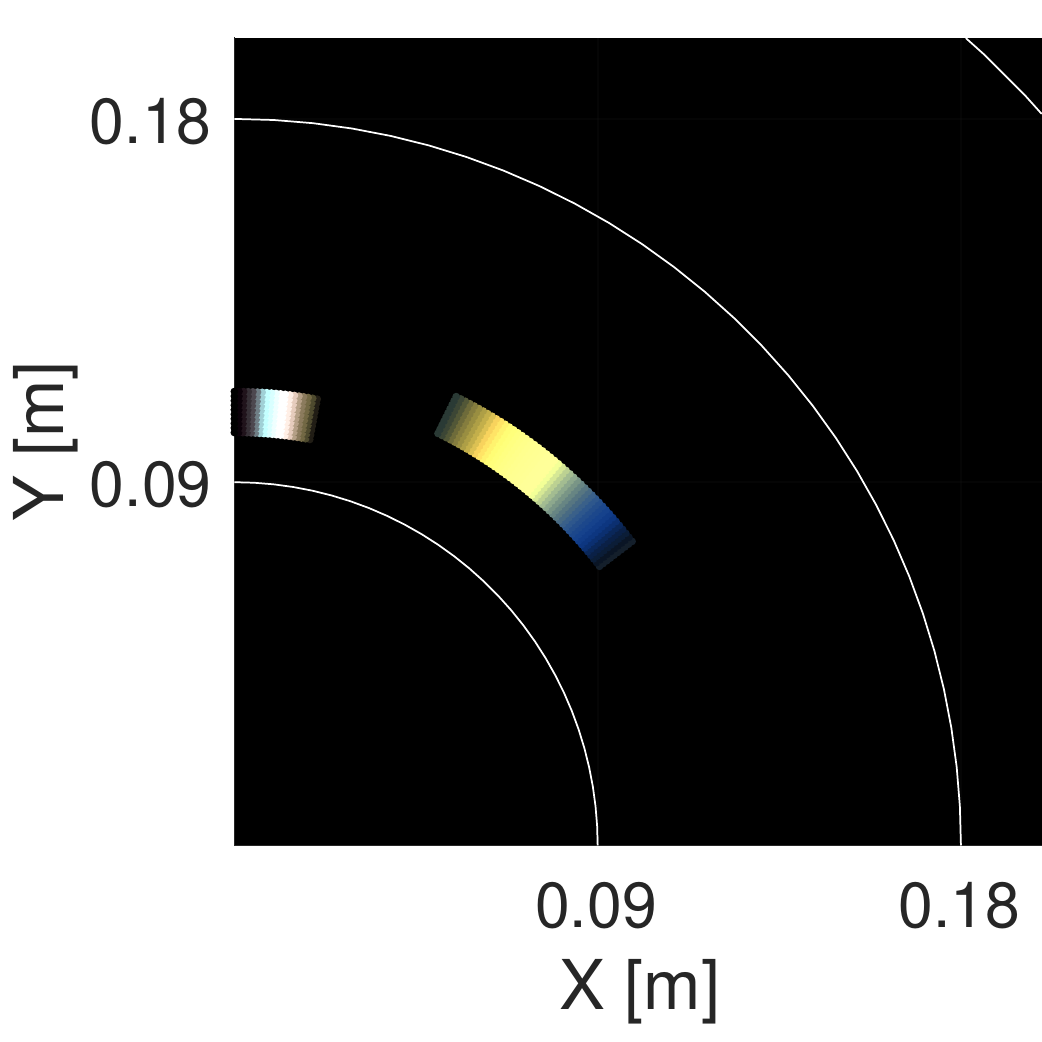}
      \captionsetup{justification=centering}
      \caption{\footnotesize ($\hat{\rhovec}_4$, $\hat{\bar{\svec}}_4$)}
      \label{fig:recon2_5Color}
    \end{subfigure}
        \caption{Demonstration of RGB alternating inversion algorithm as SNR is reduced. In this scenario, the white cylinder (width $0.025~{\rm m}$) and yellow-blue stripes (widths $0.029~{\rm m}$ and $0.027~{\rm m}$, respectively) were arranged as shown in (e), at $0.21~{\rm m}$ and $0.16~{\rm m}$ respectively. Note the increased difficulty due to the white cylinder's placement so deep, in angle, into the scene. In this location, only a small fraction of the measurement pixels are exposed to light from the cylinder. Still, only the lowest SNR reconstruction fails to place the white cylinder at a greater range than the yellow-blue object. In all cases, the two objects are resolved in great angular detail. The black arcs on the floor correspond to the ranges marked in white on the reconstructions.
        }
    \label{fig:VaryingAmbientLightColor}
\end{figure*}

In \Cref{fig:VaryingAmbientLightColor}, we demonstrate algorithm robustness to increasing amounts of ambient light. Here, the hidden scene, arranged as shown in \Cref{fig:colorAmbientSetup}, consists of the yellow-blue target and a more distant, in both range and angle, white cylinder. With an angular location close to $\Frac{\pi}{2}$, very few pixels in the measurement are exposed to light from the white cylinder making range estimation more challenging. Still, all but the lowest SNR reconstructions correctly place the white cylinder at a greater range than the yellow-blue target. All four reconstructions demonstrate high angular resolution, even resolving the sharp boundary between the yellow and blue portions of the yellow-blue target. 

\section{Discussion}
\label{sec:Disc}

We proposed and tested two inversion algorithms: one based on a more conventional linear model and the other on a more constrained, alternating approach that more directly inverts the nonlinear forward model \eqref{eq:discreteforwardmodel_nonlinear}.
Both make use of regularization to solve an ill-conditioned problem and demonstrate high angular resolution and significantly coarser range resolution in reconstruction results, owing to the conditioning of range estimation. While the linear model \eqref{eq:discreteforwardmodel_linear} enjoys simplicity, it omits the opacity assumption that is naturally embedded in the nonlinear model \eqref{eq:discreteforwardmodel_nonlinear}, thus allowing multiple nonzero pixels in a single angle. In addition, without enforcing a single range per target, this method also raises questions about how to effectively promote coincident pixels across the colour channels. The second (nonlinear) method benefits from the natural separation of range and angle estimation problems, enabling a highly effective alternating recovery algorithm.

The linear approach discretizes the hidden scene into a polar grid.
Even with fine angular discretization and sparsity-enforcing priors, estimating a range per angle when a hidden scene likely contains only a few targets is unnecessarily challenging.
In contrast, the initialization step of the alternating algorithm that we propose for solving the nonlinear problem \eqref{eq:discreteforwardmodel_nonlinear} allows us to exploit excellent angular resolution to count the number of targets and estimate only one range per detected hidden target.
In this light, the range update step can be interpreted as \textit{learning} the forward model to ultimately allow for better angular reconstructions, as demonstrated in \Cref{fig:singleTarget_Example}.
With the few unknown ranges as \textit{parameters}, the alternating approach enjoys the potential for less model mismatch than the linear inversion algorithm, because the range parameters are not discretized.

The fact that the alternating algorithm treats unknown ranges as parameters also lends itself to a natural three-channel RGB extension.
Estimating a single range per fixed angular extent enforces consensus across color channels.
In contrast, the linear inversion algorithm operates separately on three color channels and may place RGB values for the same object at different ranges or angles, as shown in \Cref{fig:LinearInversionRecons}.
One way this may ultimately be improved is by forming a reconstruction in the YUV color space.
Enforcing sparsity on component Y (i.e., the `luma', or `intensity' component) would penalize intensities at multiple ranges in the same angular bin.

Experimental results presented in \Cref{sec:experimental_results} were obtained using an experimental setup with target distances on the order of half a meter, but we believe both inversion algorithms could work with a larger experimental setup, given comparable SNR and larger camera FOV\@. CRB analysis may be extended to determine the effect of camera FOV on estimate variance for targets at a given range.
We conjecture that for a given target range, there may be an optimal camera FOV for range recovery, although generally speaking a larger camera FOV makes angular estimation more challenging.
In the alternating algorithm, this trade-off may be managed by taking one photo with a smaller FOV to use in scene $\svec$ initialization and update steps, \eqref{eq:alternating_initialization} and \eqref{eq:s_update}, and a larger FOV photo for range estimation step, \eqref{eq:rho_update}.
Though at the scale of the experiments in this paper, this was not necessary.

The alternating algorithm may be further adapted to handle the common scenario of a few hidden objects with heights known \textit{a priori}.
Imagine a scene composed of people or cars that we observed entering the hidden scene from the visible side.
If the heights of hidden objects are known and we assume a constant radiosity across height for a fixed angle, we may write the radiosity of the hidden scene as
$\Sh(\rho, \alpha, z) = \Sh(\rho,\alpha)u(\Frac{z}{\eta(\rho,\alpha)})$,
where $\Sh(\rho,\alpha)$ is the radiosity per unit height and $\eta(\rho,\alpha)$ is the known height at hidden scene point $(\rho,\alpha)$.
Now, instead of recovering $\Shbar(\rho,\alpha)$ of the hidden scene, we seek to recover radiosity per unit height $\Sh(\rho, \alpha)$.

With this additional information, our expression for incident light on the floor (originating from the hidden scene), \eqref{eq:Lh_cylindrical}, may be rewritten:
\begin{align}
    L_{\rm h}(r,\theta) &= \int_{\Frac{\pi}{2}}^{\Frac{\pi}{2}{+}\theta} \!\!\! \int_{0}^{\infty} \!\!\! \int_{0}^{\eta(\rho,\alpha)} \!
                            \frac{\Sh(\rho, \alpha)}{d^2 + z^2} \rho  \, \mathrm{d}z \, \mathrm{d}\rho \, \mathrm{d}\alpha
                            \\
                            &=
                           \int_{\Frac{\pi}{2}}^{\Frac{\pi}{2}{+}\theta} \!\!\! \int_{0}^{\infty} 
                            \frac{\Sh(\rho, \alpha){\rm arctan} \!\left(\frac{\eta(\rho,\alpha)}{d}\right)}{d} \rho \, \mathrm{d}z \, \mathrm{d}\rho \, \mathrm{d}\alpha.
    \label{eq:Lh_knownHeight}
\end{align}
In the alternating method for inverting our nonlinear model, using a vector of known target heights
${\bar{\bm{\eta}}} = [\bar{\eta}_1, \bar{\eta}_2, \ldots, \bar{\eta}_{N_{\rm t}} ]\in\mathbb{R}^{N_{\rm t}}$
to recover $S_{\rm h}$ instead of $\Shbar (\rho, \alpha)$ can be appropriately incorporated by replacing \eqref{eq:dmat_form} with
\begin{equation}
[\Dmat(\rhobarbm)]_{m,n} = \! \frac{\bar{\rho}_j{\rm arctan}\bigg(\frac{\bar{\eta}_j}{d(r_m,\theta_m,\bar{\rho}_j,\alpha_n)}\bigg)}{d(r_m,\theta_m,\bar{\rho}_j,\alpha_n)}
\label{eq:dmat_form_knownHeight}
\end{equation}
when
$\alpha_n \in \left[\bar{\alpha}_j - \Frac{\delta_{\bar{\alpha}_j}}{2}, \bar{\alpha}_j + \Frac{\delta_{\bar{\alpha}_j}}{2}\right)$.

\section{Conclusion}
\label{sec:Conc}
In this work we explore 2D reconstruction of the region hidden behind a wall using a single photograph of the floor on the visible side.
Unlike previous work, which has assumed all light sources to be in the far field, we propose a more complete forward model to describe radial falloff, enabling 2D reconstructions of the hidden scene.
Using the \CR~bound for a single target, we demonstrate the utility and difficulty of using penumbra measurements for 2D reconstruction.
We propose an alternating nonlinear inversion algorithm for 2D reconstruction and provide a comparison to a more conventional linear inversion algorithm.
Experimental results demonstrate the promise and robustness of both methods.

\section*{Acknowledgment}
The authors thank Charles Saunders for his help with data acquisition. Computing resources provided by Boston University Research Computing Service and support provided by the Draper Fellowship program are greatly appreciated.

\bibliographystyle{IEEEtran.bst}
\bibliography{./bibs/bib_nloslitreview.bib} 

\newpage

\end{document}